\documentclass{nottsthesis}

\usepackage{ucs}
\usepackage[utf8x]{inputenc}
\usepackage[T1]{fontenc}
\usepackage{amsmath}
\usepackage{amssymb}
\usepackage[english]{babel}
\usepackage{lipsum}
\usepackage{epigraph}
\usepackage{hyperref}
\usepackage{cleveref}
\usepackage{titlesec}
\usepackage{subcaption}
\usepackage{graphicx}
\usepackage{float}
\usepackage{textcomp}
\usepackage{commath}
\usepackage{tikz}
\usepackage{afterpage}

\renewcommand{\vec}[1]{\boldsymbol{#1}}
\newcommand{\D}{\mathcal{D}_t}
\newcommand{\p}{\partial}
\newcommand{\vnab}{\vec{\nabla}}
\renewcommand{\eval}[2]{\left. {#1} \right|_{#2}}
\newcommand{\be}{\begin{equation}}
\newcommand{\ee}{\end{equation}}
\newcommand{\bsub}{\begin{subequations}}
\newcommand{\esub}{\end{subequations}}
\newcommand{\bea}{\begin{eqnarray}}
\newcommand{\eea}{\end{eqnarray}}
\newcommand{\atan}{\mathrm{atan}}

\numberwithin{equation}{section}

\thesisauthor{Theo Torres}
\thesistitle{Hydrodynamic simulations of rotating black holes}
\degree{Doctor of Philosophy}
\degreedate{August 2019}
\school{Mathematical Sciences}

\begin{document}

\maketitle

\afterpage{\null\newpage}
\thispagestyle{empty}

\frontmatter
\begin{Abstract}

Wave scattering phenomena are ubiquitous in almost all Sciences, from Biology to Physics.
Interestingly, it has been shown many times that different physical systems are the stage to the same
processes. 
The discoveries of such analogies have resulted in a better understanding of Physics and
are indications of the universality of Nature. 
One stunning example is the observation that waves propagating on a flowing fluid effectively experience the presence of a curved space-time.

In this thesis we will use this analogy to investigate, both theoretically and experimentally, fundamental effects occurring around vortex flows and rotating black holes.
In particular, we will focus on light-bending, superradiance scattering, and quasi-normal modes emission.
The experimental nature of this work will lead us to study these processes in the presence of dispersive effects.

After a general and historical discussion of the field of analogue gravity, we will first present a well-established technique, the gradient expansion method, to obtain approximate solutions of dynamical equations. 
This method will be used to generalise the notion of light-rings around black holes to vortex flows.
Secondly, we will present a wave-vortex scattering experiment in which the superradiance process was observed. 
Finally, we will relate the properties of the light-rings to the characteristic modes emitted during the relaxation phase of a perturbed vortex flow. We will show that these characteristic modes can be used to develop a flow measurement technique that we call `Analogue Black Hole Spectroscopy'. We will then report on an experiment in which these characteristic modes were observed and the analogue black hole spectroscopy technique was applied successfully.

Our results strengthen the link between vortices and rotating black holes and open up new challenges to be addressed in the future.

\end{Abstract}

\afterpage{\null\newpage}

\begin{Acknowledgements}

First of all, I would like to thank my PhD supervisor, Silke Weinfurtner, for her support, guidance and trust during the years I spent working on this thesis. 
Her curiosity and broad interest for general physical processes associated to her enthusiasm, optimism, and determination created a very stimulating and exciting environment to be working in.
I am also extremely grateful for her genuine care. Her presence was very helpful and reassuring.
 
I would also like to thank her for generating very fruitful collaborations and allowing me to work with incredible people. Maur\'{i}cio Richartz, Sam Dolan, Tasos Avgoustidis, and Richard Hill, it was a real pleasure to work with you and I can only hope to continue doing so in the future. Maur\'{i}cio, thank you for struggling with us on the various experiments.
\\

I am also indebted to the administrative and technical staff of the University of Nottingham. In particular, I want to thank Tommy Napier, Andrew Stuart, Ian Taylor, and Terry Wright for their creativity, technical skills and patience to deal with our crazy ideas. Without them, this work could not have been done.
\\

Working in the Black Hole Laboratory has been a real pleasure, not only because of the physics I was studying but also because of the people I was sharing it with. 
I would like to thank my friends and colleagues: Antonin Coutant, Sam Patrick, Zack Fifer, Sebastian Erne, Cisco Gooding, August Geelmuyden, and Steffen Biermann. 
I have re-experienced the joy of being amazed by physics with them and I have learned a lot scientifically and personally by spending time with them. 
I would specifically like to thank Sam Patrick, the flowmaster, for his strong heart and for being as silly as I can be. 
I can't count the number of nonsensical moments we had together but I know that my PhD experience would have been completely different without them.
I will truly miss working with him and I wish him the best of luck in his quest for the QBS.
I would also like to thank Antonin Coutant for his support during the first stages of my PhD. 
His scientific intuitions, love for WKB, and broad knowledge of physics truly inspired me. 
I am greatly thankful to him for the many discussions we had, as friends and colleagues.
\\

I cannot stress enough how much I owe to my family, Muriel, Aymeric, Aur\'{e}lie and Michel. 
Thank you for always encouraging me to do what I like, for being proud of me and for trusting me.
\\

Finally, I want to thank my beloved Mathilde, for her affection and support, from near and far, and for always inspiring me to be a better person.

\end{Acknowledgements}

\afterpage{\null\newpage}

\tableofcontents

\mainmatter

\chapter{Introduction}\label{Intro_sec}
\epigraph{\textit{It is probably true quite generally that in the history of human thinking the most fruitful developments frequently take place at those points where two different lines of thought meet.  
[...] (I)f they are at least so much related to each other that a real interaction can take place, then one may hope that new and interesting developments will follow.}}{Heisenberg}

Before entering into the heart of the subject of this thesis, which will be the study of universal phenomena occurring around black holes and vortices, I would like to dedicate some time to taking a step back and looking at the bigger picture. 
While later we will focus on the technical machinery that science requires, I believe that it is worth taking a small detour to think about what we are doing, how we are doing it and why we are doing it this way. 

Physics is a subtle discipline in which one is, in the words of Feynman, "guessing laws" to describe Nature~\cite{feynman2017character}.  
The process of Physics,~i.e. what one does when doing physics, might go as follow: guess a law - compute its consequences - compare it with Nature. 
The word "guess", here, means that we cannot be certain that a new law is correct before comparing it with Nature.
Even though there is this element of uncertainty, the guess is usually an educated one.
The physical laws describe the world by constructing structures and telling how these structures interact with each other. For example, structures could be particles and forces and Newton’s law would tell us how these two different structures interact and connect.
These interactions create a pattern between the structures.
Since, according to Steen, "mathematics is the science of patterns"~\cite{STEEN611}, the physical laws are currently expressed using mathematics, and the two disciplines are very much connected. 
However, one should not forget the differences that exist between the two, which mostly lie in the guessing and testing parts of the process. 
Guessing a new law is a very \textit{irrational} process, as Einstein underlines: ``There is no logical path leading to these [...] laws. They can only be reached by intuition, based upon something like an intellectual love of the objects of experience"~\cite{einstein2011world}. 
We see that the process of seeking a new law (which is referred by Einstein as "The supreme task of the physicist") is not only different from a mathematical reasoning but it is also closely linked to experiments. 
Now the question is: are there still new laws to guess? 
This is a difficult question and the answer may vary depending on what you want the law for. 

Our current understanding of the world is largely based on the theory of General Relativity (GR) and Quantum Theory (QT). 
On the one hand, GR successfully describes phenomena on the macroscopic scale, from the historic perihelion of Mercury~\cite{Einstein} to gravitational waves~\cite{gwaves} and even at the sub-millimetre scale~\cite{Hoyle}.
On the other hand, QT has been tested numerous times at the microscopic scale through the measurement of the fine-structure constant~\cite{Mittleman1993} or the observation of the Higgs boson~\cite{Higgs}. 
Quantum mechanical behaviours have also been observed at the millimetre scale~\cite{Friedman}. 
So far it seems that no experiment or observation contradicts these two pillars of modern physics. 
Therefore, from a purely practical point of view, there is no need for a new law to describe the current observations. 
It is known, however, that GR and QT cannot be combined into a single theory of Quantum Gravity (QG) and therefore they cannot be both correct in their domain of applicability. 
The gravitational effect might be negligible between two electrons, it is nonetheless still present and it is therefore reasonable to believe in the existence of a theory combining QT and GR~\cite{Kiefer2007,Carlip}.
Moreover, even though no experimental observation begs the need for a quantum theory of gravity, such theory is hoped will solve some long-standing problems in systems where QT and GR are both relevant, such as the Big Bang. A natural question is then, how do we guess a law? Where is this ``intellectual love" coming from?

In order to find a new law, one needs to challenge the known concepts as well as the range of applicability of current ideas in order to make progress. 
Throughout history, this has always been done by means of experiments. 
An example of such process is the development of QT. 
At the time, physicists tried to apply their solid concept of position to particles, but they faced paradoxes coming from experimental observations. 
And, in the word of Heinsenberg: ``It was from this time on that the physicists learned to ask the right questions; and asking the right question is frequently more than halfway to the solution of the problem"~\cite{Heisenberg}. 
This revaluation of the old concept of position led to the formulation of QT and drastically changed our view of Nature. 
We therefore see that a possible way to gain knowledge about the world is to challenge our ideas by exploring extreme systems.
 
One such system is a Black Hole (BH). BHs are regions in space-time where the gravitational pull is so strong that nothing can escape from it. 
A black hole, from a mathematical viewpoint, exhibits at least one \textit{horizon} and one \textit{singularity}.
The horizon is the limiting surface, after which no object that crosses it will be able to leave the BH. 
The singularity is a point in space-time where all matter that crosses the horizon ends up.
These objects have been speculated more than a century ago~\cite{Schwarzschild}. While at the beginning their existence was very controversial, the accumulation of experimental evidence which culminated with the recent observation made by the Event Horizon Telescope~\cite{EHT}, definitely established BHs as part of our universe if one believes in GR.
We see from the definition of a BH, that it might be a candidate to look for signatures of new physics as they involve a tremendous amount of mass located in a restricted volume of space-time. 
For that reason, physicists have studied BHs in details and they have indeed shown that they are the scenes of peculiar phenomena and paradoxes. 
Amongst them, Hawking Radiation (HR) arguably stands as one of the most surprising discoveries of modern physics. In the 70's, Stephen Hawking showed that BHs are not perfectly black when one includes quantum effects~\cite{Hawking74}. 
Rather they emit particles, as a black body at a specific temperature, leading eventually to the evaporation of the BH. 
This process is at the heart of many open questions such as the information loss problem~\cite{Hawking_info,Mathur} or the trans-Planckian problem~\cite{THOOFT,Jacobson91}. 
In a few words, the trans-Planckian problem comes from the fact that Hawking's original derivation of HR involves waves with a wavelength smaller than the Planck scale. 
In this regime, the validity of the theory used by Hawking is not guaranteed and compromises the existence of HR. 
In addition, it seems that the process of BH evaporation is not a unitary process, information going through the event horizon will not be retrieved after the evaporation, this is the basis of the information loss problem.
Another remarkable process is the superradiance effect around rotating BHs. 
Superradiance is a radiation enhancement process, during which waves incident on a rotating BH can extract some of its energy. 
This effect was predicted more than forty years ago, and is linked to instabilities, known as \textit{black hole bombs}~\cite{pressBHB,CardosoBHB}. 
We see that BHs and their associated fundamental effects are a perfect playground for physicists to get inspiration for deepening our understanding of scientific laws.  

However, despite the enormous progress in technology which allowed physicists to observe gravitational waves or even take a picture of a BH, the current experimental apparatuses are not sensitive enough to probe fundamental effects such as the HR emitted by an astrophysical BH. 
For a solar mass BH, the temperature of the HR is 8 order of magnitude lower than the cosmic microwave background.
Fortunately, and surprisingly, the process of emission by a BH is not just observable around BHs. 

Indeed, it was realised in the 80's by Bill Unruh~\cite{Unruh:1980cg} and later rediscovered by Matt Visser~\cite{Visser:1993ub} that the necessary ingredients for HR and for other effects occurring in curved space-time can be found in a much more common system: sound waves in water. 
This stunning fact is the basis of \textit{Analogue Gravity} (AG) and offers new possibilities to challenge our physical concepts and to look for new inspiration~\cite{Carlip:2001wq}. 
Indeed, not only does AG open the doors to experiment and to test these fundamental processes, it also suggests a translation of the concepts from the gravitational language to the condensed matter language. 

The aim of this work is to extend the field of AG, by exploring both experimentally and theoretically fundamental processes occurring around rotating BHs and vortex flows. 
We will start the discussion by reviewing the history of AG, its achievements and current limitations. 

\section{Analogue gravity}

\begin{figure}[!h]
\centering
\includegraphics[scale=0.5]{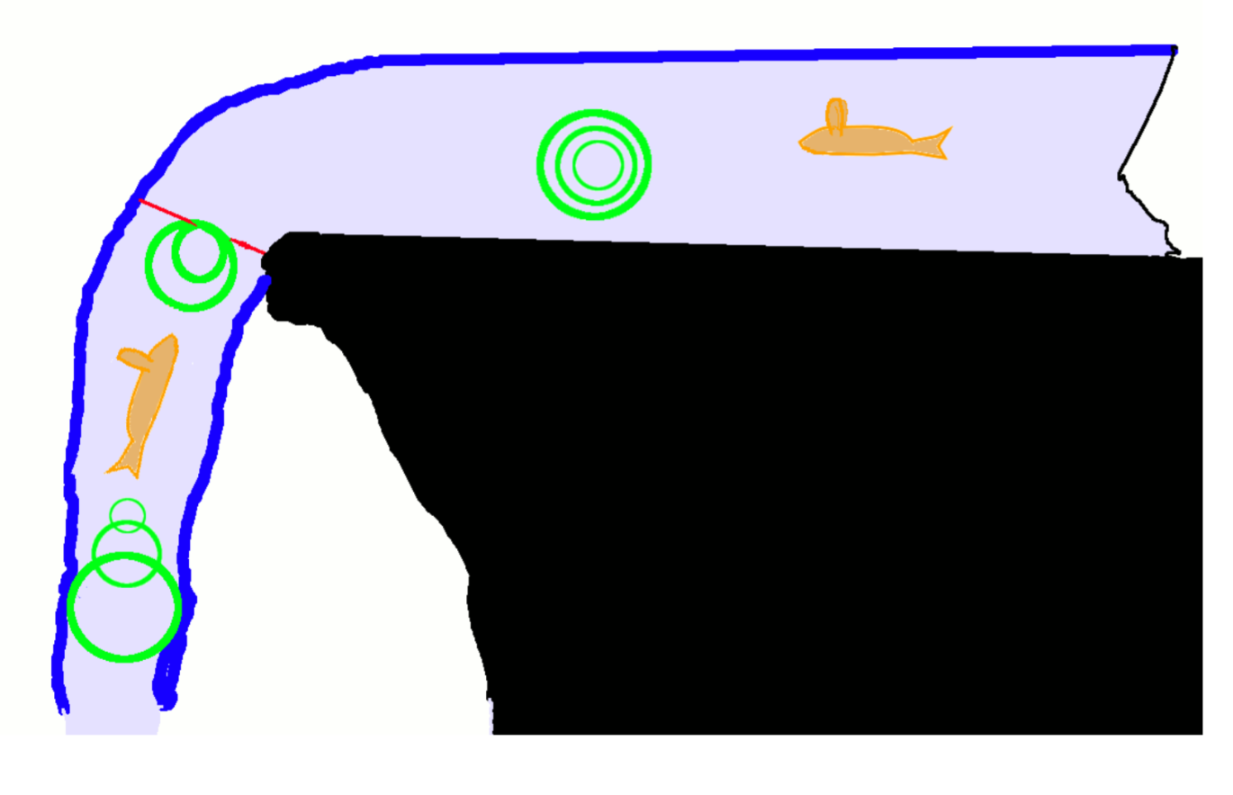}
\caption{Drawing by Unruh representing the analogy. The picture is taken from~\cite{Unruh:2014hua}. The green circles represent the sound waves propagating in the waterfall. The red line is the analogue horizon, at this point the flow is flowing at the speed of sound. Note, the appendage on the head of the fish are ears and not fins.}\label{Unruh_fish}
\end{figure}

The idea of AG came to Unruh when giving a colloquium on black holes.
In order to convey an intuitive picture of an horizon, he imagined the following system (see Fig.~\ref{Unruh_fish}). 
Imagine a waterfall with a group of fish. The fishes can talk to one another by emitting sound waves, which propagate in water at a finite speed of approximately $1500m/s$. 
When a fish is far away from the edge of the waterfall and the water is not flowing too fast (i.e. slower than the speed of sound) then the fish is able to talk to any of its friends as the sound it emits is able to reach any point in the stream. 
However, as the fish gets closer to the edge, the water runs faster. 
There will be a point where the speed of the fluid is equal to the speed of sound. Passed that point a fish will not be able to talk to its friends located upstream as the sound is not able to propagate. 
The transition from a fluid being subsonic (flowing at a speed slower than the speed of sound) to supersonic (flowing at a speed faster than the speed of sound) is the analogue of a BH horizon.
This simple yet powerful analogy, led Unruh to show that sound waves propagating in an idealized fluid obey the same equation as a massless scalar field on an effective curved space-time whose properties are given by the fluid flow~\cite{Unruh:1980cg}. In his seminal paper, Unruh considered a irrotational fluid flow and neglected non-linear and dispersive effects. This regime will be referred to as the \textit{analogue regime} throughout the thesis.

It is interesting to note that the effective metric derived by Unruh is very general and can be used to mimic various space-times. However, the emphasis was specifically laid on the ability to mimic horizons and to observe the analogue of the BH evaporation process. As we will see, this set the direction of the newly born field of AG, which remained unchanged for a long time.
In order to place this work in context, we will go through the history of AG. In 2011, a review on the subject was published~\cite{Barcelo05} which contains a very detailed history of the field until the year 2010. We will therefore only mention the major ideas and papers from before 2010 and invite the reader seeking more information to look at the review article. We will instead focus on the work done after the year 2010, highlighting the major results and different development of the field.

\subsection{1980 to the late 90's: Analogue horizons and the trans-Planckian problem}

As previously mentioned, Unruh's original paper~\cite{Unruh:1980cg} focuses on the ability to mimic the BH evaporation process with sound waves in fluids. 
He underlies nonetheless that while the analogy is mathematically rigorous, it is at the time more of a "theoretical laboratory" rather than a true proposal for an experimental detection of analogue HR. 
Unruh's paper remained forgotten, until the early 90's, when Jacobson entered the theoretical laboratory to tackle the trans-Planckian problem linked to BH evaporation~\cite{Jacobson91}. 
Jacobson's paper as well as the rediscovery of the analogy by Visser~\cite{Visser:1993ub}, stimulated other to think about the trans-Planckian problem via the scope of AG.  
From this emerged a series of theoretical papers which main results can be summarised as follow:

\begin{itemize}
\item Unknown Planck scale physics do not affect the existence of BH radiation~\cite{Jacobson91,Unruh_numeric,Jacobson:1995ak,Brout95,Corley96}.
\item HR is a purely kinematical effect linked to the presence of an horizon and does not depend on the underlying dynamics~\cite{Visser:1997yu}. This contrasts the strong link between HR and the law of BH thermodynamics~\cite{Hawking1975} which directly follows from the underlying Einstein equations.
\end{itemize} 

During this first period, AG helped shining new light on the role of high frequency modes in HR and helped to reveal the necessary ingredients for the BH evaporation to take place.

\subsection{The late 90's to 2008: Going beyond the horizon in ${}^3\mathrm{He}$}

So far, the investigations have been purely theoretical. 
However, by the end of the 90's, researchers started to imagine new effects linked with HR which could be observed in a realistic experiment. One example is the \textit{black hole laser}, in which perturbations trapped between a black hole and a white hole horizon are amplified via the Hawking process~\cite{Corley:1998rk}. 
The first one to be considered as a good candidate for analogue experiments is the superfluid ${}^3\mathrm{He}$ since it provide non-dissipative motion and the Hawking temperature was not too small compared to the temperature of the system to be observable~\cite{Jacobson_volovik}. 
Soon after, people started to develop other set-ups in order to investigate experimentally analogue fundamental processes.
Amongst them are:
\begin{itemize}
\item \underline{Optical systems}: They were introduced by Leonhardt and Piwnicki as a promising experimental system~\cite{Leonhardt:1999fe,Leonhardt:2000fd} due to the technological apparatus developed at the time. 
They were however strongly criticised~\cite{Visser:2000pk,Unruh:2003ss} for not having the correct quantum behaviour.
The optical system was later re-explored and shown to be a suitable analogue simulator~\cite{Schutzhold:2001fw,Schutzhold:2004tv}. 

\item \underline{Bose-Einstein condensate}: Garay, Anglin, Cirac, and Zoller showed that Bose-Einstein condensates can be set-up in stable configurations in order to constitute a sonic analogue BH~\cite{Garay:1999sk,Garay:2000jj}. Their study was then extended by Barcelo, Liberati, and Visser~\cite{Barcelo:2000tg} which established Bose-Einstein condensate (BEC) as a very strong candidate to experiment on analogue HR~\cite{Barcelo:2001ca}.

\item \underline{Surface gravity waves}: In 2002, Sch\"{u}tzhold and Unruh established that surface gravity waves can also be used as a simulator for phenomenon occurring around BHs~\cite{SCH02}. We will come back in details to this system in the following section as this is the analogue system we will use in this study.
\end{itemize}
We can already see, via the broad class of systems exhibiting the presence of horizons and capable of mimicking HR, one strength of AG. Namely, it establishes a link between the seemingly disparate disciplines and therefore progress made in one field can potentially benefit all the other. Note that this list is by no mean complete, for a comprehensive discussion we report the reader to~\cite{Berti_review}.

Alongside with the development of novel analogue systems to mimic the BH evaporation process, people have started to realise that AG can be used to investigate other fundamental processes occurring in curved space-times. 
\\
In 2003, Basak and Majumdar, followed by others, explored the analogue of the superradiance effect~\cite{basak,basak2,Slatyer:2005ty,Federici:2005ty}. Also starting in 2003, people have expand the analogy beyond the realm of BH physics by investigating analogue cosmology~\cite{Fedichev03,Barcelo:2003et,Weinfurtner:2004mu}.
\\
In 2004, Berti, Cardoso, Lemos, and Yoshida investigated the response to perturbations of analogue BHs and their associated Quasi-Normal Modes (QNMs)~\cite{Berti:2004ju,Cardoso:2004fi}. 
\\
In 2005, Balbinot and collaborators, focused on the backreaction process linked with HR in analogue BHs~\cite{PhysRevLett.94.161302,Balbinot:2006ua,Balbinot:2006cy}. Their work was followed by Maia and Sch\"{u}tzhold~\cite{Maia:2007vy}.

It is also interesting to note that the initial achievements of AG about the trans-Plackian problem in BH inspired similar work in cosmology~\cite{Kempf:2000ac,Niemeyer:2000eh,Kempf:2001fa}.

Finally, AG continued to shine light on HR and to reveal its essential features~\cite{Visser:2001kq,Unruh:2004zk}.

\subsection{2008 to today: The experimental age}

As we have seen, starting from the late 90's, AG branched off its initials goal and researchers started to use the analogy not only to investigate HR but other curved space-time effects, such as superradiance or inflation. 
This diversity of AG continues to increase from 2008, as it is the year that the first experiments on analogue systems were conducted. 
We will therefore divide this period of AG history in subsections. 
We will first focus on the quest for the observation of HR. 
We will then mention the various theoretical developments related to other effects done in parallel. 
Finally, we will report on the various recent analogue experiment which do not focus on HR.

\subsubsection{The quest for Hawking Radiation}
 
The experimental area started when Rousseaux and collaborators reported on the observation of negative phase velocity waves in a water tank experiment~\cite{Rousseaux:2007is}. 
While the results presented were inconclusive, this experiment showed a true interest from experimentalists to investigate analogue horizons.
This was reinforced by Philbin and co-workers who set-up an analogue event-horizon in an optical system~\cite{Philbin:2007ji}. 
In their experiment, an ultrashort light pulse travel in a medium and changes its refractive index and therefore the speed of light in the media. Perturbation, or what is called a probe, is then sent towards the pulse.  A horizon is created when the velocity of the pulse matches the velocity of the probe.  
A couple of years later, Lahav, Steinhauer and their colleagues also experimentally demonstrated the creation of an analogue BH horizon in a BEC~\cite{Lahav:2009wx}. 
With these experiments came the very concrete question of how to recognise practically HR. It quickly appeared that correlations of the emitted waves is the way to identify HR~\cite{Balbinot:2007de,PhysRevA.80.043603}.

Alongside these experiments, researchers started to simulate particular systems, most specifically BEC, and to develop theoretical tools in order to described present and future experimental results~\cite{Carusotto:2008ep,Coutant:2009cu,Macher:2009nz,Finazzi:2010nc}.
\\

Arguably one of the major result in this period is the observation of the stimulated Hawking process by Weinfurtner, Unruh and collaborators, in a water tank experiment~\cite{Silke_exp} in Vancouver (see~\cite{Weinfurtner:2013zfa} for a longer version of the paper). 
By placing a submerged obstacle in a one-dimensional open channel flow and studying the scattering of surface waves with the flow, they were able to measure the mode conversion process at the heart of HR. 
The Vancouver experiment stimulated and continues to inspire many theoretical studies in order to explain their observation. 
For example, an undulation, i.e. a standing wave located above the obstacle was observed in the experiment. 
This undulation and its link to HR has been studied in depth by Coutant, Parentani and colleagues~\cite{Coutant:2011in,Coutant:2012mf,Coutant:2012zh}. 
Also, it was realised that the flow was not supersonic above the obstacle (this is because a large undulation will form when the flow is supersonic). 
This effect was studied recently by Coutant and Weinfurtner and they have shown that HR still persists in subcritical flow as the system will exhibit a \textit{complex} horizon~\cite{Coutant16}. 
We can see that the experimental results obtained deeply helped to shine new light on our existing concepts.
However, since the system used was entirely classical, Weinfurtner \textit{et al} could only probe the stimulated Hawking emission.  
Even though, one can argue that the difference between the stimulated and the spontaneous HR is only the source of the initial state (coming from quantum fluctuations in the spontaneous case) and that the Vancouver experiment is a true observation of HR~\cite{Unruh:2014hua}, the measurement of spontaneous Hawking emission was still not observed at that time.

The same year as the Vancouver experiment, Belgiorno, Faccio and colleagues claimed to have observed the long-expected spontaneous HR~\cite{Belgiorno:2010wn}.
However, after years of discussion, it is not entirely convincing that their observation is truly the result of spontaneous HR~
\cite{Schutzhold:2010am,Belgiorno:2010hk,Unruh:2012tz}. 
In 2015, the Vancouver experiment was also repeated, and the results confirmed by Euv\'{e}, Michel, Parentani, and Rousseaux ~\cite{euve2015wave}.
\\

The next major experimental result in the quest for the spontaneous HR was made by Steinhauer in 2014, when he reported the observation of the BH laser effect in a BEC~\cite{Steinhauer:2014dra}. 
While his experiment was performed in a system capable of simulating the quantum HR, it would seem that the signal Steinhaeur observed was either seeded by classical perturbation or involved some non-linear effects~\cite{Michel:2015pra,Tettamanti:2016ntx}. In addition, there is still doubt as whether the black hole laser effect was ever present in the experiment (see~\cite{steinhauer2017self,Wang:2016jaj} for a discussion on the subject). 
Even though his measurement did not provide a conclusive measurement of spontaneous HR, they showed that a quantum simulator of BH horizon was technologically feasible.
A year later, Steinhauer published another experimental paper in which he claims to have observed the spontaneous emission of HR in a BEC~\cite{Steinhauer:2015saa}. 
In this experiment, Steinhauer set-up a BEC of ${}^{87}\mathrm{Rb}$ atoms. 
The condensate was confined along one-dimension by a trap.
The analogue horizon was formed by means of a short-wavelength laser which modifies the trapping potential in order to give it a step-like shape on which the condensate is flowing (in reality, the condensate is held fixed and the trap is moved along the condensate). 
The observation of spontaneous HR was made via the measurement of the correlation function between the Hawking partners~\cite{Steinhauer:2015ava}.
However the characterisation of the quantum nature of a measurement is a subtle issue~\cite{finke2016observation} and Steinhauer's experiment lead to a scientific debate~\cite{Michel:2016tog,Leonhardt:2016qdi,Steinhauer:2016ftg,Robertson:2017ysi} and the final word on these experiments has still not be settled. 

Simultaneously, Euv\'{e}, Rousseaux and collaborators reported the observation of the stimulated Hawking effect through a measurement of correlated noise in a water tank experiment in Poitiers~\cite{Euve15}. 
In this experiment, they have used a similar set-up as the Vancouver experiment with a submerged obstacle aimed at reducing the undulation. 
However, such undulation cannot completely be removed and its implication in scattering experiments have been studied~\cite{Michel:2018oyz}. 
The Poitiers as well as the Vancouver experiment were followed by theoretical studies by the AG~\cite{Robertson:2016ocv,Robertson:2017nly,Coutant18prd1} but also by the fluid dynamics communities~\cite{Churilov17}. 
Also, in 2015, Nguyen, Amo, and colleagues demonstrated experimentally the existence of an analogue BH horizon in polariton fluid~\cite{nguyen2015acoustic}.

At the end of 2018, Steinhauer and his group published a new paper reporting on the observation of a thermal spectrum linked to HR~\cite{MunozdeNova2019}. 
In this work, build on previous experiments and theoretical studies, the thermality of the Hawking spectrum is tested. 
The temperature seems to agree with numerical simulation. 
However, one should be careful when linking the thermality of HR to BH thermodynamics in the analogue context since the underlying dynamics are utterly different, as pointed out earlier~\cite{Visser:1997yu}.

The latest result concerning experimental HR came at the beginning of 2019 when Drori, Leonhardt, and colleagues claimed to have observed the stimulated Hawking process in an optical system~\cite{Drori:2018ivu}. 
This claim is based on the measurement of the power spectrum of the probe, similar to the analysis used in the Vancouver experiment. 
Their work is a promising step towards the observation of the spontaneous HR in an optical system.
\\

As we can see, the quest for the observation of the analogue Hawking effect brought a new look on this process and stimulated many discussions and studies, both theoretical and experimental, and from various communities. 
It is certain that further experimental results will come from various set-ups. 
For example, Rousseaux's team started to experiment on co-current surface waves~\cite{euve2018scattering} (so far, the experiments involving water waves and HR were made on an analogue white hole, which is the time reverse of a BH).

Even though each result can and must be contested\footnote{"We are trying to prove ourselves wrong as quickly as possible, because only in that way can we find progress." R. Feynman}, there is no doubts that every single experiment stimulated the field of AG and helped to deepen our understanding of HR, BEC, water waves, optics and other condensed matter systems.
It is also very interesting to see that philosophers have started to think about the analogical reasoning and the epistemology of AG~\cite{Thebault:2016udp,10.1093/bjps/axv010,Crowther:2018gdf,gryb2018universality}, concluding that AG provides reasonable arguments of the universality of HR.

\subsubsection{Exploring other space-times}

At the time the experiments started, it was well known theoretically that analogue system could be used to explore other curved space-time effects.
However, we have seen that the first analogue experiments developed were aimed at answering the original question of Unruh about a possible "experimental black hole evaporation?".
This gap between the broad range of effects AG offers to study and the single goal of observing HR motivated this thesis. 
\\

In 2016, myself, Patrick, Coutant, Tedford, Richartz and Weinfurtner have performed the first analogue experiment aimed at observing an effect other than HR, namely the superradiance effect~\cite{Superradiance}. This experiment was based on the many theoretical studies of the superradiance effect in analogue systems (see for e.g. ~\cite{Richartz:2009mi,Richartz:2012bd,Dolan_scattering,Richartz:2014lda}) and will be the subject of Chapter~\ref{Superradiance_sec}. A year after, Vocke, Faccio and co-workers were able to set-up an analogue rotating black hole in an optical system~\cite{Vocke17}. This is a very exciting and promising set-up to gain more experimental insight in the superradiance process. In this system, perturbations can only propagate for a very short time, making the analysis performed in~\cite{Superradiance} very difficult (if at all possible) to implement. Observing superradiance in the optical system will therefore require a novel understanding of the effect~\cite{Prain:2019jqk}. Such interpretation should predict consequences that are observable in a short-life time system.
\\

In 2017, the possibility of detecting stimulated Unruh radiation has been discussed in \cite{Leonhardt:2017lwm}. The detection of Unruh radiation in a BEC was recently reported in~\cite{Hu:2018psq}.
\\

In 2018, Eckel, Campbell and colleagues, started to experiment on cosmological analogues in a BEC~\cite{Eckel:2017uqx}. They have set-up the analogue of an expanding universe in the form of an expanding ring-shaped BEC and they have measured the redshift associated to this expansion of the phonon modes. This is the first experiment to investigate cosmological scenarios and is also the result of a long theoretical investigation~\cite{PhysRevA.68.053613,PhysRevLett.91.240407,PhysRevA.70.063615,PhysRevA.76.033616}. In 2019, Wittemer, Sch\"{u}tzhold, Schaetz, and colleagues reported on the observation of particle pair created during an analogue inflation in an ion trap~\cite{wittemer2019particle}.
\\

At the end of 2018, we performed another experiment aiming at studying the relaxation process of analogue BHs and we reported on the observation of light-ring modes~\cite{Torres_QNM}. In addition to this observation, we have used the analogy in order to develop a new fluid flow measurement. This will be the subject of Chapter~\ref{Ringdown_sec}.

Finally, the most recent AG experiment to my knowledge was conducted by Goodhew, Patrick, Gooding, and Weinfurtner~\cite{backreaction}. In this experiment, they have investigated the backreaction process of waves incident on a vortex flow. This backreaction manifests itself in the form of a change in the height of the water which was measured and compared with theoretical estimate.
\\

We can see that starting from 2016, the range of phenomena investigated by AG experiments greatly increased. 
These experiments are still rather recent, but they nonetheless show the potential of the field of AG to probe fundamental effects and deepen our understanding by challenging our concepts. 

\subsection{Remarks on analogue gravity}

We conclude this section with two remarks about AG.

The first one is about the word analogue. Formally, an analogy is a process of transferring information from a subject (a source) to another (a target). 
There exist many analogies in Physics and they are often of great use to better understand certain effects. 
The strength of an analogy lies in its capacity of putting forwards the essential ingredients needed for a phenomenon to happen. 
It can even occur that the source and the target systems are completely equivalent. 
In such cases, one might talk about an identification rather than the analogy.
A very practical example is the mechanical-electrical analogy~\cite{olson1958dynamical}.
Indeed, it is known that some mechanical systems can be represented and studied by means of electrical circuits. 
The famous example is the one of a mass attached to a spring with some friction and an RLC (resistor, inductor and capacitor) circuit. 
The two systems are described by the same second order differential equation and one can therefore use any one of the two systems to learn about the other. 
However, it is clear that the two systems are different, and the identification is only valid in a certain regime.
Similarly to the mechanical-electrical analogy, the fluid-gravity analogy is strengthened by the fact that \textit{under specific assumptions}, the target and the source systems are described by the same equations. 
In this analogue regime, we can therefore talk about an identification between gravity and condensed matter. 
It is important to keep in mind that this identification is only rigorous if one works in the analogue regime. 
Outside this regime, the analogy can still be used as a powerful tool (as it was the case in the many studies mentioned above) but one should be careful when choosing the concepts he/she is transferring from the source to the target system.

The second remark is about the word gravity and on interconnectedness of the perturbation and the space-time. 
It is clear that AG proposes a way to mimic gravitational effects \textit{on the propagation of fields}.
Without the waves, there is no space-time.
This seems to undermine the underlying geometry in comparison with GR in which the space-time is the main actor. 
It is worth noting however, that for practical reasons a space-time without matter is not particularly physically relevant. 
Actually one can define a space-time via an action describing how matter propagates on a manifold~\cite{Schuller}. 
In~\cite{Schuller}, the authors show that the dynamics of the space-time is dictated by the way matter propagate on this space-time and the assumption that the matter equation should be predictive and quantizable. 
In particular, they show that the Einstein equation can be derived from the Maxwell equations. 
Therefore, understanding how field propagate on a space-time can lead to new information about the underlying dynamic of the space-time.

\section{Surface gravity waves as an analogue model}\label{surface_wave_analogy_sec}

\subsection{Surface gravity waves}

We have seen that AG is a vast field, which can be studied via various condensed matter system.
We will now focus on one such system and derive the analogy for surface water waves on a flowing fluid.
The equation presented here is at the heart of the analogy and we shall take great care when deriving it.
We start by considering a fluid described by a velocity field $\vec{v}$, with pressure $p$ and constant density $\rho$. From the continuity equation, this assumption implies that the fluid velocity is divergence free:
\begin{equation}
\vec{\nabla}.\vec{v} = 0.
\end{equation} 
This equation is valid in the entire domain occupied by the fluid. 
Secondly, we assume the fluid to be inviscid, therefore, its dynamics is governed by the Euler equation:
\begin{equation}
\D \vec{v} = - \frac{\vec{\nabla} p}{\rho} + \vec{g} + \vec{F},
\end{equation}
where $\vec{g}$ is the gravitational acceleration and $\vec{F}$ represents additional forces acting on the fluid.
We finally assume the flow to be irrotational:
\begin{equation}
\vec{\nabla} \times \vec{v} = 0.
\end{equation}
This implies that our system can be described by one scalar function, the velocity potential $\phi$, such that:
\begin{equation}
\vec{v} = \vec{\nabla} \phi.
\end{equation}
Using the vector identity\footnote{$\vec{\nabla}\left(\vec{A}.\vec{B} \right) = \left(\vec{A} . \vec{\nabla} \right) \vec{B} + \left(\vec{B} . \vec{\nabla} \right) \vec{A} + 
\vec{A} \times \left( \vec{\nabla} \times \vec{B} \right)\vec{B} \times \left( \vec{\nabla} \times \vec{A} \right)$} for the curl free velocity field, the Euler equation, can be simplified to the Bernoulli equation:

\begin{equation}
\frac{\p \phi}{\p t} + \frac{(\vec{\nabla} \phi)^2}{2} = - \frac{p}{\rho} - gz - V,
\end{equation}
where we have introduced the potential $V$ corresponding to the force \mbox{$\vec{F} = - \vec{\nabla} V$}.

In addition to the Bernoulli equation and the continuity equation, which are valid everywhere in the fluid, we need to impose some boundary condition. We impose that the bottom of the tank is a hard wall and that water does not penetrate:
\begin{equation}
\left. v_z \right|_{z=0} = \eval{\frac{\p \phi}{\p z}}{z=0}  = 0.
\end{equation}
We further impose that a particle at the surface stays at the surface:
\begin{equation}
\eval{v_z}{z = h} = \eval{\frac{\p \phi}{\p z}}{z=h}  = \D h,
\end{equation}
where $\D$ is the material derivative.
Finally, we also impose that the pressure vanishes at the free surface:
\begin{equation}
\eval{p}{z=h} = 0.
\end{equation}
Since the Bernoulli equation is valid everywhere, it is certainly valid at the free surface. We can therefore evaluate it, at the upper boundary $z = h$. Using the boundary condition for the pressure, it simplifies to:
\begin{equation}
\eval{\frac{\p \phi}{\p t}}{z=h} + \eval{\frac{(\vnab \phi)^2}{2}}{z=h} = -gh - \eval{V}{z=h}
\end{equation}
This last equation completes our set of equations to derive the dynamics of surface waves.

We now consider small perturbations (the waves) of a stationary and irrotational background flow These perturbations correspond to small displacement of the water height $\xi$. The background flow is given by $\vec{v}_0$ and the water depth $h_{0}$ which is assumed to be constant. The background quantities therefore satisfy, by choosing $V$ such that:
\begin{equation}
\frac{\p }{\p z} \vec{v}_0 = 0,
\quad \text{and} \quad
\frac{\vec{v}_0^2}{2} = -gh_0 - V.
\end{equation}
The perturbations are considered to be irrotational, such that the velocity perturbations are also given by a potential $\delta \phi$. The perturbations satisfy the linear continuity equation:
\begin{equation} \label{conti}
\vnab^2 \delta \phi = 0,
\end{equation}
and the perturbed boundary conditions:
\begin{equation}\label{BC1}
\eval{\frac{\p \delta \phi}{\p z}}{z=0} = 0,
\end{equation}
\begin{equation} \label{BC2}
\eval{\frac{\p \delta \phi}{\p z}}{z=h_0} = \D \xi,
\end{equation}
and
\begin{equation} \label{BC3}
\D \delta \phi = - g \xi,
\end{equation}
where $\D$ is the material derivative computed with the background flow, i.e. $\D = \p_t + \vec{v}_0 . \vnab$.

Eqs.~\eqref{conti} to \eqref{BC3} are four relations governing the system. 
Two of these are valid at the free surface, one is valid at the bottom of the tank, and one is valid everywhere in the fluid.
We therefore need to relate the boundary condition at the bottom to the ones at the top, this is done by integrating in the vertical direction, through the bulk, the continuity equation.
To do so, we separate the gradient operator, into its horizontal and vertical component: $\vnab = \vnab^{\parallel} + \p_z \vec{e}_z$.
The continuity equation becomes:
\begin{equation}
{\vnab^{\parallel}}^2 \delta \phi + \p_z^2 \delta \phi = 0.
\end{equation}
We look for solutions to this equation in term of its power series in $z$, i.e. we look for solution of the form:
\begin{equation}
\delta \phi = \sum_{n=0}^{+ \infty} \delta \phi_n(x,y) \frac{z^n}{n!}.
\end{equation} 
Plugging this expression into the continuity equation and equating each term of the expansion to zero, we obtain a recursive relation between the various $\delta \phi_n$:
\begin{equation}
{\vnab^{\parallel}}^2 \delta \phi_n + \delta \phi_{n+2} = 0.
\end{equation}
Therefore, we can deduce all the term of the power series by knowing only the first two terms $\delta \phi_0$ and $\delta\phi_1$.
From the boundary condition, given by Eq.~\eqref{BC1}, we directly see that $\delta\phi_1 = 0$, and therefore $\delta\phi_{2n+1} = 0$ for all integer $n$. This implies that the entire series is determined by $\delta \phi_0$ such that:
\begin{equation}
\delta \phi = \sum_{n=0}^{+\infty} \frac{z^{2n}}{2n!} (iz\vnab^{\parallel})^{2n} \delta\phi_0 = \cosh\left( iz\vnab^{\parallel} \right) \delta\phi_0(x,y),
\end{equation}
where we have recognized the power series of the hyperbolic cosine.
We can now plug this expression in the two boundary conditions at the free surface and we get:
\begin{eqnarray}\label{coupled_we}
i\vnab \sinh(ih_0\vnab) \delta \phi_0  &=& \D \xi\\
\D \cosh(ih_0 \vnab ) \delta \phi_0 &=& - g \xi.
\end{eqnarray}
Note here that we have dropped the superscript $\parallel$, but it is understood that $\vnab = \vnab^\parallel$ as we have integrated out the vertical component.
Taking the material derivative of the second equation, we finally obtain the equation of motion for $\delta \phi_0$:
\begin{equation} \label{wave_equation_dispersive}
\D^2 \delta \phi_0 - ig\vnab\tanh(-i h_0 \vnab) \delta \phi_0 = 0.
\end{equation}

We note that the above wave equation is written for the velocity potential $\delta \phi_0$ (to simplify notation, we will refer to the velocity potential perturbation $\delta \phi_0$ simply as $\phi$ in the following). 
However, the direct observable quantity is the height perturbation $\xi$, which satisfies a more complicated wave equation, which comprises derivatives of the background flow (The time dependence and hence the frequency content stay however unchanged). It is straightforward to relate the velocity potential perturbation to the height perturbation via the relations given in Eqs.~\eqref{coupled_we}. In the following we will work with the velocity potential, but we should keep in mind that the observable quantity is the height variation $\xi$. 

It is useful to note that Eq.~\eqref{wave_equation_dispersive} can be obtained as the Euler-Lagrange equations for the following Lagrangian~\cite{Coutant:2012mf,Richartz:2012bd,Coutant:2016vsf}:
\be\label{Lagrangian}
\mathcal{L} = |(\p_t + \vec{v}_0.\vnab) \phi |^2 - |f(\vnab)\phi|^2,
\ee
where $f^2(\vnab) = ig\vnab\tanh(-ih_0\vnab)$. Note that the last term in Eq~\eqref{Lagrangian} is different from the one in~\cite{Coutant:2012mf}. The two expressions are equivalent if one considers the water surface to be flat.

\subsubsection{Symmetries and conserved quantities}

The symmetries of a Lagrangian are associated to conserved quantities via Noether's theorem.
The Lagrangian given in Eq.~\eqref{Lagrangian} is invariant under time-translation \mbox{$t \to t + t_0$} and \mbox{$\phi \to \phi e^{i\alpha}$}.

The infinitesimal transformation $t \to t + \epsilon$ leads to a change in the field of the form:
\be
\phi(t+\epsilon,\vec{x}) \approx \phi(t,\vec{x}) + \epsilon\p_t\phi(t,\vec{x}). 
\ee

According to Noether's theorem, the associated conserved quantity is:
\be
J_t^\nu = -\frac{\p \mathcal{L}}{\p \p_\nu\phi} \p_t \phi.
\ee
$J_t$ is called the energy current.

The infinitesimal transformation $\phi \to \phi e^{i\epsilon}$ leads to a change in the field of the form:
\be
\phi(t,\vec{x}) e^{i\epsilon} \approx \phi(t,\vec{x}) + i\epsilon\phi(t,\vec{x}). 
\ee

According to Noether's theorem, the associated conserved quantity is:
\be
J^\nu = -\frac{\p \mathcal{L}}{\p \p_\nu\phi}i\phi.
\ee
$J$ is called the norm current.

In stationary systems, the field can be assumed to have the simple time dependence $\phi(t,\vec{x}) = \phi(\vec{x}) e^{-i\omega t}$. 
In this case the time derivative becomes $\p_t = -i \omega$, and we see that the two currents $J_t$ and $J$ relate via $J_t = -\omega J$.
This can be understood by the fact that, in stationary systems, a shift in the phase of the wave can be written as a time translation.

Assuming that the field takes the form of a plane wave $\phi = A e^{-i\omega t + i \vec{k}.\vec{x}}$, we get an expression for the norm of the wave, $J^0$:
\be
J^0 = (\omega - \vec{v}_0.\vec{k}) A^2.
\ee

Note that in fluid mechanics, this quantity is usually referred to as the wave action. 
It is linked to the energy of the wave, $E = J_t^0$, simply by a factor $\omega$.
We see here that it is possible to have negative norm modes when waves propagate on top of a flowing fluid if it satisfies the condition:
\be
(\omega - \vec{v}_0.\vec{k}) < 0.
\ee

\subsection{Shallow water limit and the analogue metric}

We now turn our attention to the case of surface waves of very long wavelength. Assuming that the wavelength is significantly larger than any other length scale in the system, Eq.~\eqref{wave_equation_dispersive} simplifies to:
\be\label{wave_equation_shallow}
\D^2 \phi - c^2\Delta \phi = 0,
\ee
where we have introduce the propagation speed of the wave $c^2 = \sqrt{gh_0}$.

The key observation is that this equation can be recast in the form:
\be\label{analogue_KG}
\frac{1}{\sqrt{-g}} \p_\mu \left( \sqrt{-g} g^{\mu \nu} \p_\nu \phi  \right) = 0,
\ee
where $g$ is the analogue metric tensor defined by its components (for the inverse/contravariant metric):
\be\label{analogue_metric}
g^{\mu\nu} = \begin{bmatrix}
    -1     &      -v_0^x        &      -v_0^y      \\
   -v_0^x  &  c^2 - (v_0^x)^2   &    v_0^x v_0^y   \\
   -v_0^y  &    v_0^x v_0^y     &   c^2 - (v_0^y)^2 
\end{bmatrix}
\ee
and for the covariant metric:
\be\label{analogue_metric_inv}
g_{\mu\nu} = \frac{1}{c^2}\begin{bmatrix}
    -(c^2 - \vec{v_0}^2) & -v_0^x & -v_0^y \\
           -v_0^x        &    1   &    0   \\
           -v_0^y        &    0   &    1
\end{bmatrix}
\ee

\subsection{Draining bathtub vortices vs rotating black holes}

We have just seen that surface gravity waves propagating on top of a flowing fluid experience the presence of an effective space-time. 
This space-time is described by the metric given in Eq.~\eqref{analogue_metric}. 
The various components of this metric are given by the fluid velocity and one can therefore imagine setting-up different fluid flows in order to study the propagation of fields on different space-times.
In this thesis, we will focus on one particular type of flow, namely the \textit{Draining bathtub flow} (DBT flow). 
The DBT flow is a two-dimensional, axisymmetric, incompressible and irrotational flow and the free surface is assumed to be flat.
It is characterised by two parameters $C$ and $D$, representing the circulation and the drain respectively.
The background velocity of the DBT is given, in polar coordinates $(r,\theta)$, by:
\begin{equation} \label{DBT}
\vec v_0 = - \frac Dr \vec e_r + \frac Cr \vec e_\theta.
\end{equation}

The line element corresponding to this fluid flow is:
\be\label{DBT_ds}
ds_{\mathrm{DBT}}^2 = -c^2 dt^2 + \left( dr + \frac{D}{r} dt\right)^2 + \left( r d\theta - \frac{C}{r} dt \right)^2.
\ee

Since our aim is to investigate the propagation of fields around rotating BHs (also known as Kerr BHs), we shall compare the DBT line element to the one of a rotating BH. 
Before comparing the two metrics, we note first that the effective space-time corresponding to the DBT flow has one spatial dimension less than the Kerr space-time.
Therefore, the DBT flow can at best mimic a subspace of the Kerr space-time. 
A Kerr BH of mass $M$ and angular momentum $\mathcal{J}$ is given by the Kerr metric.
The equatorial slice of the Kerr space-time is given, in Boyer-Lindquist coordinates, by the following line element~\cite{Visser:2007fj}:
\bea \label{Kerr_ds}
ds_{\mathrm{Kerr}}^2 &=& -\left( 1 - \frac{2M}{r}\right)dt^2 - \frac{4Ma}{r} dt d\theta +\left[ \frac{dr^2}{1 - 2M/r + a^2/r^2} \right] \nonumber \\
& & + \left[ r^2 + a^2 + \frac{2Ma^2}{r} \right] d\theta^2,
\eea
where $a = \mathcal{J}/M$ and the speed of light has been set to one.
This line element exhibits two important surfaces. 
The first one is the static limit and corresponds to the surface which sets the term in front of $dt^2$ to $0$. 
Beyond this surface, no particle can remain at rest with respect to an observer at infinity.
The second is the horizon and is defined as the surface that makes the term in front $dr^2$ diverge.
The horizon is the surface past which nothing can escape the black hole.
For a detailed study of the Kerr metric, we refer the reader to~\cite{Chandrasekhar:1985kt}.

To compare the DBT space-time and the Kerr space-time, it is useful to write the DBT line element in Boyer-Lindquist type coordinates.
Defining the coordinates:
\be
dt' = c\left( dt -\frac{Dr dr}{(r^2c^2 - D^2)} \right) + \frac{C}{c} d\theta \quad\text{and}\quad d\theta' = d\theta - \frac{CDdr}{r(r^2 c^2 - D^2)},
\ee
and substituting them in Eq.~\eqref{DBT_ds}, we can write the line element of the DBT space-time as:
\bea\label{DBT_ds_BL}
ds_{\mathrm{DBT}}^2 &=& -\left( 1 - \frac{r_e^2}{r^2} \right) dt^2 + \left( 1 - \frac{r_h^2}{r^2} \right)^{-1} dr^2 - 2C\frac{r_e^2}{r^2} dt d\theta  \nonumber \\
& & + \left( r^2 + \frac{C^2}{c^2} - \frac{C^2 r_e^2}{c^2 r^2}\right) d\theta^2,
\eea
where
\be
r_e = \frac{\sqrt{C^2 + D^2}}{c} \quad \text{and} \quad r_h = \frac{D}{c},
\ee
and where we have dropped the prime.
In this form, we see that the two line elements given by Eqs.~\eqref{DBT_ds_BL} and~\eqref{Kerr_ds} share some similarities.
In particular, the DBT metric also exhibits a static limit, located at $r = r_e$, as well as an horizon at $r=r_h$.
Both metrics also asymptote the same form as $r \to \infty$, namely:
\be
ds_\infty^2 = -dt^2 + dr^2 + (r^2 + K ) d\theta^2 + \mathcal{O}\left(\frac{1}{r}\right),
\ee
where $K$ is a constant which is different for the DBT and the Kerr metric but does not affect the form of $ds_\infty^2$. 
However, it is known that the two metrics do not exactly represent the same space-time\footnote{
Note that one can reproduce the equatorial slice of Kerr by fine-tuning the equation of state of the system. This is not the case considered here.}~\cite{Visser:2004zs}.
This can be seen in the way the two metrics approach their asymptotic form.
For the DBT metric, we have:
\be
ds_{\mathrm{DBT},\infty}^{2} = ds_\infty^2 + \frac{r_e^2}{r^2} dt^2 + \frac{r_h^2}{r^2}dr^2 - 2C\frac{r_e^2}{r^2}dtd\theta - \frac{C^2r_e^2}{c^2r^2}d\theta^2 + \mathcal{O}\left(\frac{1}{r^4} \right),
\ee
while for the Kerr metric:
\be
ds_{\mathrm{Kerr},\infty}^2 = ds_\infty^2 + \frac{2M}{r}dt^2 + \frac{2M}{r}dr^2 - \frac{4Ma}{r}dtd\theta + \frac{2Ma^2}{r}d\theta^2 + \mathcal{O}\left(\frac{1}{r^2} \right).
\ee
The fact that the fall off of two metrics are different will have some consequences on the possible effects occurring in both space-times. 
For example, the DBT flow will allow for the existence of the analogue of the Aharonov-Bohm effect as we will see in Chapter~\ref{Superradiance_sec}. 
Despite these differences, we will see in the following chapters that both space-times share some remarkable properties.

\chapter{Rays and waves}\label{Rays_sec}
\epigraph{\textit{While we must by
 no means neglect the physical correctness of predictions (if we did, it would
 not be physics at all), we must look round for postulates so simple that the
 logical process of deduction from them can be carried out with rigour.}}{Synge}

\section{Introduction} 

We have seen that in practice, one does not solve the most general equations describing a system. 
\thispagestyle{empty}
It is the art of physics to identify the relevant terms and effects needed to describe the desired phenomena. 
This allowed us to go from the Euler and continuity equations to the wave equation describing surface gravity waves on an irrotational flow. 
However, even after this simplification, one still does not end up with an equation solvable analytically. 
We therefore need to develop techniques to solve our equation in order to make predictions to compare the theory with experiments. 
In this chapter, we will describe one such method, called the asymptotic series expansion, or the gradient expansion. 
We will start by reviewing the foundation for this well-established technique before applying it to the case of gravity waves incident on a vortex flow.

 \subsection{Asymptotic series: From waves to rays}
 
The gradient expansion method consists in looking for solutions which are rapidly oscillating with respect to the typical scale of changes of the system.  
This can be translated via the ansatz:
\begin{equation}\label{ansatz_eik}
\phi(x,t) = A(x,t) e^{iS(x,t)/\epsilon}
\end{equation}

$S(x,t)$ is the local phase of the wave and $A(x,t)$ the local amplitude, and $\epsilon \ll 1$. Since we have a small parameter at our disposition, we can look for $A(x,t)$ and $S(x,t)$ using perturbation techniques and expanding them in power series of $\epsilon$:

\begin{equation}
A(x,t) = \sum_{n=0}^{+\infty} A_n(x,t) \epsilon^n,
\quad \text{ and } \quad 
S(x,t) = \sum_{n=0}^{+\infty} S_n(x,t) \epsilon^n.
\end{equation}

These series are called asymptotic series and should be manipulated carefully. Indeed, Poincar\'e demonstrated that in general such series diverges when $n \rightarrow \infty$ at fixed $\epsilon$. 
However, it asymptotically converges, in the sense that the partial sum will converge to the true solution as $\epsilon \rightarrow 0$. That is:
\begin{equation}
\lim_{\epsilon \rightarrow 0} \frac{A(x,t) - \sum_{n=0}^{N} A_n(x,t) \epsilon^n }{A_N(x,t) \epsilon^N} = 0
\quad
\forall N \in \mathbb{N}
\end{equation}
and similarly for $S$. 

A natural question is to ask ourselves, why should we use asymptotic series which are usually divergent rather than convergent Taylor series? 
A first practical answer is that in the limit we are interested in, i.e. $\epsilon \rightarrow 0$, the asymptotic series usually converges more rapidly than the Taylor series. 
We therefore need to compute less terms when using asymptotic series and still get an accurate physical picture in this limit. 
This can be demonstrated at hand of a well-known example, the Bessel functions.

\begin{figure}
\includegraphics[scale = 1, trim = 2cm 0cm 0cm 0cm]{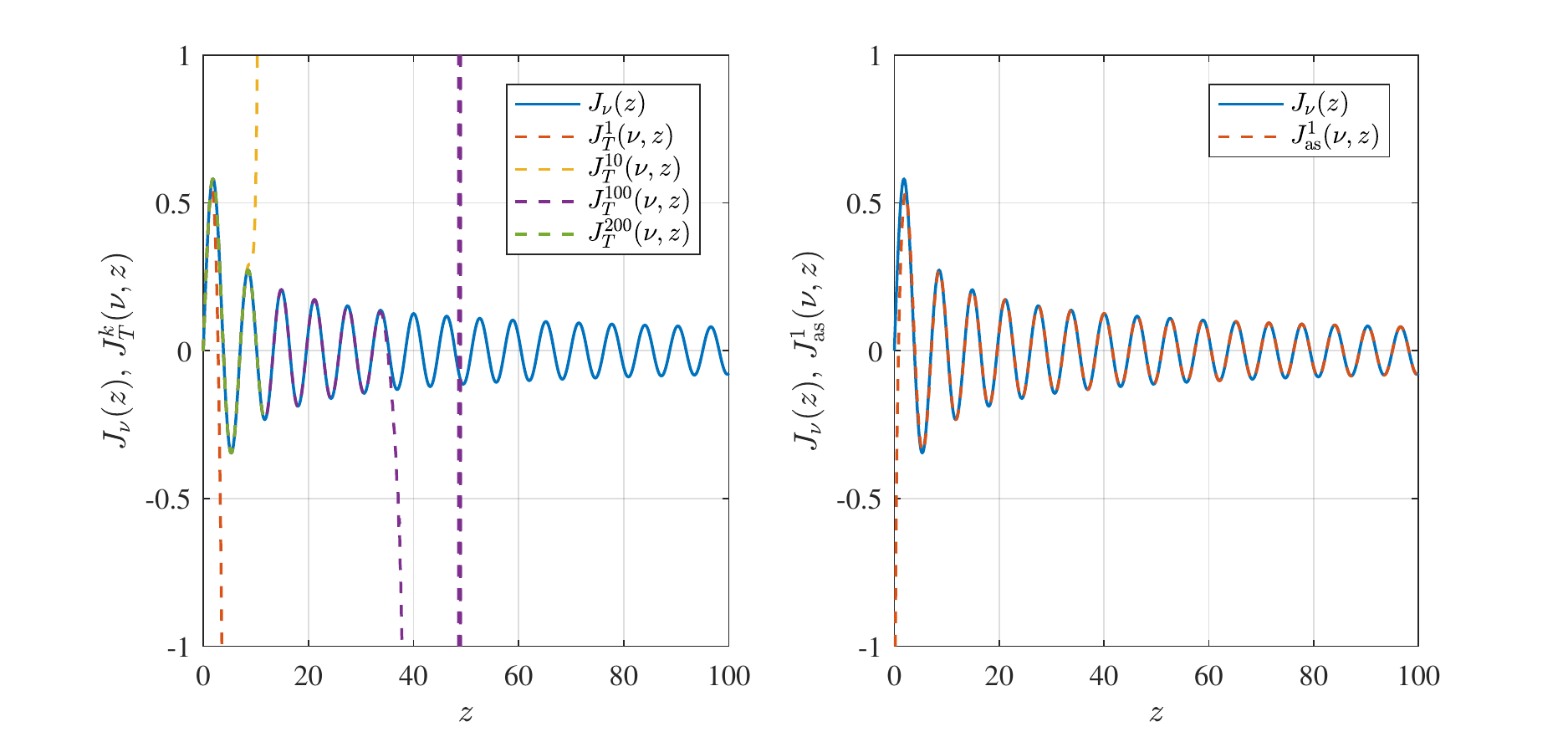}
\caption{ \textbf{Comparison between Bessel function and its various expansions.}
The left panel depicts the Bessel function $J_\nu(z)$ with $\nu =1$ and its approximation using Eq.~\eqref{Bessel_Taylor} for $k = 1, 10,100,$ and $200$.
 We see that no matter until which $k$ the sum is performed; the Taylor series cannot reproduce the behaviour of the Bessel function. 
 Even worst, it cannot be computed numerically. The right panel presents the Bessel function $J_\nu(z)$ and the first term of its asymptotic expansion $J_{\mathrm{as}}^1(\nu,z) = \sqrt{2/\pi}cos(\alpha)/\sqrt{z}$. 
 We see that the first term only of the asymptotic series matches very well the Bessel function and reproduces its behaviour for high values of $z$. 
We note also that the asymptotic agrees well for small values of the argument.}\label{Bessel_expansion}
\end{figure}

Bessel functions $J_\nu(z)$ are solutions to the following ordinary differential equation (ODE), (known as Bessel's equation):
\begin{equation}
z^2 \frac{d^2 f}{dz^2} + z \frac{d f}{dz} + (z^2 - \nu^2)f = 0.
\end{equation}
This equation arises in many physical situations, one of such is the free wave equation in cylindrical coordinates. 
The free wave equation in two spatial dimensions is:
\begin{equation}
\p_t^2 \phi - c^2 \vnab^2\phi = 0.
\end{equation}
This equation can be separated by looking for solutions of the form:
\begin{equation}
\phi = f(r) e^{-i\omega t + i m \theta},
\end{equation}
which leads to the ODE for the radial profile $f$:
\begin{equation}
z^2 \frac{d^2 f}{dz^2} + z \frac{df}{dz} + (z^2 - m^2) = 0,
\end{equation}
with $z = \omega r/c$.
This equation has the form of Bessel's equation and its solutions are given by the Bessel functions $f = J_m(z)$. 
Bessel functions have a well-known series expansion:
\begin{equation} \label{Bessel_Taylor}
J_\nu(z) = \sum_{k=0}^{+\infty} J_T^k(\nu,k) = \sum_{k=0}^{+\infty} \frac{(-1)^k}{k!\Gamma(\nu + k +1) } \left( \frac{z}{2} \right)^{2k+\nu},
\end{equation}
where $\Gamma$ is the Gamma function.
However, this series is not valid for practical numerical computation if one is interested in the high frequency or large radius regime which are of most interest for physical phenomena. 
In such regime, it is more useful to use the asymptotic expansion of the Bessel function:
\begin{equation} \label{Bessel_as}
J_\nu(z) \sim \left( \frac{2}{\pi z} \right)^{1/2} 
\left( 
\cos(\alpha) \sum_{k=0}^{+\infty} (-1)^k \frac{a_{2k}(\nu)}{z^{2k}}   
- \sin(\alpha) \sum_{k=0}^{+\infty} (-1)^k \frac{a_{2k+1}(\nu)}{z^{2k+1}}  
\right),
\end{equation}
with $\alpha = z - \frac{1}{2}\nu\pi - \frac{1}{4}\pi$ and $a_k(\nu) = \frac{(4\nu^2 - 1^2)(4\nu^2 - 3^2)...(4\nu^2 - (2k - 1)^2)}{k! 8^k}$ with \mbox{$a_0(\nu) =1$}.
We can see in Fig.~\ref{Bessel_expansion} that the first term of the asymptotic expansion is enough to describe the asymptotic behaviour of the Bessel function when the Taylor expansion cannot even be computed for large values of the argument. 
We also note that the asymptotic expansion accurately describes the Bessel function even for small values of the argument.

In the following, we will restrict our attention to the first term of the asymptotic expansion.
We make this choice as we have seen that the first term in the asymptotic series is a reasonable approximation of the full solution. 
Moreover, the leading order term of the asymptotic series has a nice physical interpretation and is therefore of great use to gain insight and interpret specific phenomena. 
Indeed, it is known that in the high frequency limit, also known as the geometrical limit, rapidly oscillating waves can be thought of as fictitious particle. 
The term geometrical limit, comes from optics, were the law of geometrical optics (law of refraction, law of reflection, etc...) appear from the high frequency limit of the wave description of light. 
The correspondence between high frequency waves and particles is easily seen in the free wave equation. 
For simplicity, we restrict ourselves to the one-dimensional case. The wave equation, after performing a time Fourier transform is:
\begin{equation}
- \omega^2 \tilde{\phi}(\omega,x) - c^2 \p_x^2 \tilde{\phi}(\omega,x) = 0,
\end{equation}
where
\begin{equation}
\phi(t,x) = \frac{1}{\sqrt{2\pi}}\int_{- \infty}^{+\infty} \tilde{\phi}(\omega,x) e^{i\omega t} d\omega.
\end{equation}
Assuming that $\omega \gg 1$, we can express $\tilde{\phi}$ in term of its asymptotic series. Keeping only the first term, we get that
\begin{equation}
\tilde{\phi}(\omega,x) = \frac{A(x)}{\sqrt{2\pi}}e^{-i\omega S(x)}.
\end{equation}
The prefactor and the minus sign are only there for convenience and can be reabsorbed in the definition of $A$ and $S$.
From this expression we can reconstruct the solution in position space:
\begin{equation}
\phi(t,x) = A(x) \delta\left( t - S(x) \right).
\end{equation}
We see that the first term in the asymptotic expansion corresponds to a Dirac delta, a discontinuous solution, that can be interpreted as a particle which position at time $t$ satisfies $S(x) = t$.
In this case, finding $S(x)$ is straightforward as it is given by:
\begin{equation}
\left(\p_x S \right)^2 = c^{-2}.
\end{equation}
We therefore find that the trajectory of the fictitious particle is $x = ct$, a straight line.

In general, $S$ satisfies a first-order non-linear partial differential equation (PDE), as we will see, known as the eikonal equation. 
This PDE can be solved using the method of characteristics. 
The characteristics are the rays which the fictitious particle follow. As we will see in the next section, the process can be thought backwards, and it is possible to reconstruct waves from the rays.

 \subsection{The Hamiltonian approach: From rays to waves}
 
We have seen via the use of asymptotic series that high frequency waves can be seen as fictitious particles. 
These particles will have a world line and can be described via a Hamiltonian like any particle. 
The trajectories of such particles are called rays and are the characteristic of the initial dynamical equation. 
We now see that wave solutions can be constructed as a system of rays. 
This construction of waves via the Hamiltonian method has been beautifully explained in \cite{synge1963hamiltonian}. 
The discussion below is not claiming at any new result but is simply intended at recalling this elegant technique before using it to the case of surface waves incident on a vortex flow.

We start by recalling the basic principle of Hamiltonian mechanics. 
In this formulation, a system is described by a set of variables $(x_1,...,x_n)$ and their conjugate variables $(k_1,...,k_n)$. 
The space spanned by $(\vec{x},\vec{k})$ is called the \emph{ray phase space}. 
At the heart of the Hamiltonian formulation is the \emph{Hamiltonian function} or simply the \emph{Hamiltonian} $\mathcal{H}(\vec{x},\vec{k})$ which is a function on the ray phase space. 
A ray is then defined as being a parametrized curve $(\vec{x}(\tau),\vec{k}(\tau))$ on ray phase space satisfying an optimisation criteria, \emph{Hamilton's principle}, namely:
\begin{equation}\label{Ham_principle}
\delta \left( \int_{\tau_1}^{\tau_2} 
\left[
\vec{k}(\tau')\frac{d\vec{x}(\tau ')}{d \tau} - \mathcal{H}(\vec{x}(\tau '),\vec{k}(\tau ')) 
\right] d\tau '
\right) = 0.
\end{equation}
If we choose the usual time $t$ to parametrize our curve, we recognise that the integrand is nothing else than the Lagrangian associated to the Hamiltonian $\mathcal{H}$ via a Legendre transform.
Independent variations with respect to $\vec{x}$ and $\vec{k}$ imply that a ray satisfies:
\begin{equation}
\frac{d\vec{x}}{d \tau} = \vnab_k \mathcal{H}
\quad  \text{ and } \quad
\frac{d\vec{k}}{d \tau} = -\vnab_x \mathcal{H}.
\end{equation}
The subscript indicates along which coordinate the gradient is taken.
These equations are Hamilton's equations. They imply that the Hamiltonian is constant along a ray. Indeed, using Hamilton's equations, we have that:
\begin{equation}
\frac{ d \mathcal{H}}{d\tau} = \vnab_x\mathcal{H} \frac{d \vec{x}}{d\tau} + \vnab_k \mathcal{H} \frac{d\vec{k}}{d\tau} = 0.
\end{equation}
An important feature of this formulation is that rays must be independent of the parametrization. 
This means that one can multiply the Hamiltonian $\mathcal{H}$ by any non vanishing function $N(\tau)$ and one will obtain the same set of rays, instead parametrized by $\tilde{\tau}$ where \mbox{$d\tilde{\tau} = N(\tau)d\tau$}.
Substituting $\mathcal{H}\rightarrow N\mathcal{H}$ in Eq.\eqref{Ham_principle} and varying with respect to $N$, we see that the re-parametrization condition imposes the constraint that $\mathcal{H}(\vec{x},\vec{k}) = 0$ along rays (we anticipate that this condition will be the dispersion relation).

From this definition of rays, we can now construct a wave as a collection of rays. To properly define a wave, these rays must satisfy a circulation condition:
\begin{equation}
\int_{C} \vec{k} d\vec{x} = 0,
\end{equation}
along every closed circuit $C$.
This implies that the integrand is a perfect differential, i.e.:
\begin{equation}
\exists \psi(\vec{x}) 
\quad \text{ such that } \quad
\vec{k}d\vec{x} = d\psi. 
\end{equation}
This function $\psi$ is \emph{the phase} of the wave and the wave fronts are defined as surfaces of constant phase.
It directly follows from the circulation condition that:
\begin{equation}
\vec{k} = \vnab_x \psi.
\end{equation}
Substituting $\vec{k}$ for the phase gradients in the constraint $\mathcal{H}(\vec{x},\vec{k}) = 0$, we have that the phase function $\psi$ satisfies the first order PDE:
\begin{equation}
\mathcal{H}(\vec{x},\vnab_x \psi) = 0,
\end{equation}
which is the Hamilton-Jacobi equation.

\section{Ray tracing equations} \label{ray_tracing_sec}

We now derive the equations for the leading order term and the next-to-leading term in the gradient expansion. These equations are respectively called the eikonal equation and the transport equation.

 \subsection{Eikonal equation}
 We now focus on obtaining the ray equations for a general linear PDE. 
 Eventually we will be interested in the wave equation describing surface waves on a current, i.e.~ Eq.~\eqref{wave_equation_dispersive}, but we will keep the discussion general by considering the following differential equation:
 \begin{equation}
\D^{2}\phi +F(-i\vnab ) \phi - 2 \nu \gamma(-i\vnab ) \D \phi = 0.
 \end{equation}
Here we have introduced a pair of functions $F$ and $\gamma$, encoding respectively the dispersion relation and dissipation. The parameter $\nu$ quantifies the amount of dissipation. 
The only assumption of $F$ and $\gamma$ are that they shall be analytic and even. 
We will restrict ourselves to two spatial dimensions, but the principle can be applied to more dimensions without difficulties.
We seek solution that are rapidly oscillating and that we assume that the background flow
changes over a scale significantly larger than the wavelength. 
We therefore look for solutions of the form given by Eq.~\eqref{ansatz_eik}
and consider the rescaling:
\begin{subequations}
\begin{eqnarray}
\p &\rightarrow & \epsilon\p, \\
\nu &\rightarrow & \epsilon \nu.
\end{eqnarray}
\end{subequations}
The second line translates the fact that we consider dissipation to be weak (over a scale of the order of the gradient scale). 
We impose $S$ and $A$ to be real valued functions so that dissipation will be incorporated in the amplitude equation later on.

Assuming that $F$ is analytic, it can be expanded as a Taylor series:
\begin{equation}
F(-i\epsilon \vnab )= \sum_{n=1}^{+\infty} T_{2n} (-1)^n \epsilon^{2n} \Delta^{n} , \label{F_series}
\end{equation}
where $\Delta \equiv \vnab^2 = \partial_x^2 + \partial_y^2$ is the 2D Laplacian. To expand the term $F(-i\vnab ) \phi$, we look at each term of the Taylor series in Eq.~\eqref{F_series} separately. Using the binomial formula, we obtain 
\begin{eqnarray}
\Delta^{n} \phi &=& \sum_{k=0}^{n} \binom{n}{k} \partial_{x}^{2k}\partial_{y}^{2n - 2k} \phi , \label{Delta_expansion} \\
&\simeq& \frac{(-1)^n}{\epsilon^{2n}} \sum_{k=0}^{n} \binom{n}{k} (\partial_x S)^{2k} (\partial_{y} S)^{2n - 2k} A e^{i S/\epsilon} , \nonumber \\
&\simeq& \frac{(-1)^n}{\epsilon^{2n}} \left((\partial_x S)^2 + (\partial_y S)^2\right)^n A e^{i S/\epsilon} , \nonumber
\end{eqnarray}
where $\simeq$ means that we have kept only the leading term in $\epsilon$. Re-summing the Taylor series given in Eq.~\eqref{F_series}, we see that 
\begin{equation} \label{action_of_F}
F(-i\epsilon \vnab ) \phi \simeq A e^{i S/\epsilon} F(\vnab S). 
\end{equation}
The same applies for the material derivative, that is 
\begin{equation} \label{action_of_D}
\mathcal{D}_t^{2}\phi \simeq -A e^{iS/\epsilon}(\mathcal{D}_t S)^{2}. 
\end{equation}
Combining the last two equations, it follows that the phase obeys the Hamilton-Jacobi equation:
\begin{equation} \label{H-J_eq}
\left( \D S\right)^2 - F(\vnab S) = 0.
\end{equation}
From this we recognise the Hamiltonian function, by introducing the local wave vector $\vec{k} = \vnab S$ and local frequency $\omega = - \p_t S$
\begin{equation}
\mathcal{H}(\vec{x},\vec{k}) = -\frac{1}{2} \left( \omega - \vec{v}_0.\vec{k} \right)^2 + \frac{1}{2} F(\vec{k}).
\end{equation}
The condition $\mathcal{H} = 0$ expresses the dispersion relation of the system.
Eq.~\eqref{H-J_eq} can be solved via the method of characteristic. To find the characteristic of the equation, we solve Hamilton's equations:
\begin{subequations}
\label{Ham_eq}
\begin{eqnarray}
\dot t &=& -\frac{\partial \mathcal{H}}{\partial \omega}, \qquad \dot \omega =  \frac{\partial \mathcal{H}}{\partial t} , \label{Ham_eq_om}\\
\dot{\vec{x}} &=& \vnab_k \mathcal{H} \quad \text{and} \qquad \dot{\vec{k}}=-\vnab_x \mathcal{H}, \label{Ham_eq_x}
\end{eqnarray}
\end{subequations}
where the dot denotes derivation with respect to $\tau$ which parametrises the curve. The unusual sign in Eq.~\eqref{Ham_eq_om} is due to the fact that the conjugate variable to $t$ is defined to be $-\omega$ instead of $\omega$.

 \subsection{Transport equation}
 
The previous equation for the phase $S$ was obtained by looking at the leading order term in $\epsilon$ in our gradient expansion. 
At this level, the amplitude of the wave doesn't play any role. 
To derive an equation for the wave amplitude, we look at the next-to-leading order term in the expansion. 
Let us first examine the next-to-leading-order term of the action of $\D^{2}$ on $\phi$. As $\D^{2}$ contains two derivatives, the next-to-leading-order term of its action on $\phi$ is attained when only one derivative hits the exponential. 
\begin{equation}
\mathcal{D}_t^{2} \phi \simeq  \frac{-i \epsilon}{A} \mathcal{D}_t\left(A^2 \Omega_0 \right)
\end{equation}
where we have introduced $\Omega_{0}=\omega - \vec{v_0}.\vnab S$ the co-moving angular frequency of the wave.
We get the action of the operator $F(-i\epsilon\vnab )$ the same way. 
To obtain the next to leading order term of $\Delta^n \phi$ (in $1/\epsilon^{2n-1}$), we must count the number of terms in the expansion of Eq.~\eqref{Delta_expansion} such that the exponential $e^{iS/\epsilon}$ is derived exactly $2n-1$ times. The last derivative of $\Delta^n$ will then act on $A$ or $\vnab S$. Doing so, we obtain 
\begin{eqnarray}
\Delta^n \phi &\simeq& \frac{(-1)^{n-1}}{\epsilon^{2n-1}}i \bigg( 2n \vnab A \cdot (\vnab S)^{2n-1}  + \frac{2n(2n-1)}{2} A \cdot (\vnab S)^{2n-2} \Delta S \bigg) e^{iS/\epsilon}. \nonumber \\ && 
\end{eqnarray}
Re-summing the Tayor series, we recognise the derivative with respect to the momenta of the function $F$. Gathering the terms, the action of $F(-i\epsilon\vnab )$ can be rewritten as a total derivative: 
\begin{eqnarray}
F(-i\epsilon \vnab )Ae^{iS/\epsilon} \simeq  -\frac{i\epsilon}{2 A} \vnab . \left( A^2 \nabla_k F(\nabla S) \right).
\end{eqnarray}
Finally, a similar calculation yields the action of the damping operator,
\begin{equation}
2\epsilon \nu\gamma(-i\vnab)\D \phi \simeq 2i\epsilon \nu\Omega_0 A \gamma(\vnab S).
\end{equation}
Combining the action of the three operators we obtain the equation for the amplitude:
\begin{equation} \label{transport_eq}
\partial_t ( \Omega_0 A^2 ) + \vnab \cdot ( \Omega_0 A^{2} \vec{v}_g^{\mathrm{lab}}) = - 2 \nu \Omega_0 A^2 . \gamma(\vnab S), 
\end{equation}
where $\vec{v}_g^{\mathrm{lab}}$ is the group velocity in the laboratory frame, i.e. $\vec{v}_g^{\mathrm{lab}} = \vec{\nabla}_k \omega = \vec{v}_g^{\mathrm{fluid}} + \vec{v_0}$.
Note that in order to solve this transport equation, one need to solve for the rays first. 
Once the phase is known along the rays, we can compute the amplitude along each characteristic curve. 
In the absence of dissipation, the transport equation  Eq.~\eqref{transport_eq} tells us that the wave action is transported with the velocity $\vec{v}_g^{\mathrm lab}$ and is conserved in a tube delimited by characteristic curves. 
If the characteristic curves converge (diverge) then the wave action will increase (decrease).
 
 \section{Waves on a vortex}
 
 We will now apply the general ray-tracing equations derived above to study the specific case of waves incident on irrotational draining vortices. 
 The interaction between waves and vortices has attracted attention in various fields of physics~\cite{Vivanco00}, such as acoustics~\cite{Fetter64,Nazarenko95,Pagneux01,Kopiev10}, ocean surface waves~\cite{Buhler,Buhler05}, and wave generation by turbulence~\cite{Lund89,Cerda93,Fabrikant94,Umeki97}. The problem was studied with the use of singularities in the complex plane by~\cite{Vandenbroeck87,Stokes08,Moreira12}. The stability analysis of rotating fluids is also a rich topic, in particular due to the possibility of \emph{over-reflection} mechanisms~\cite{Acheson76}, which was studied both experimentally \cite{Vatistas94,Jansson06} and theoretically \cite{Tophoj13,Mougel17}. 
 
 The wave-vortex interaction is the setting for two intriguing physical analogies. Firstly, as we have already seen in Chapter~\ref{Intro_sec}, a draining vortex constitutes an analogue of a rotating black hole.
 Secondly, waves on a vortex give rise to a hydrodynamical analogue of the Aharonov-Bohm effect of quantum mechanics~\cite{Berry,Coste99,Coste99b,Vivanco99,Sonin02,Dolan11}. 
We now postpone the discussion of the Aharonov-Bohm effect until Chapter~\ref{Superradiance_sec} to focus on the black hole/fluid analogy. 
 In the analogue regime, where the flow is irrotational and the wavelength of the waves are much larger than the fluid depth, we recall that the wave equation for the velocity potential can be written as the Klein-Gordon equation for a massless scalar field on an effective curved space-time:
 \begin{equation}
 \frac{1}{\sqrt{-g}} \p_\mu \left( \sqrt{-g} g^{\mu \nu} \p_\nu \phi  \right) = 0
\end{equation}
Following the steps of section~\ref{ray_tracing_sec}, we get that the eikonal equation for the Klein-Gordon equation is:
\begin{equation}
g^{\mu \nu} \left(\p_\mu S \right)  \left(\p_\nu S \right)= 0.
\end{equation}
From which we recognise the Hamiltonian:
\begin{equation}
\mathcal{H} = P_\mu P^\mu.
\end{equation}
We note there that the word Hamiltonian is a misnomer as $P_\mu P^\mu$ is not the Hamiltonian of a free particle in the sense of its energy (simply because energy is an observer dependent concept while $P_\mu P^\mu$ is invariant). 
However, we still call it a Hamiltonian in the sense that Hamilton's equation with this particular function will provide us with the equations of motion. 
These equations are in fact the geodesic equations, defined as the path of minimum length on a manifold with a specific metric $g^{\mu \nu}$.  
The asymptotic expansion is however more general as it can be applied to system that are not described by a metric and therefore allows us to extend some concepts beyond the analogue regime where one does not have access to the analogue metric.
We will focus in the following on one of these concepts, namely the concept of \emph{light-rings} (LRs).

A ubiquitous feature of a black hole is the existence of photon orbits called `light-rings'. Outside a Schwarzschild black hole, for example, there is a `photon sphere', at $3/2$ times the event horizon radius, comprising the (zero-measure) set of light rays (null geodesics) that orbit around the black hole perpetually. 
These orbits comprise a separatrix in phase space, dividing between rays that fall into the black hole, and rays that escape. 
Analysis of the properties of LRs yields a heuristic understanding of many black hole phenomena. 
The orbital frequency and Lyapunov exponent $\Lambda$ of the LRs -- just two parameters -- play a dominant role in (semi-classical approximations to) the spectrum of damped resonances known as black hole quasi-normal modes 
~\cite{Goebel72,Cardoso_Lyapu,Dolan_QNM,Yang:2012he,Konoplya17}; 
the absorption cross section 
~\cite{decanini2011universality, Macedo:2013afa}; and wave-interference phenomena such as `orbiting' and `glories' in the scattering cross section 
~\cite{Matzner:1985rjn, Dolan:2017rtj}. 

The aim of this section is to show that there generically exist circular orbits around a draining vortex, which are the analogue of black hole light rings. We apply the ray-tracing methods to show that scattering of waves in a draining vortex can be well-understood in terms of the rays of a frequency-dependent effective Hamiltonian. We will discuss the presence of circular orbits and its consequences in terms of resonances appearing in the time-dependent response of the system. 
 \subsection{Setup} \label{orbit_subsec}
 
 We consider the simplest model of an irrotational vortex with a drain, namely the DBT flow given by Eq.~\eqref{DBT}.
\begin{figure}[!h]
\centering
\includegraphics[trim=0.7cm 0 0 0]{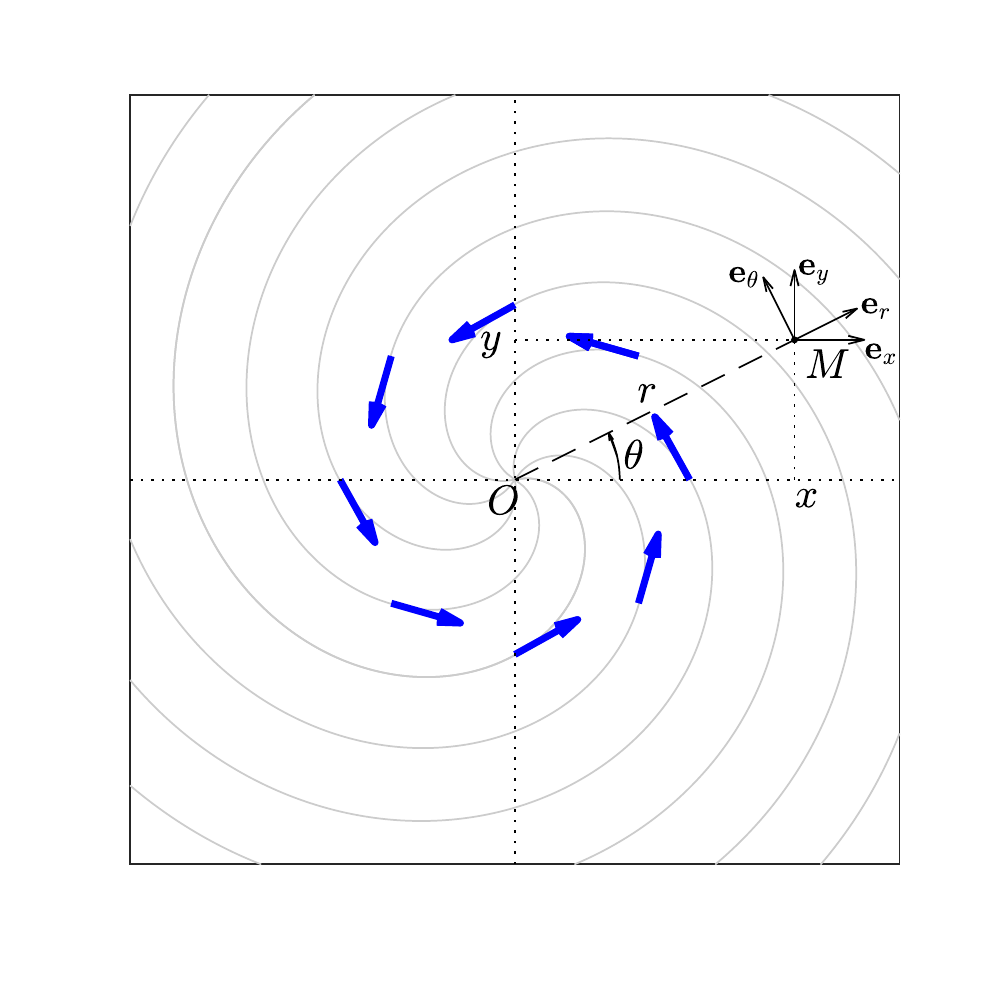}
\caption{Schematic of the flow and associated coordinate systems, from the perspective of an observer looking down on the wavetank. The vortex considered here is rotating counter-clockwise when $C>0$ (as shown by the bold blue arrows). The light grey lines represent streamlines of the flow. A point $M$ on the surface is located using either Cartesian coordinates $(x,y)$ or polar coordinates $(r,\theta)$.
}
\label{schematic}
\end{figure}
We recall that this model assumes that the surface is flat and
that $C$ and $D$ are the circulation and draining constants (positive), respectively.
The schematic of the flow and the coordinate systems are presented in Fig.~\ref{schematic}. 
This model is not only simple to analyse, it also provides a realistic description for a wide variety of situations. The reason is that far from the centre, and far from the edges of the fluid tank, the free surface tends to be flat and the flow to resemble that of Eq.~\eqref{DBT}.

\subsection{Circular orbits\label{subsec:circorbits}}
Although we shall have in mind the flow profile of Eq.~\eqref{DBT}, our results stay conceptually identical for a large class of flows. To see this, the following discussion is done using a general symmetric flow of the form  
\begin{equation}
\vec v_0 = v_r(r) \vec e_r + v_\theta(r) \vec e_\theta. 
\end{equation}

Using the symmetry of the vortex flow, we switch to polar coordinates. We build the conjugate variables as before, that is, $k_r = \partial_r S$, and $m = \partial_\theta S$. As we shall see, depending on the context, $m$ will be a real number (interpreted as an angular momentum), or an integer (interpreted as the azimuthal number of the wave). We also note that $\omega$ is conserved since the flow is stationary. For simplicity, we assume $\omega > 0$ (but negative values can be obtained via the symmetry $\omega \to - \omega$ and $m \to -m$).

A circular orbit is an equilibrium point in the radial direction. This means that it is a critical point of the Hamiltonian for $(r,k_r)$. It thus satisfies the conditions
\bsub \label{CO_eq} \bea 
\partial_r \mathcal H &=& 0, \\
\partial_{k_r} \mathcal H &=& 0,
\eea \esub
In the bathtub profile of Eq.~\eqref{DBT}, Eqs.~\eqref{CO_eq} give 2 relations between 4 unknowns. For a fixed pair $(\omega , m)$, Eqs.~\eqref{CO_eq} determines a unique pair $(r_\star, k_{r \star})$. Furthermore, imposing the Hamiltonian constraint $\mathcal H = 0$ gives a relation between $\omega$ and $m$ on the circular orbit, analogous to a dispersion relation, which we write as
\be \label{LR_disp}
\omega = \omega_{\star}(m). 
\ee
This relation can be read in either ways. At fixed angular frequency $\omega$, it gives the value of $m$ such that one has a circular orbit; at fixed azimuthal number $m$ it yields the particular wave frequency corresponding to the circular orbit.

\subsubsection{Shallow water regime}
\label{LR_shallow_Sec}
In the limit of shallow water, where all modes propagate at the same speed, the  radius of the circular orbit is \emph{independent} of the angular frequency $\omega$, and furthermore the relationship between $\omega$ and $m$ is linear (see e.g.~\cite{Dolan12} or \cite{Dempsey2017}): 
\be \label{om-shallow}
\omega_\star = \hat{\Omega}^{\pm} m .
\ee
Here $\hat{\Omega}^\pm \equiv \dot{\theta} / \dot{t}$ is the orbital frequency of the circular orbit. In general, there exist two circular orbits, one co-rotating ($\dot{\theta} > 0$, $\hat{\Omega}^+ > 0$) and one counter-rotating  ($\dot{\theta} < 0$, $\hat{\Omega}^- < 0$) relative to the circulating flow. The frequencies are given by
\be
\hat \Omega^{\pm} = \pm \frac{c^2 \sqrt{C^2+D^2}}{B_{\pm}^2} . 
\ee
The radii of these circular orbits (that we shall now refer to as `orbital radii') are then given by 
\be \label{radius_rel}
r_\star^{\pm} = \frac{B_{\pm}}{c},
\ee
where $c = \sqrt{gh}$ is the propagation speed of shallow water waves, and a $+$ ($-$) sign denotes the co-rotating (counter-rotating) case. The parameter $B_{\pm}$ is defined as
\be \label{B_param}
B_{\pm} = \left(2(C^2 + D^2) \mp 2C\sqrt{C^2 + D^2} \right)^{1/2}. 
\ee 
As we shall see, $B_\pm$ also governs the orbital radius for deep water waves. An immediate conclusion of Eq.~\eqref{radius_rel} is that the co-rotating circular orbit is closer to the centre of the vortex, while the counter-rotating orbit is further out. Hence, the counter-rotating orbit is in general more visible (this is further confirmed in Chapter~\ref{Ringdown_sec} when studying damped resonances).

\subsubsection{Deep water regime}
For further insight into circular orbits in the presence of dispersion, it is instructive to look at another simple case where analytic formulae can be obtained. Here we consider deep water without capillarity, and thus a group velocity given by
\be 
v_g(k) = \frac12 \sqrt{\frac gk}. 
\ee
In this regime it is possible to derive a simple formula for the radius of the orbits as a function of the frequency.
We can directly express the norm of the wave vector in terms of the radius as
\be \label{p_deep}
|\vec{k}| = \frac{g r_{\star}^2}{4B_\pm^2}.
\ee
(with $\pm$ indices omitted on $k$ and $r_\star$ for clarity). 
With the other conditions for the circular orbits, Eq.~\eqref{CO_eq}, we can express both $k_{r \star}$ and $m$ as functions of $r_{\star}$, via.,
\be \label{pr_m_deep}
k_{r \star} =\frac{D g r_{\star}^2}{4 B_\pm^3} \quad\text{and} \quad m = \pm \frac{g r_{\star}^3}{4 B_\pm^2}\sqrt{ 1 - \frac{D^2}{B_\pm^2}}.
\ee

As we have seen, there are two different light rings (represented by the $\pm$). In order to keep track of this sign we introduce $\epsilon_m = \text{sign}(m)$. Using it, we can express $m$ as a function of $r$ as 
\be
m = \epsilon_m \frac{g r_{\star}^3}{4 B_\pm^2}\sqrt{ 1 - \frac{D^2}{B_\pm^2}}. 
\ee
On a circular orbit, we can use the Hamiltonian constraint to express the orbital frequency in terms of the other parameters: 
\be
\omega_\star = -\frac{D}{r_\star} k_{r\star} + \frac{C}{r_\star} \frac{m}{r_\star} + \sqrt{g k}.
\ee
Substituting Eq.~\eqref{p_deep} and Eq.~\eqref{pr_m_deep} in the above equation, we obtain 
\be \label{om_star_app}
\omega_\star = \frac{gr_\star}{2B_\pm} \left( - \frac{D^2}{2B_\pm^2} + \epsilon_m \frac{C}{2B_\pm^2}\sqrt{B^2 - D^2} + 1 \right).
\ee
We now rewrite our flow parameters $C$ and $D$ as 
\be \label{new_CD}
C= R \sin(\alpha) \quad \text{and} \quad D = R \cos(\alpha),
\ee
and $B_\pm$ becomes 
\be \label{new_B}
B=\sqrt{2}R\sqrt{1 - \epsilon_m sin(\alpha)}.
\ee
Inserting Eq.~\eqref{new_CD} and Eq.~\eqref{new_B} into Eq.~\eqref{om_star_app} we get 
\be
\omega_\star = \frac{gr_\star}{8B_\pm} \left( - \frac{\cos^2(\alpha)}{1 - \epsilon_m \sin(\alpha)} + \epsilon_m \frac{\sin(\alpha) \sqrt{2 - 2\epsilon_m \sin(\alpha) - \cos^2(\alpha)}}{ 1 - \epsilon_m \sin(\alpha)} 
+ 4 \right). \nonumber
\ee
Using $\cos^2(\alpha) = 1 - \sin^2(\alpha) $, we can simplify this expression:
\be
\omega_\star = \frac{gr_\star}{8B_\pm} \left( \frac{ -1 + \sin^2(\alpha) }{1 - \epsilon_m\sin(\alpha)} + \epsilon_m \sin(\alpha)  + 4 \right), 
\ee
which further simplifies into 
\be
\omega_\star = \frac{3gr_\star}{8B_\pm}.
\ee
We can inverse this relation to obtain $r_\star$ as a function of $\omega_\star$:
\be \label{rad_deep}
r_\star = \frac{8B_\pm}{3g} \omega_\star ,
\ee
 or express $r_\star$ in terms of $m$ to obtain the effective dispersion relation:
\be \label{effective_dr}
\omega_\star(m)=\frac{3}{8}\left( \frac{4g^2}{\sqrt{B_\pm^2-D^2}} \right)^{1/3} m^{1/3}.
\ee

Note that the wave angular frequency $\omega_\star$ scales linearly with $m$ in shallow water, but with $m^{1/3}$ in deep water. This is a direct consequence of the dispersive nature of the waves. 

\subsubsection{General case}
For any non-linear dispersion relation, Eq.~\eqref{CO_eq} implies the relation  
\be \label{Implicit_rad_circ}
r_\star = \frac{B_{\pm}}{v_g^f}. 
\ee
This equation shows that it is the change in the group velocity $v_g^f$ that shifts the location of the circular orbit.
Here $v_g^f$ is evaluated with the local wave vector on the orbit $k^{2} = k_{r \star}^2 + m^2/r_\star^2$, and hence the equation above is implicit. To find $r_\star$ and $\omega_\star(m)$ explicitly, one must insert Eq.~\eqref{Implicit_rad_circ} into Eqs.~\eqref{CO_eq} and the Hamiltonian constraint. 

\begin{figure}
\centering
\includegraphics{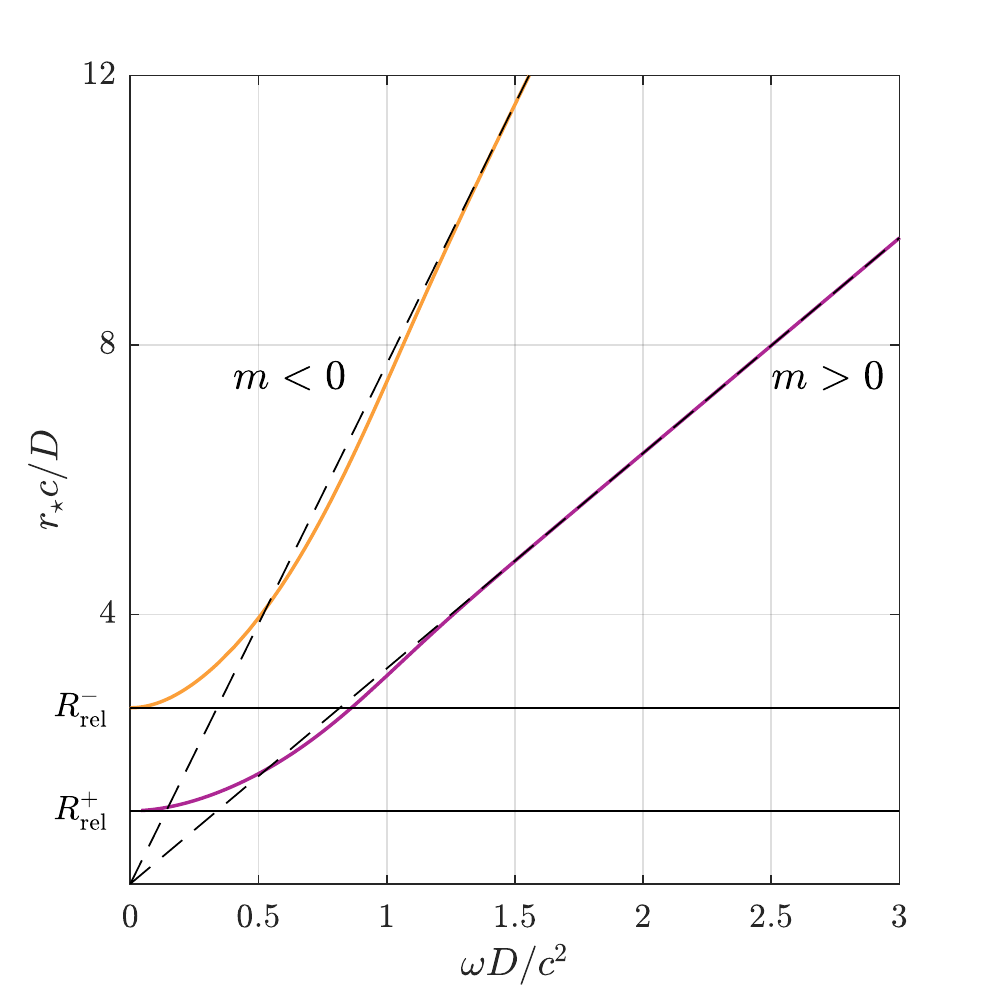}
\caption{Radius of the unstable orbit as a function of the angular frequency of the co- and counter-rotating rays (in purple and orange respectively) and comparison with the shallow and deep water regime (in solid and dashed black respectively). We have chosen the flow parameters such that $C/h_0c = D/h_0c = 0.9$ (chosen to make the transition from shallow to deep water more visible), where $c = \sqrt{gh_0}$ the celerity of shallow water waves.  $R_{\mathrm rel}^\pm$ is the value of $r_\star^\pm$ in the shallow water (or relativistic) regime. (We recall that when the frequency varies, the corresponding orbital $m$ also does, so that Eq.~\eqref{LR_disp} is satisfied.)
}
\label{radius_orbit}
\end{figure}

Fig.~\ref{radius_orbit} shows numerical-obtained solutions for the orbital radius as a function of wave frequency, for a vortex with $C/h_0c = D/h_0c = 0.9$. 
The values of the parameters are simply chosen to illustrate our discussion. 
The two branches correspond to the co-rotating ($+$) and counter-rotating ($-$) orbits. At low frequency (long wavelength), the orbital radius is anticipated by the shallow-water result, Eq.~\eqref{om-shallow}. 
At higher frequencies, the circular orbits migrate away from the vortex core, as anticipated in Eq.~\eqref{rad_deep}.

Even though the ray structure will qualitatively be similar for various values of the parameters C and D, one can distinguish two interesting limits. First, when $C \ \gg D$, we can see that the radius of the co-rotating orbit shrink to zeros as $B_{+} \to 0$. At the opposite, when $C \ll D$, the two circular orbits asymptotically approach one another as $B_{+} \to B_{-}$.
 
\begin{figure}[!]
\centering
\includegraphics{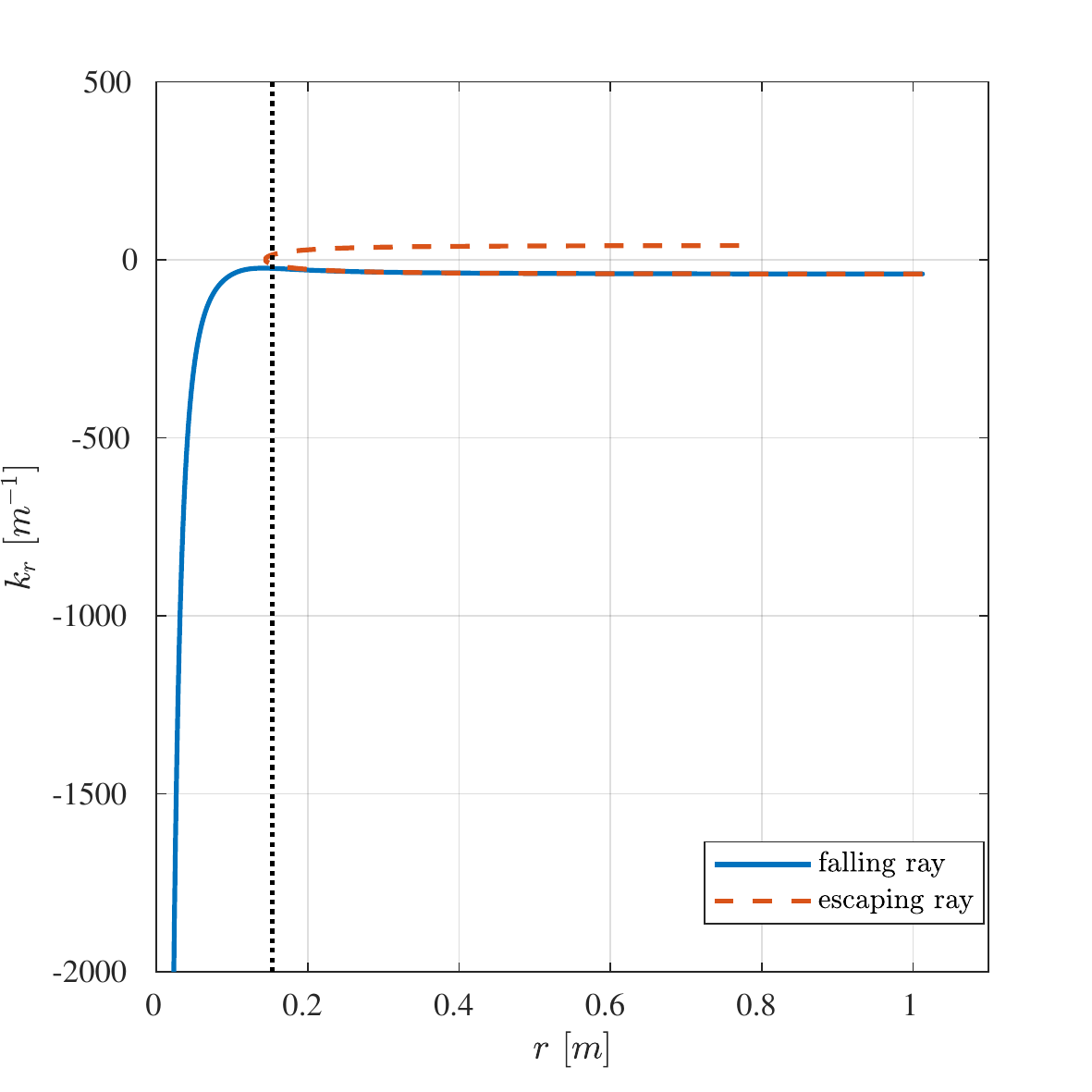}
\caption{Radial wave vector as a function of the ray falling onto the vortex. The dashed red line represents an ray escaping to infinity while the blue curve corresponds to a ray falling in the vortex. We can see that the two rays start at the same radius with the same radial wave-vector. The black dotted line represents the radius of the outer orbit. As we can see $k_r$ diverges as $r \to 0$ for the falling ray. This effect happens after the ray crosses the outer orbit.
}
\label{k_vs_r}
\end{figure}

Moreover, it is instructive to discuss the case of a vortex without a drain. By taking the limit $D \to 0$ in Eqs.~\eqref{radius_rel} and \eqref{rad_deep}, we see that the co-rotating orbit degenerates to $r=0$, but the counter-rotating one still exists at a nonzero radius (e.g. $C/2c$ in the shallow water regime). 
For sound waves, it was shown in \cite{Nazarenko94} and \cite{Nazarenko95} that rays approaching $r=0$ of an ideal non-draining vortex ($D=0$ in Eq.~\eqref{DBT}) see their wavelength decrease to zero, and ultimately reach the dissipative scale.
As such rays do not escape, it suggests that the scattering will be similar to a vortex with small drain (that is $D \neq 0$ but $D \ll C$). In particular, we expect to have a counter-rotating orbit, as well as the associated resonance effects. This behaviour is also present in the current system both in the non-dispersive (similar to the case studied in \cite{Nazarenko94,Nazarenko95} and in the dispersive regime. Fig.~\ref{k_vs_r} shows that the wavelength of a ray collapsing to the centre tends to $0$.

\subsection{Ray scattering}
We now study a family of characteristics (rays) encroaching upon the vortex from the right ($x \to +\infty$) with an angular frequency $\omega$ and various impact parameters $b$, which is the ray equivalent of sending a plane wave. Here, $m$ is a continuous parameter which is given in terms of the impact parameter via the relation 
\be \label{impact_param}
b = \frac{m}{k_{\mathrm in}} + \frac{C}{v_g^{\mathrm in}}. 
\ee
Here $k_{\mathrm in}$ is the \textit{incoming} wavenumber satisfying the dispersion relation (or the Hamiltonian constraint) at infinity for a given $\omega$ and $v_{g}^{\mathrm in}$ is the group velocity at infinity (as the flow is negligible at infinity, the group velocity in the fluid frame and in the laboratory frame are identical). We numerically solve Hamilton's equations, Eqs.~\eqref{Ham_eq}, to find the characteristics. A standard fourth order Runge-Kunta (RK4) scheme is used. To ensure the validity of our numerical simulation, the step size is chosen such that along the rays the adimensional quantity $|\mathcal{H}/ \omega^2|$ is smaller than $ 10^{-7} $. Our solution therefore satisfies the Hamiltonian constraint. 
To numerically compute the amplitude of the wave, we use Eq.~\eqref{transport_eq} with $\nu = 0$. This shows that the flux of wave action is conserved along a tube of rays. Numerically, we use this conservation to obtain the change in amplitude between neighbouring points along two neighbouring rays, by ensuring that the product of the distance between the rays and the wave action is constant. This means that if the rays converge (diverge), the amplitude will increase (decrease). 

\begin{figure}[!]
\centering
  \includegraphics[scale=1.3]{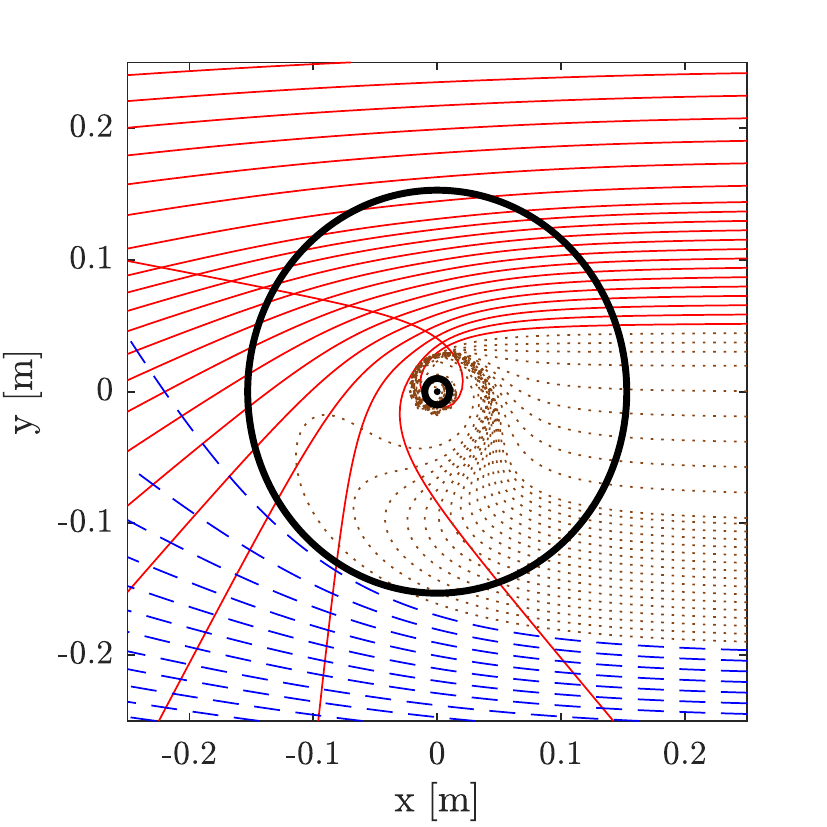}
  \caption{Congruence of rays incoming from far right on a draining vortex flow. The flow parameters are $C=1.6\times10^{-2} \text{m}^{2} \text{s}^{-1}$, $D=1\times10^{-3} \text{m}^{2} \text{s}^{-1}$ and $h_0=0.06 \text{m}$. They correspond to the experimental flow of the experiment presented in Chapter~\ref{Superradiance_sec}. The angular frequency of the wave is $\omega =19.8 \; \mathrm{rad/s}$. For these parameters, we have $h k_{\mathrm{in}} \simeq 2.4$ and hence are in the deep water regime. The red (and blue) curves represent the co-rotating (and counter-rotating) rays escaping to infinity. The dotted brown lines are the rays falling in the hole. The two black circles represent the orbital radius for co- and counter- rotating rays, respectively at $r=1$ cm and $r=15.3$ cm.}
  \label{characteristics}
\end{figure}

Fig.~\ref{characteristics} depicts a congruence of rays at fixed angular frequency \mbox{$\omega = 19.8 \; \mathrm{rad/s}$} impinging on a draining bathtub vortex with parameters \mbox{$C=1.6\times10^{-2} \text{m}^{2} \text{s}^{-1}$}, \mbox{$D=1\times10^{-3} \text{m}^{2} \text{s}^{-1}$} and $h_0=0.06 \text{m}$. 
These values are chosen to correspond to the laboratory study presented in Chapter~\ref{Superradiance_sec}. 
We can clearly distinguish two types of rays. The first type are rays that are able to escape to infinity (co-rotating in red and counter-rotating in dashed blue). The impact parameter for those rays are beyond some critical values $b_{\mathrm c}^{\pm}(\omega)$. We note that $|b_{\mathrm c}^{-}(\omega)|>|b_{\mathrm c}^{+}(\omega)|$, implying that co-rotating rays are able to go closer to the vortex than counter-rotating ones. The second type of rays are those that fall into the vortex core (in dotted brown). 
We note from Eq.~\eqref{impact_param} that the impact parameter depends on the group velocity and therefore on the frequency of the wave. This behaviour differs from the shallow water regime where the group velocity matches the phase velocity and is constant for all frequency.

\begin{figure}[!]
\centering
  \includegraphics[scale=1.3]{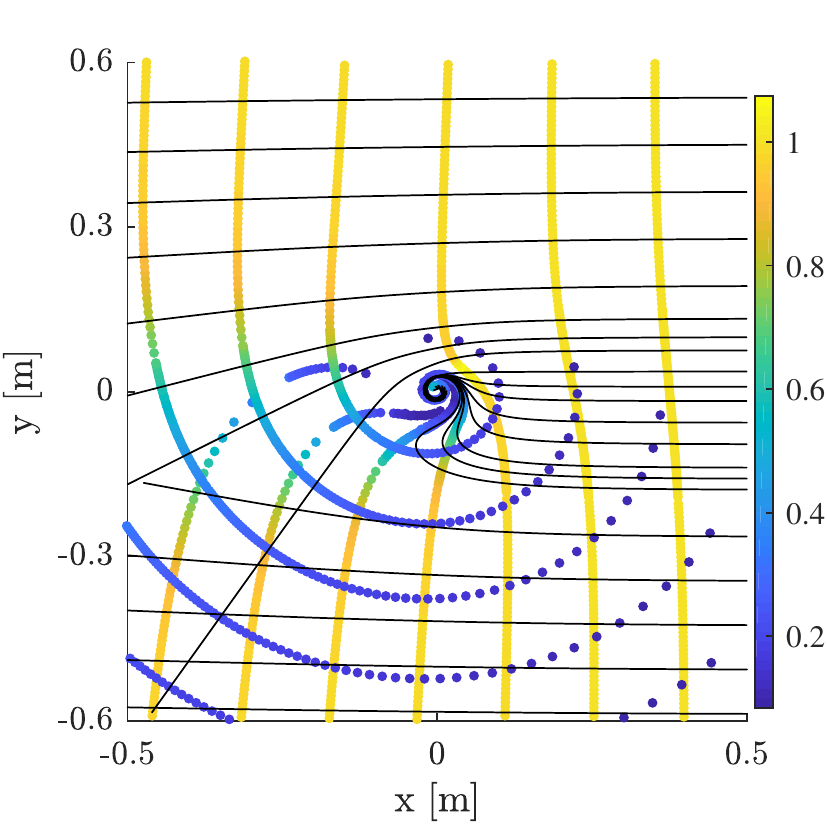}
  \caption{Reconstructed eikonal wavefront from the phase along the rays. The wave and flow parameters are similar to the ones used in Fig.~\ref{characteristics}. The colour scale represents the amplitude of the wave normalized to one initially. The black line are some rays from Fig.~\ref{characteristics}.}
  \label{wavefront}
\end{figure}

In the process of obtaining the rays by solving Hamilton's equations, we also compute the phase of the wave along each trajectory. It is then possible to reconstruct the eikonal wave fronts by finding the constant phase points along each ray (see \cite{Dolan_Dempsey} for a shallow-water study). The wave fronts are presented in Fig.~\ref{wavefront}. 
The coloured dots represent the location of the constant phase points and the colour scale gives the amplitude of the wave at these points. Far from the vortex, where the flow becomes negligible, we can see that the wave fronts are orthogonal to the rays (black lines). On the contrary, closer to the vortex, we clearly observe the non-orthogonality of the wave fronts and the rays where the flow becomes more rapid.

When varying the angular frequency $\omega$, the rays display a similar pattern as in Fig.~\ref{characteristics}, but the orbital radii $r_\star^\pm$ vary. When increasing the angular frequency, they interpolate between their shallow water value to the deep water behaviour of Eq.~\eqref{rad_deep}. To illustrate this, we plotted the dependence of the orbital radii with the frequency in Fig.~\ref{radius_orbit}.

\subsection{Experimental validation}

\begin{figure}
\centering
\includegraphics[trim=0.7cm 0 0 0]{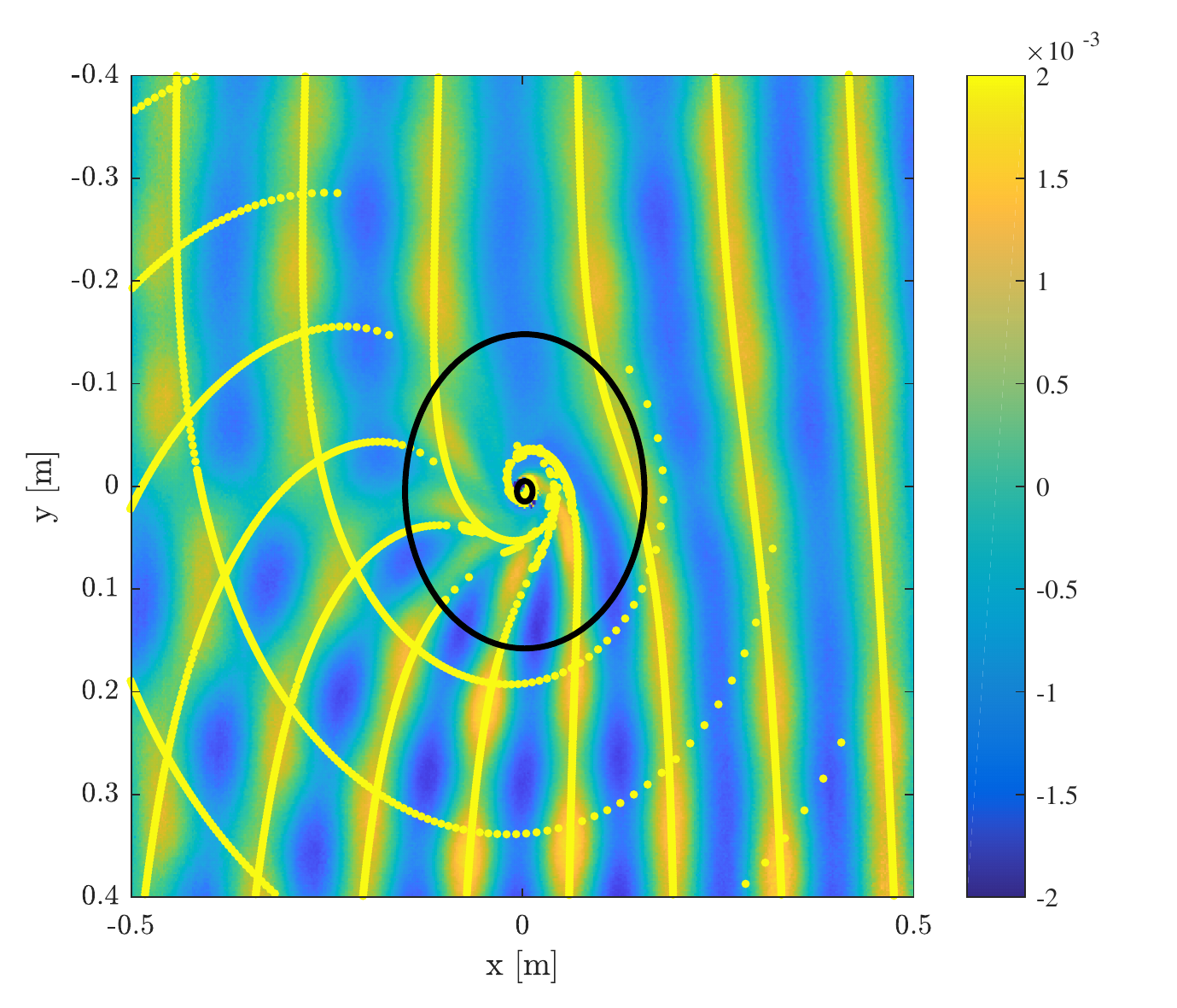}
\caption{Comparison between eikonal wave front computed numerically and experimental data. The bright yellow dots represent the eikonal wave front reconstructed from the phase along the rays. The two black circles are the unstable orbits, also present in Fig.~\ref{characteristics}. The background image is a measurement of the free surface of the water. The flow parameters and wave frequency are the same as in Fig.~\ref{characteristics}. The frequency of the eikonal wave ($3.15$ Hz) is chosen so as to obtain the best fit with the $3.27$ Hz wave of Fig.~\ref{wave_charac} in Chapter~\ref{Superradiance_sec}. The two frequencies agrees within error bars (about $4\%$). The colorbar represents the amplitude of the wave in metres.
}
\label{wavefront_exp}
\end{figure}

In Fig.~\ref{wavefront_exp} we compare our ray-tracing results with experimental data taken from the wavetank  experiment of~\cite{Superradiance} and which will be detailed in Chapter~\ref{Superradiance_sec}. 
The circulation-to-draining ratio in this experiment was approximately $C/D \simeq 16$. 
One can see that, broadly, the eikonal wave fronts agree well with the experiment. 
We observe small deviations in two regions: close to the centre of the vortex, and after the wave has propagated through the vortex (on the left of the image). 
The eikonal approximation we used is expected to degrade close to the centre of the vortex, where the flow varies more rapidly. 
Moreover, cumulative errors might explain the shift between our simulation and the data after the wave has propagated away from the vortex (on the far left side of Fig.~\ref{wavefront_exp}). 

The good agreement between our numerical solution and the experimental data suggests that circular rays are present in the system. While the inner one (co-rotating with the vortex) is very close to the centre, in a region where the vortex model of Eq.\eqref{DBT} becomes questionable, the outer one (counter-rotating) lies in a region where our method provides an accurate description. To estimate the order of the error, we compare the gradient scale of the background to the frequency of the wave: $|\partial_r v_\theta| / \omega$. 
Near the outer LR $r_\star^+$, this ratio reduces to the inverse of the azimuthal number $m_\star$ and is approximately $0.1$. 
As a last remark, we point out that within our approximation, the location of the rays are usually more accurate than the phase itself. This is because the phase is rapidly varying, but the local momentum is slowly varying (see the discussion in section 4.3.4 of \cite{Buhler}). Hence, this fact and the good agreement in figure~\ref{wavefront_exp} strengthens our conclusion regarding the presence of circular orbits around the vortex of Fig.~\ref{wavefront_exp}. 
This experiment is an observation of the analogue light bending effect in dispersive systems.

\section{Summary and discussion}

In this chapter, we have reviewed the technique of gradient expansion and its application to physical systems. 
We have seen that high frequency waves can be thought of as effective particles. 
The world-line of these effective particles are given by the characteristics of the waves.
In the shallow water limit, we have seen that these characteristics correspond to the geodesics of the underlying effective space-time revealed by the fluid-gravity analogy.

We have applied the ray-tracing method to a realistic dispersive, dissipative wave equation. 
We argued that fundamental features of vortex wave scattering can be understood with reference to a pair of (frequency-dependent) circular rays (rings), which are analogous to the LRs of black holes.

In addition, we have reconstructed the wave fronts of a plane wave scattering on a
vortex using the ray-tracing approximation. This allows us to test our methods against
experimental results. The agreement with the data
suggests the presence of an outer light ring (for counter-rotating modes) and raises the
challenge of observing it. This will be the subject of Chapter~\ref{Ringdown_sec}.


%
\chapter{Superradiance scattering}\label{Superradiance_sec}

\epigraph{\textit{It would seem that the physical intuition ought not only provide the mathematician with interesting and challenging conjectures, but also show him the way toward a proof and toward possible generalisation.}}{Kac}

\section{Introduction}

As we have seen in the previous section, the scattering of surface waves incident on a vortex flow can be chiefly understood by means of a Hamiltonian function. 
This description, based on an asymptotic expansion, has the advantage that it provides us with predictions that can be directly verified against experimental data. 
The downside is that it does not capture all phenomena at place during waves-vortex scattering. 
One of such phenomena is the so-called rotational superradiance effect and it is the subject of the present section.

Superradiance is a radiation enhancement process, during which a wave scattering of a rotating object will extract some of the object's energy. 
This process was originally derived by Zel'Dovich by considering electromagnetic radiations incident on a rotating cylinder~\cite{zeldovich1,zeldovich2}.
However, it was soon realised that this effect was not only restricted to the specific Zel'Dovich setup but can be applied to a broad class of a system~\cite{Brito:2015oca}.
This is because the superradiance effect does not rely on the underlying dynamical equations governing the system but rather on two conditions that can be satisfied in a large variety of physical situations.
The two fundamental conditions for superradiance to happen are:
\begin{itemize}
\item The system needs to allow for the presence of negative energy modes. When considering rotational superradiance, the scatterer must be a rotating object\footnote{We note that other amplification processes exist when considering waves scattering of a discontinuous media, known as over-reflection~\cite{Acheson76}, but we will focus on rotational superradiance here.}.
\item The system needs to provide an absorption mechanism.
\end{itemize}

One interesting system to satisfy these conditions are rotating BHs. 
Indeed, the ergosphere provides a mechanism to allow for the presence of negative energy and the event horizon act as a perfect absorber. It was indeed shown, soon after Zel'Dovich's seminal paper that rotating BHs can be the stage of the superradiance effect~\cite{misner,staro1,staro2}.
Since then, a long list of systems spanning the various fields of physics have been found (and are still looked for) to exhibit this effect. Amongst them are
water waves on a cylinder~\cite{Cardoso_detecting},
orbital angular momentum beam incident on a disk~\cite{Cisco18},
or binary BHs~\cite{Wong:2019kru}.

Despite its long theoretical history, the rotational superradiance effect has only been observed recently in an experiment we have performed, and which will be the subject of this section. We will start by reviewing the superradiance effect in vortex flow and show that this system is indeed a suitable candidate to observe this effect. We will then describe the experimental setup and protocol we have used to perform this experiment. We will finally present our results, attesting of the first observation of the rotational superradiance scattering effect.

\subsection{Superradiant scattering in a vortex flow}

Here, we present the superradiance effect from an hydrodynamical point of view.
Our starting point is the wave equation for shallow water waves, Eq.~\eqref{wave_equation_shallow}:
\begin{equation}
\D^2 \phi - c^2\Delta \phi = 0.
\end{equation} 
The flow velocity profile is the usual DBT model:
\be
\vec v_0 = - \frac Dr \vec e_r + \frac Cr \vec e_\theta, 
\ee

As seen in Chapter~\ref{Intro_sec}, long surface gravity waves obey a wave equation analogous to massless scalar field on a curved space-time. 
In the case of the DBT flow, we have also seen that the metric of this effective space-time shares some similarities with the Kerr metric.
In particular it exhibits the analogue of a black horizon at $r_h = D/c$ and an ergosphere at $r_e = \sqrt{C^2 + D^2}/c$.

Using the symmetry of the system, namely rotational symmetry and time invariance, we can separate the wave equation by assuming the following ansatz for the solution:
\be
\phi(t,r,\theta) = e^{-i\omega t + i m \theta}R(r).
\ee
Substituting this into the wave equation, one gets:
\be\label{we1}
P(r) \p_r^2 R + Q(r) \p_r R + T(r)R = 0,
\ee
where $P,Q$, and $T$ are defined as:
\bea
P(r) &=& \frac{D^2}{r^2} - c^2 \\
Q(r) &=& -\frac{D^2}{r^3} - \frac{c^2}{r} + 2i\frac{D\omega}{r} - 2i \frac{CDm}{r^3} \\
T(r) &=& 2i\frac{CDm}{r^4} + \frac{c^2m^2}{r^2} - \frac{C^2m^2}{r^4} + 2\frac{Cm\omega}{r^2} - \omega^2.
\eea

If we further assume that $R = \tilde{R}Z(r)$, where $Z$ is given in \cite{Berti:2004ju} and introduce the tortoise coordinate $r_\star$ via:
\be
\frac{dr_\star}{dr} = \left( 1 - \frac{D^2}{c^2r^2}\right)^{-1},
\ee
the wave equation, given by Eq.~\eqref{we1}, simplifies to:
\be \label{WE_tortoise}
\frac{d^2\tilde{R}}{dr_\star^2} - V(r) \tilde{R} = 0
\ee
with the effective potential $V$ defined as:
\be \label{potential_shallow}
 V(r) = -\left[ \left(\omega - \frac{Cm}{r^2} \right)^2 - \left( c^2 - \frac{D^2}{r^2}\right)\left( \frac{m^2 - 1/4}{r^2} + \frac{5D^2}{4r^2}\right) \right].
\ee
Note here that $r$ is understood as $r = r(r_\star)$.
This equation can be solved in two asymptotic regimes, namely at the horizon and at infinity.

At infinity, the potential simply reduces to:
\be
V(r) \rightarrow -\omega^2 \text{ for } r \rightarrow \infty.
\ee
Therefore a solution at infinity of Eq.~\eqref{WE_tortoise} is:
\be
R_\infty(r_\star) = A e^{-i\omega r_\star} + B e^{i \omega r_\star} = A f_1(r_\star) + Bf_2(r_\star) \text{ for } r_\star \rightarrow \infty.
\ee
We can see that this consists in a radially incoming and outgoing wave. 
In particular, $(f_1,f_2)$ forms a basis of the solution space admitting the asymptotic form of plane waves at infinity. 
Similarly, at the horizon the potential reduces to the simple form:
\be
V(r) \rightarrow -\tilde{\omega} = -\left(\omega - \frac{mC}{r_h^2} \right)^2 \text{ for } r \rightarrow r_h,
\ee
such that a solution is given by:
\be
R_h(r_\star) = C e^{-i\tilde{\omega} r_\star} + D e^{i\tilde{\omega} r_\star}  = C g_1(r_\star) + D g_2(r_\star) \text{ for } r_\star \rightarrow -\infty.
\ee
Here again $(g_1,g_2)$ forms a basis of the solution space admitting the simple plane wave behaviour at $r_\star \rightarrow -\infty$.
Since $(f_1,f_2)$ and $(g_1,g_2)$ are bases of the solution space, we can express any solution $f$ to Eq.~\eqref{WE_tortoise} as a linear combination of these functions:
\be
 f = \alpha_\infty f_1(r_\star) + \beta_\infty f_2(r_\star) = \alpha_{H} g_1(r_\star) + \beta_h g_2(r_\star).
\ee
The coefficients $\alpha_\infty,\beta_\infty, \alpha_h,\beta_h$ correspond to the amplitude of the ingoing/outgoing waves both at infinity and at the horizon.
It is possible to relate these amplitude by noticing that Eq.~\eqref{WE_tortoise} is a second order ODE in $r_\star$ with real coefficients which does not involve first order derivative. 
Therefore the Wronskian $W$, of any solution $f$, $W = f\p_{r_\star}\bar{f} - \bar{f} \p_{r_\star}f$ is conserved (here $\bar{f}$ denotes the complex conjugate of $f$)\footnote{
Suppose $f$ satisfies $f'' = Af$, with $A$ a real function. We have $W' = f'\bar{f}' + f \bar{f}'' - \bar{f}' f' - \bar{f} f'' = f\bar{Af} - \bar{f}Af = 0$ since $\bar{A} = A$.
}.
Calculating the Wronskian at the horizon using the asymptotic forms of $g_1$ and $g_2$ and at infinity using the asymptotic forms of $f_1$ and $f_2$, we get the following relation:
\be
\omega \left(¦\alpha_\infty¦^2 - ¦\beta_\infty¦^2 \right) = \Omega \left(¦\alpha_h¦^2 - ¦\beta_h¦^2 \right).
\ee
Imposing the condition that no waves can be outgoing at the horizon we have that $\alpha_h = 0$. Defining the reflection and transmission coefficient as:
\be
R = \frac{\alpha_\infty}{\beta_\infty} \text{ and } T = \frac{\beta_h}{\beta_\infty},
\ee
we have the relation:
\be
¦R¦^2 = 1 - \frac{\omega - m\Omega_h}{\omega} ¦T¦^2, 
\ee
where $\Omega_h = C/r_h^2$ is the angular frequency of the flow at the horizon.
We therefore see from the above expression that if the frequency of the incident wave is tuned such that
\be
\omega - m\Omega_h < 0,
\ee
then the reflection coefficient will be greater than one and the wave will have extracted energy. 
This is the superradiance effect in a vortex flow and the above condition is the superradiance condition.
From this condition, it is also straightforward to see that superradiance can only occurring when $m>0$.
It is possible to solve numerically Eq.~\eqref{WE_tortoise} in order to extract the reflection coefficient spectrum. 
This is done with the flow parameters $C=D=1$. The depth of the water is chosen such that $c=1$. 
The boundary conditions are implemented at the horizon where we impose the presence of purely ingoing modes.
Once the solution $f$, to the wave equation is known, we extract the reflection coefficient by taking linear combination of $f$ and its derivative far away.
Explicitly, we have:
\be
R = \abs{\frac{i\omega f_\infty - f_\infty'}{i\omega f_\infty + f_\infty'}}, 
\ee
where $f_\infty$ and $f_\infty'$ are the values of the solution and its derivatives at the far end of the radial grid used in the simulation $r_N$ where $r_N = 100 r_h$. 

\begin{figure}
\includegraphics[scale=1]{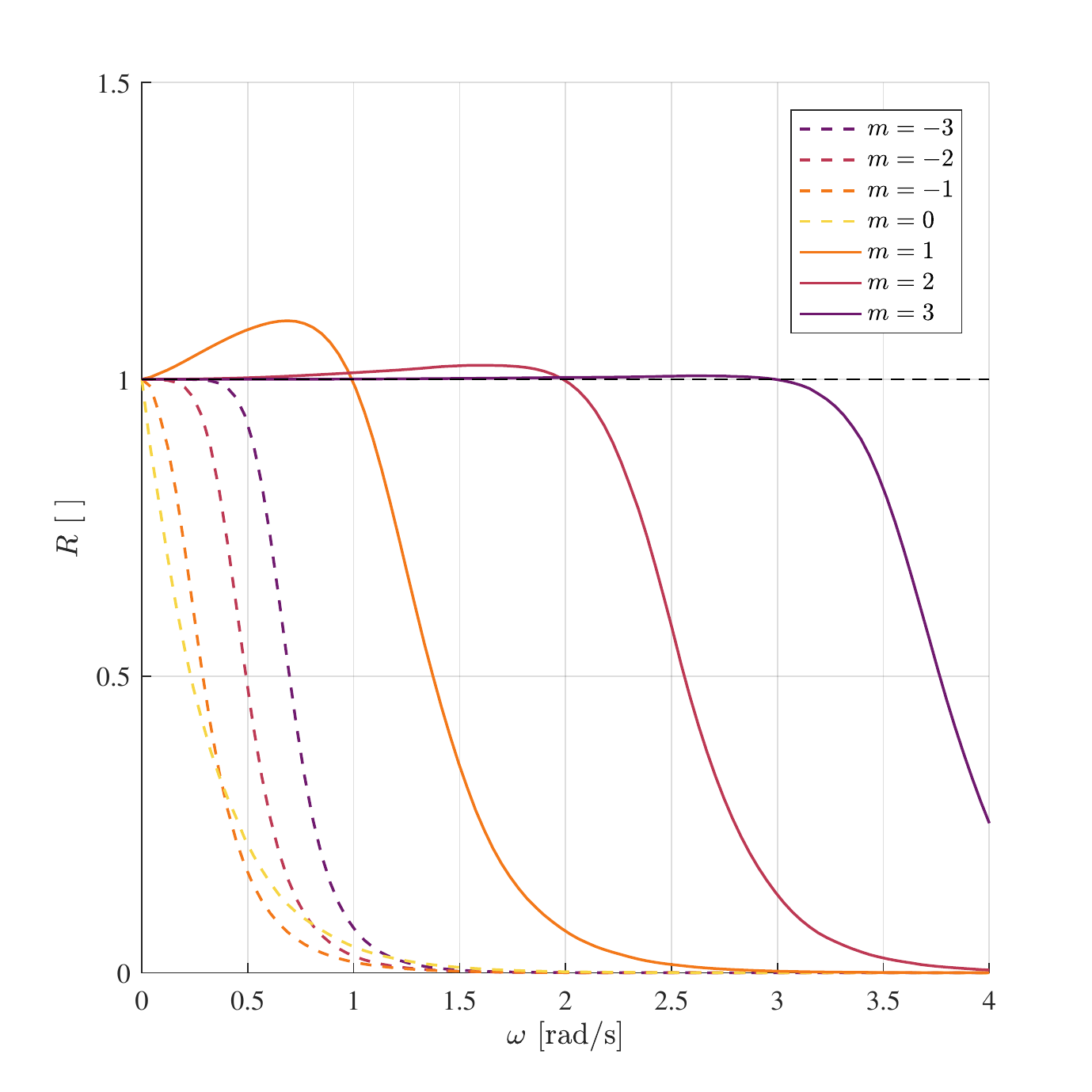}
\caption{Reflection coefficient spectrum for various azimuthal numbers $m$. The solid lines represent the spectrum of the co-rotating modes, $m>0$, while the dashed lines represent the spectrum of counter-rotating modes. The flow parameters are $C=D=1$ and the propagation is set to be $c=1$. This is such that $\Omega_h =1$. We can see that the positive $m$ modes are amplified when their frequency satisfies the superradiance condition, i.e. $\omega < m\Omega_h$.}
\end{figure}

\section{Experimental implementation}
We now turn our attention to an experimental implementation of a system capable of exhibiting the superradiance effect.
\subsection{Experimental setup}\label{Exp_setup_sec}
\begin{figure}
\centering
\includegraphics[scale=0.8, trim = 2cm 0 0 0]{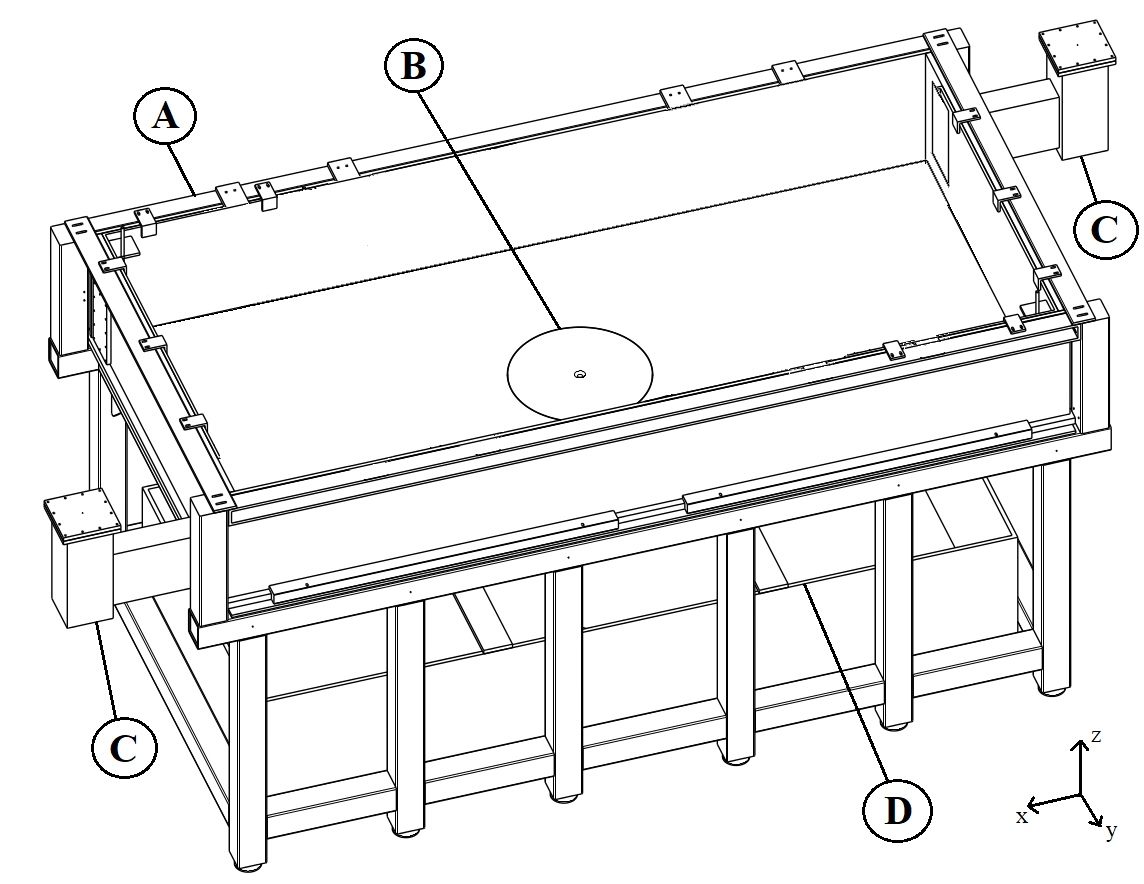}
\caption{Technical drawing of the watertank. \textbf{(A)} is an acrylic glass water tank with a width of $1.5~m$, a length of $3~m$ and a height of $0.4~m$. \textbf{(B)} is an interchangeable centre plate with various drain hole in its centre. \textbf{(C)} are the water inlets from which the water is fed into the tank. \textbf{(D)} shows a bottom tank containing $2000~L$ of water from which water is pumped to tank \textbf{(A)} via \textbf{(C)}. A picture of the real tank is shown in Fig.~\ref{Experimental_apparatus_pic}.}\label{Experimental_apparatus_tech_draw}
\end{figure}

\begin{figure}
\centering
\includegraphics[scale=0.25, trim = 4cm 0 0 0]{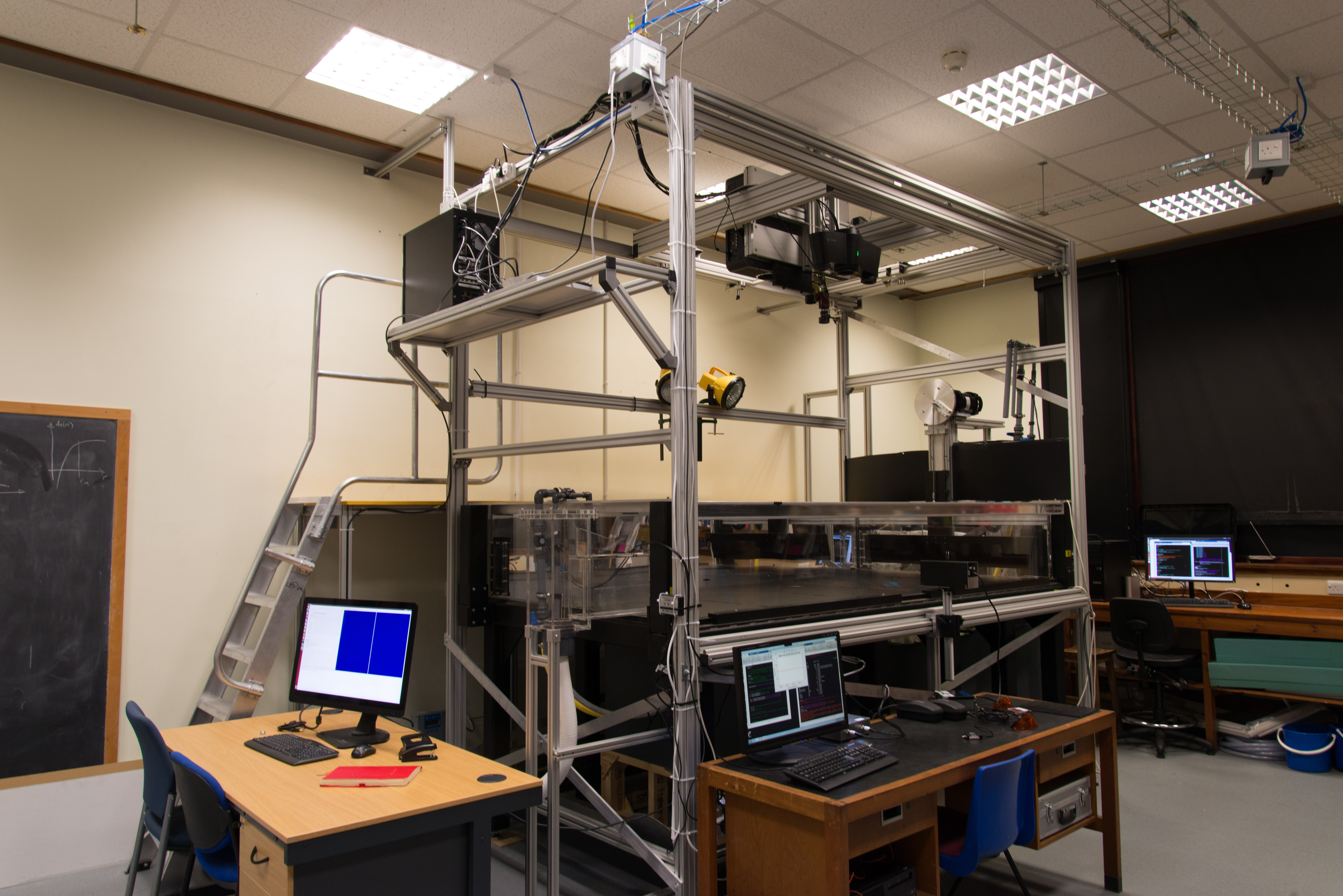}
\caption{Picture of the experimental setup corresponding to the technical drawing presented in Fig.~\ref{Experimental_apparatus_tech_draw}.}\label{Experimental_apparatus_pic}
\end{figure}

We conducted our experiment in a $3$~m long by $1.5$~m wide rectangular tank, as shown in Figs.~\ref{Experimental_apparatus_tech_draw} and \ref{Experimental_apparatus_pic}. 
The side walls of the upper tank are made of acrylic glass while the entire tank rests on a steal table. 
In the middle of the upper tank (A), there is an interchangeable plate (B) allowing us to adjust the size of the drain hole. 
For this experiment, we have used a centre plate with a $4$~cm diameter hole. 
Water is pumped from the bottom tank (D) to the upper tank (A) via a Grunfos\textregistered ~pump through an inlet (C). 
While the tank is equipped with two inlets, in this experiment we only used the inlet on the opposite side of the wave generator. 
The inlet is positioned asymmetrically in order to give angular momentum to the flow and create a rotating and draining vortex flow. 
The inlet is partially filled with open cell reticulated foam in order to reduce turbulence and noise when feeding the water. The water depth is kept constant at $6.25\pm05~\mathrm{cm}$. 

A wave generator is mounted on one side of the tank (see Fig.~\ref{Experimental_apparatus_pic}). 
The wave generator is made of a $1.5$~m long and $20$~cm high aluminium plate oscillating up and down the water. 
The plate's motion is generated by a motor capable of creating oscillation ranging from $0$ to $4.5$ Hz. 
The plate is aligned with the wall of the tank so that plane waves are sent along the $x$ direction.

On the opposite side of the wave generator is placed an absorption beach to minimize the amount of reflection. 
We have checked that the amount of reflection from the beach is about $5\%$ (see Table~\eqref{beach_ref}).

\begin{table}
\caption{Reflection coefficient from the beach}\label{beach_ref}
\centering
\makebox[\textwidth]{\begin{tabular}{||c||c|c|c|c|c|c|c||}
\hline
Frequency (Hz) & $2.7$ & $2.9$ & $3.1$ & $3.3$ & $3.5$ & $3.7$ & $4$ \\
\hline
Reflection (\%) & $4.7$ & $4.4$ & $5.7$ & $4.3$ & $4.0$ & $4.1$ & $2.9$\\
\hline
Relative Error ($\pm$\%) &$0.2$ & $0.1$ & $0.3$ & $0.4$ & $0.2$ & $0.5$ & $0.2$\\
\hline
\end{tabular}}
\end{table}

\begin{figure}
\begin{subfigure}{.5\textwidth}
  \includegraphics[trim = 2cm 0 0 0]{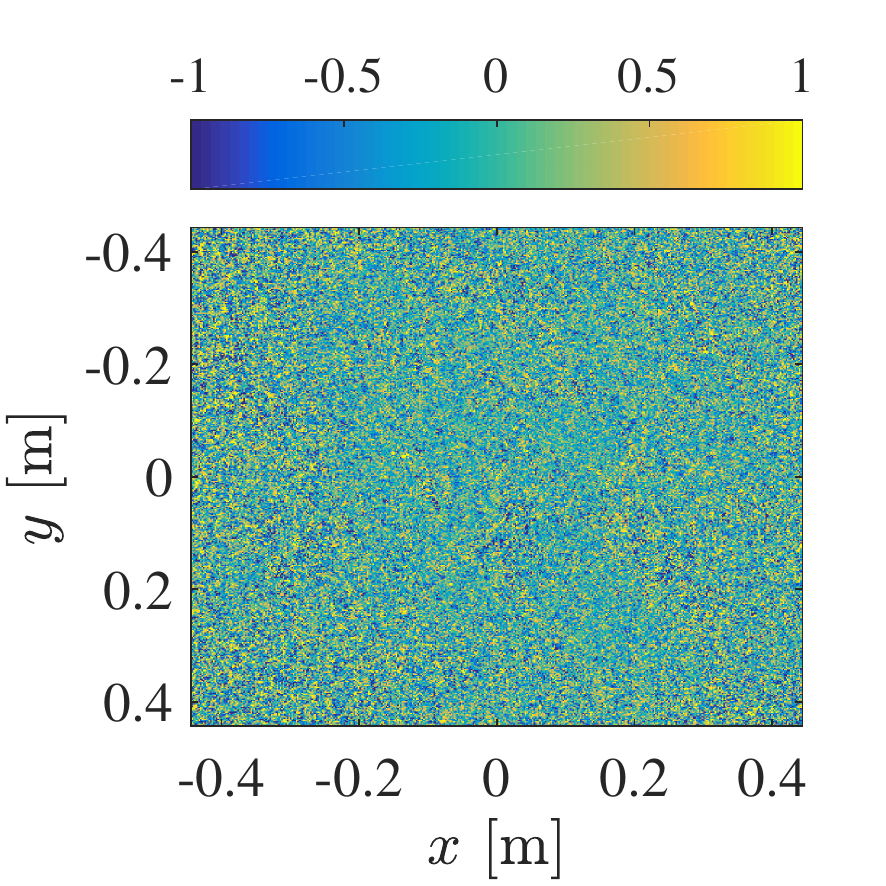}
  \caption{Measurement of a flat air-water \\ interface.}
  \label{noise_surface}
\end{subfigure}%
\begin{subfigure}{.5\textwidth}
  \includegraphics{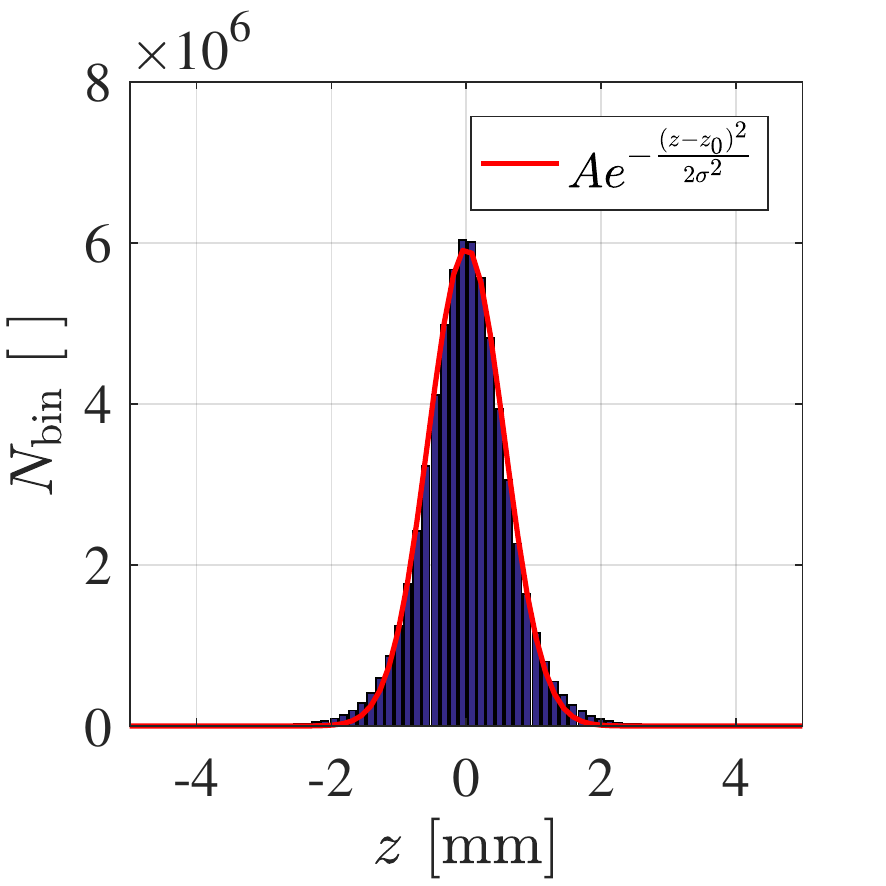}
  \caption{Histogram of the water elevation $z$ measured by the sensor.}
  \label{histo}
\end{subfigure}
\caption{The left panel shows a flat water surface at one instant of time as measured by our sensor. The colour represents the amplitude of the water elevation and is measured in millimetres. The right panel is a histogram of the water elevation $z$ measured during $13.2$~s. We can see that the noise measured by the sensor follows a Gaussian distribution centred around $0$ with standard deviation $\sigma = 4.0\times 10^{-4}$~m.}\label{resol_sensor}
\end{figure}

Above the centre plate is placed our sensor to record the free surface. 
We have used a high-speed 3D air-fluid interface sensor~\cite{sensor}. 
This sensor is composed of one pattern projector and two high-speed cameras. 
The camera is placed at about $1.2$~m above the free surface. 
Since water is a transparent liquid, we added fluorescein sodium salt in the water to visualize the free surface.
Optical filters on the projector and cameras are then used in order to remove direct reflections of the projected pattern. 
Specifically, a blue filter is placed in front of the projector so that the pattern projected is mostly composed of blue light. 
This blue light is absorbed by the fluorescein and re-emitted into green light. 
Green filters in front of the camera allow for the observation of the re-emitted pattern only while filtering direct blue reflection of the pattern (see Fig.~\ref{exp_run_pic}). 
With this sensor we can record fluctuations of the surface down to $0.4$ mm with a 1-$\sigma$ confidence (see Fig.~\ref{resol_sensor}).

\begin{figure}
\centering
\includegraphics[scale=0.4, trim = 0cm 0 0 0]{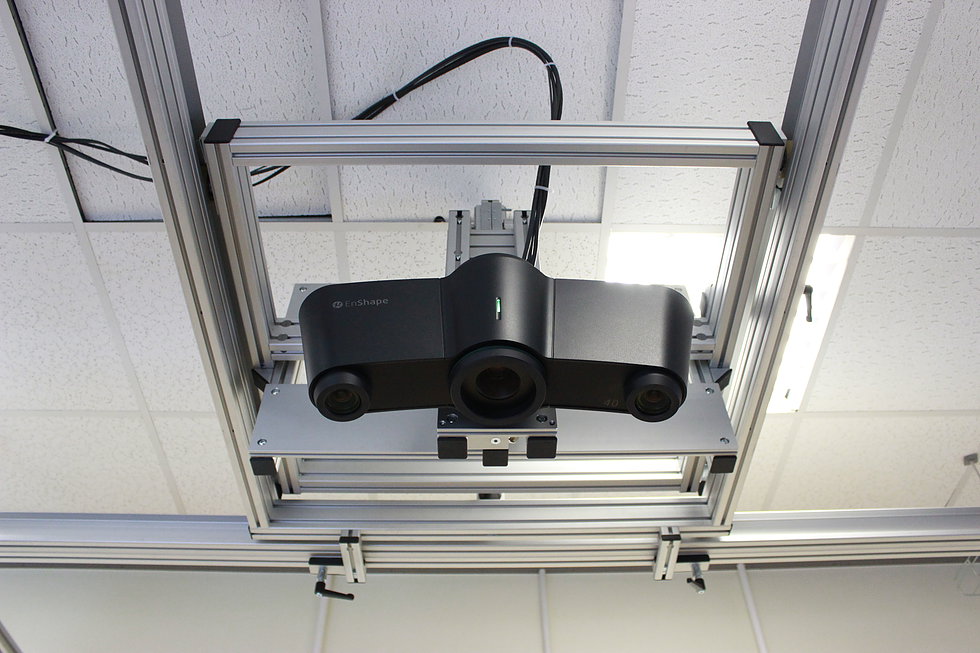}
\caption{High-speed 3D air-fluid interface sensor. The pattern projector is in the middle while the two cameras are on each side of the projector.}\label{sensor_pic}
\end{figure}

\subsection{Measuring the flow: Particle Imaging Velocimetry} \label{PIV_subsec}

In order to measure the velocity field of the vortex flow, we have performed a Particle Imaging Velocimetry (PIV) measurement.
The flow is seeded with flat paper particles of mean diameter $d = 2~ \mathrm{mm}$. 
The particles are buoyant which allows us to evaluate the velocity field exclusively at the free surface. 
The amount by which a particle deviates from the streamlines of the flow is given by the velocity lag $U_s = d^2(\rho-\rho_0)a/18\mu$ \cite{PIVthesis}, where $\rho$ is the density of a particle, $\rho_0$ is the density of water, $\mu$ is the dynamic viscosity of water and $a$ is the acceleration of a particle. 
For fluid accelerations in our system this is at most of the order $10^{-4}~ \mathrm{m/s}$, an order of magnitude below the smallest velocity in the flow. 
Thus, we can safely neglect the effects of the velocity lag when considering the motions of the particles in the flow.

The surface is illuminated using two light panels positioned at opposite sides of the tank. The flow is imaged from above using a Phantom Miro Lab 340 high speed camera at a frame rate of $800~ \mathrm{fps}$ for an exposure time of $1200 ~\mu\mathrm{s}$. The raw images are analysed using \textit{PIVlab}~\cite{PIVlab,PIVlab2} by taking a small window in one image and looking for a window within the next image which maximizes the correlation between the two. By knowing the distance between these two windows and the time step between two images, it is possible to give each point on the image a velocity vector. This process is repeated for all subsequent images and the results are then averaged in time to give a mean velocity field. 

\begin{figure}
\includegraphics[scale=1]{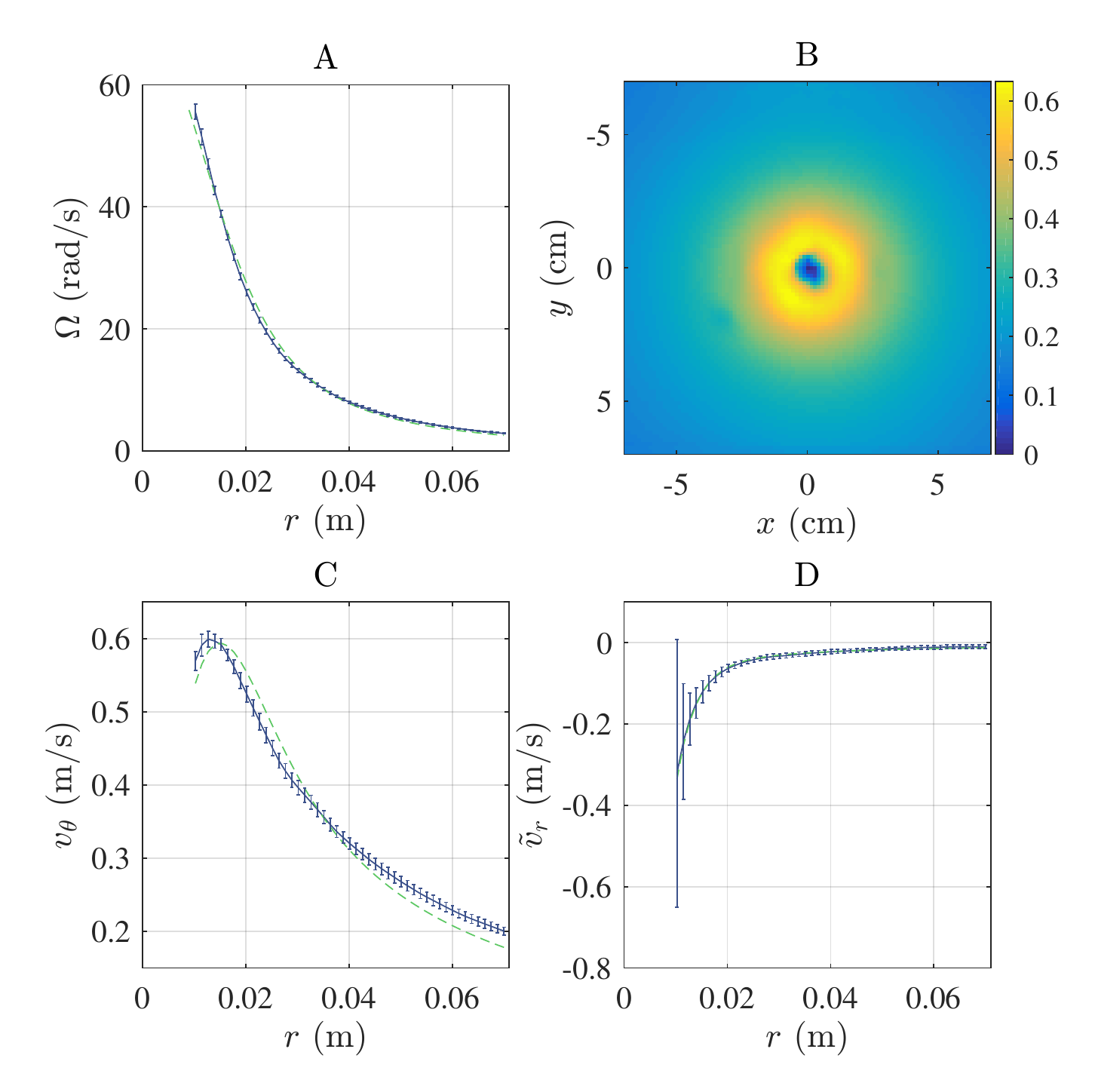}
\caption{PIV measurements of the velocity field averaged of $10$ experiments. The error bars correspond to standard deviations across the measurements. \textbf{(A)} Angular frequency profile as a function of $r$. \textbf{(B)} Norm of the velocity field of the background flow. \textbf{(C)} $v_\theta$ profile as function of $r$.  \textbf{(D)} $\tilde v_r$ profile as function of $r$. The profiles are fitted with a model of the Lamb vortex type in Eq.~\eqref{lamb}, dashed-green line.}\label{PIV_pic}
\end{figure}

The resulting velocity field is decomposed onto an $(r,\theta)$-basis centred about the vortex origin to give the components $v_r$ and $v_{\theta}$. The centre is chosen so as to maximize the symmetry. In Fig.~\ref{PIV_pic} 
we show the norm of the velocity field on the free surface. We see that our vortex flow is symmetric to a good approximation. To quantify the asymmetry of the flow, we estimate the coupling of waves with $m \neq m'$ through asymmetry. The change of the reflection coefficient due to this coupling is of the order of $|\tilde v^l/v_g|$, where $\tilde v^l$ is the angular Fourier component of azimuthal number $l=m-m'$. This ratio is smaller than $3\%$ in all experiments. To obtain the radial profiles of $v_r$ and $v_{\theta}$, we integrate them over the angle $\theta$. In panel (C) and (D) of Fig.~\ref{PIV_pic} we show $v_{\theta}$ and the inward velocity tangent to the free surface, $\tilde v_r = -\sqrt{v_{r}^2+v_{z}^2}$, as functions of $r$. 

We compare the data for $v_{\theta}$ with the Lamb vortex \cite{Lautrup2011},
\begin{equation} \label{lamb}
v_{\theta}(r,h) = \frac{\Omega_0\, r_0^2}{r}\left[1 - \exp\left(-\frac{r^2}{r_0^2}\right)\right], 
\end{equation}
where $\Omega_0$ is the maximum angular velocity in the rotational core of characteristic radius $r_0$. 
(For $v_{\theta}$ we have $\Omega_0=69.4\;\mathrm{rad/s}$ and  $r_0=1.34\;\mathrm{cm}$, and for $v_{r}$ we have $\Omega_0=-4.52\;\mathrm{rad/s}$ and  $r_0=1.39\;\mathrm{cm}$.) 
Outside the vortex core, this model reduces to the characteristic $1/r$ dependence of an incompressible, irrotational flow depending only on $r$. By observing that $v_{\theta}$ and $v_r$ exhibit similar qualitative behaviour, $v_r$ is also fitted with a model of the form of Eq.~\eqref{lamb}. 
Panels (C) and (D) of Fig.~\ref{PIV_pic} show that Eq.~\eqref{lamb} captures the essential features of the measured velocity profiles. The angular velocity of the flow is given by $\Omega(r) = v_{\theta}/r$ which is shown in panel (A) of Fig.~\ref{PIV_pic}. From this plot we note that $\Omega$ reaches large enough values.

\subsection{Experimental protocol: Wave scattering experiment}

\begin{figure}
\includegraphics[scale=0.1, trim=1cm 0 0 0]{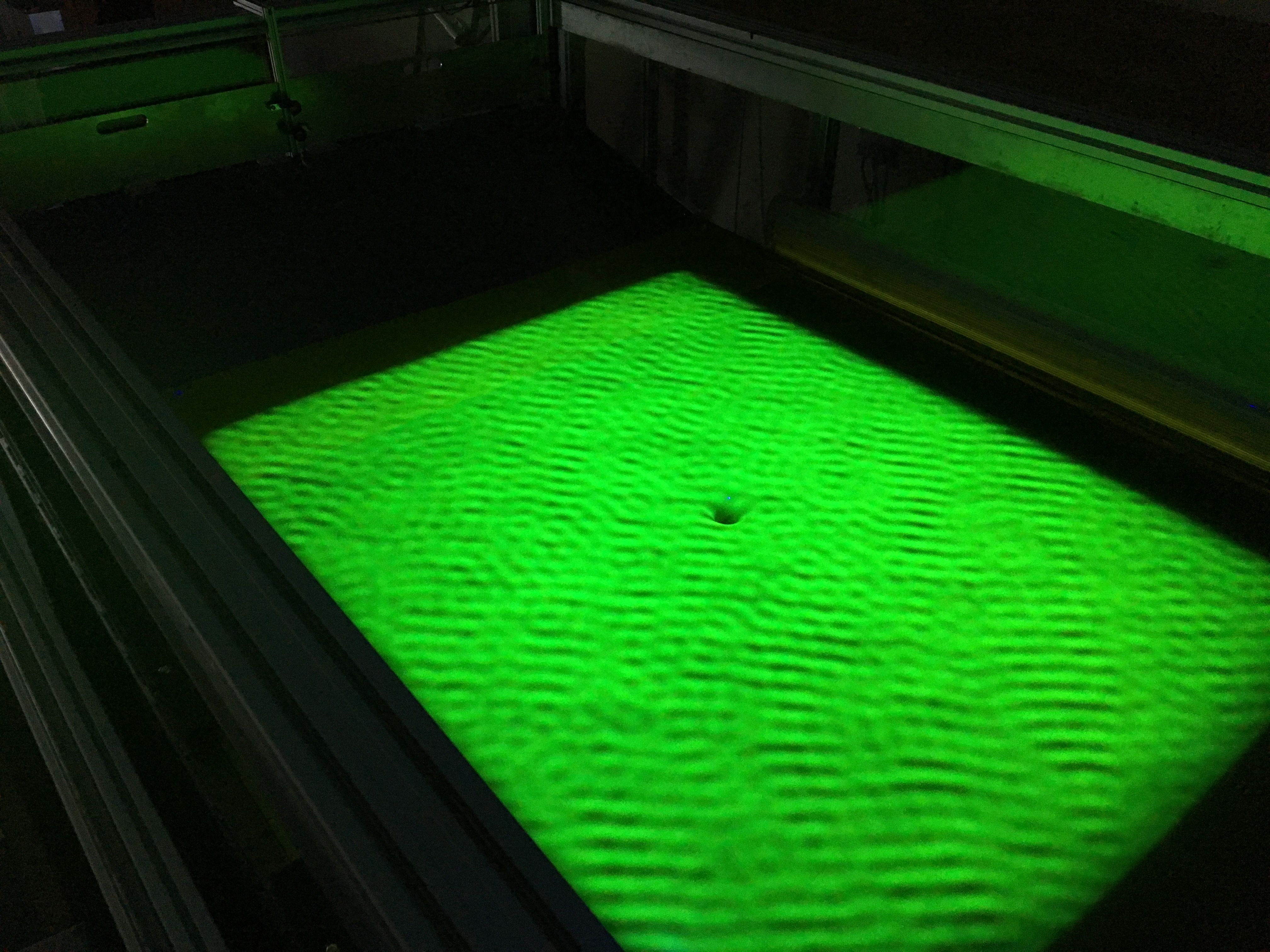}
\caption{Picture of the vortex flow illuminated by the sensor. The random pattern is projected with blue light from the sensor onto the water surface. The blue light is absorbed by the dye in the water and re-emitted in green light that is then recorded by the camera.}\label{exp_run_pic}
\end{figure}

Using the set-up described above, we have performed the following experiment. 
Water is pumped continuously in from one corner of the tank and is drained through the $4$~cm diameter hole in the centre. 
The water circulates in a closed circuit. 
We first establish a stationary rotating and draining flow by setting the flow rate of the pump to $37.5 \pm 0.5$~$l$/min and waiting until the depth of the water, far away from the vortex is stable at $6.25\pm0.05$~cm. 
We then generate plane waves from one side of the tank, with an excitation frequency ranging from $2.87$~Hz to $4.11$~Hz. 
We recorded the free surface of the water in a region of $1.33$~m $\times$ $0.98$~m over the vortex during $13.2$~s.
This procedure was repeated 6 times for each frequency and 15 times for the frequency $f = 3.7$~Hz.
We note that after every experiment, the water height dropped by a few millimetres. We therefore waited in between experiment for the water height to go back to its original value of $6.25\pm0.05$~cm. 
This consistent height dropped can be understood as a backreaction process and has been studied in detail in \cite{backreaction}. 
Pictures of the experiment running and of the vortex flow are shown in Fig.~\ref{exp_run_pic}.

\subsection{Data analysis of the scattering process}

\begin{figure}
\includegraphics[scale=1]{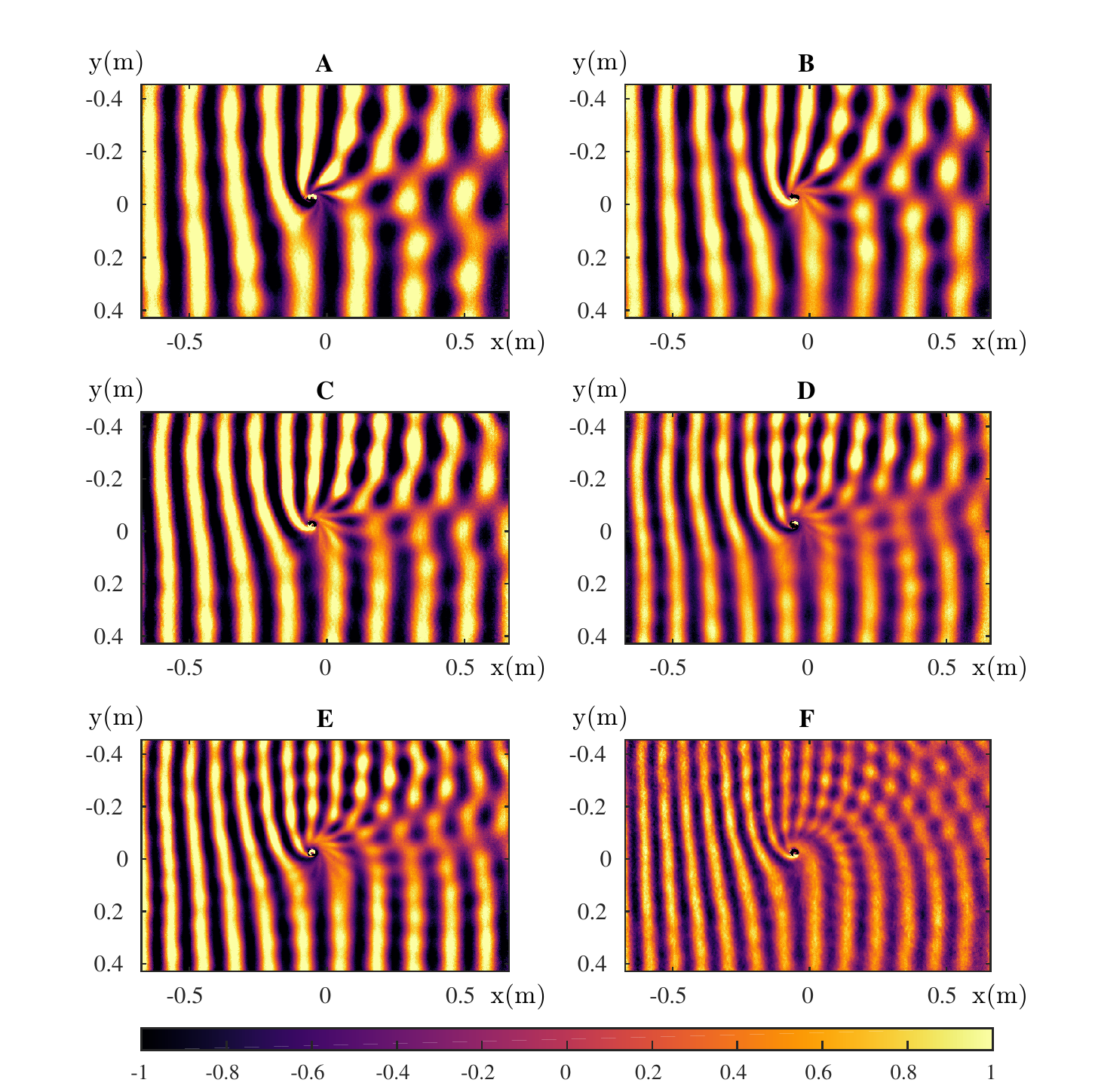}
\caption{Wave characteristics of the surface perturbation $\xi$, filtered at a single frequency, for six different frequencies. The frequencies are $2.87 ~ \mathrm{Hz}$ \textbf{(A)}, $3.04 ~ \mathrm{Hz}$ \textbf{(B)}, $3.27 ~\mathrm{Hz}$ \textbf{(C)}, $3.45 ~\mathrm{Hz}$ \textbf{(D)}, $3.70 ~ \mathrm{Hz}$ \textbf{(E)}, and $4.11 ~\mathrm{Hz}$ \textbf{(F)}. The horizontal and vertical axis are in metres ($\mathrm{m}$), while the color scale is in millimetres ($\mathrm{mm}$). The patterns show the interfering sum of the incident wave with the scattered one. The waves are generated on the left side and propagate to the right across the vortex centred at the origin.}\label{wave_charac}
\end{figure}

For each experiment, we record the free surface of the water in a region of $1.33~ \mathrm{m} \times 0.98~ \mathrm{m}$ over the vortex during $13.2~ \mathrm{s}$. 
From the sensor we obtain $248$ reconstructions of the free surface. 
These reconstructions are triplets $X_{ij}$, $Y_{ij}$ and $Z_{ij}$ giving the coordinates of $640 \times 480$ points on the free surface. 
Because of the shape of the vortex, and noise, parts of the free surface cannot be seen by our sensor, resulting in black spots on the image. 
Isolated black spots are corrected by interpolating the value of the height using their neighbours. This procedure is not possible in the core of the vortex and we set these values to zero. 
The free surface measured by the sensor are presented in Fig.~\ref{wave_charac}. Panel (C) in Fig.\ref{wave_charac} was the one used in Chapter~\ref{Rays_sec}.

To filter the signal in frequency, we first crop the signal in time so as to keep an integer number of cycles to reduce spectral leakage. We then select a single frequency corresponding to the excitation frequency $f_0$. After this filter, we are left with a 2-dimensional array of complex values, encoding the fluctuations of the water height $\xi(X_{ij},Y_{ij})$ at the frequency $f_0$. $\xi(X_{ij},Y_{ij})$ is defined on the grids $X_{ij}$, and $Y_{ij}$, whose points are not perfectly equidistant (this is due to the fact that the discretization is done by the sensor software in a coordinate system that is not perfectly parallel to the free surface).

To select specific azimuthal numbers, we convert the signal from Cartesian to polar coordinates. For this we need to find the centre of symmetry of the background flow. We define our centre to be the centre of the shadow of the vortex, averaged over time (the fluctuations in time are smaller than a pixel). To verify that this choice does not affect the result, we performed a statistical analysis on different centre choices around this value and added the standard deviation to the error bars. Once the centre is chosen, we perform a discrete Fourier transform on the irregular grid $(X_{i j}, Y_{i j})$. We create an irregular polar grid $(r_{i j}, \theta_{i j})$ and we compute 
\begin{equation}
\varphi_{m}(r_{ij})= \frac{\sqrt{r_{ij}}}{2\pi} \sum_{j} \xi(r_{ij},\theta_{ij}) e^{-im\theta_{ij}} \Delta\theta_{ij},
\end{equation}
where $\Delta\theta_{ij} =(\Delta X_{ij} \Delta Y_{ij})/(r_{ij} \Delta r_{ij})$ is the line element along a circle of radius $r_{ij}$. 

\begin{figure} 
\includegraphics{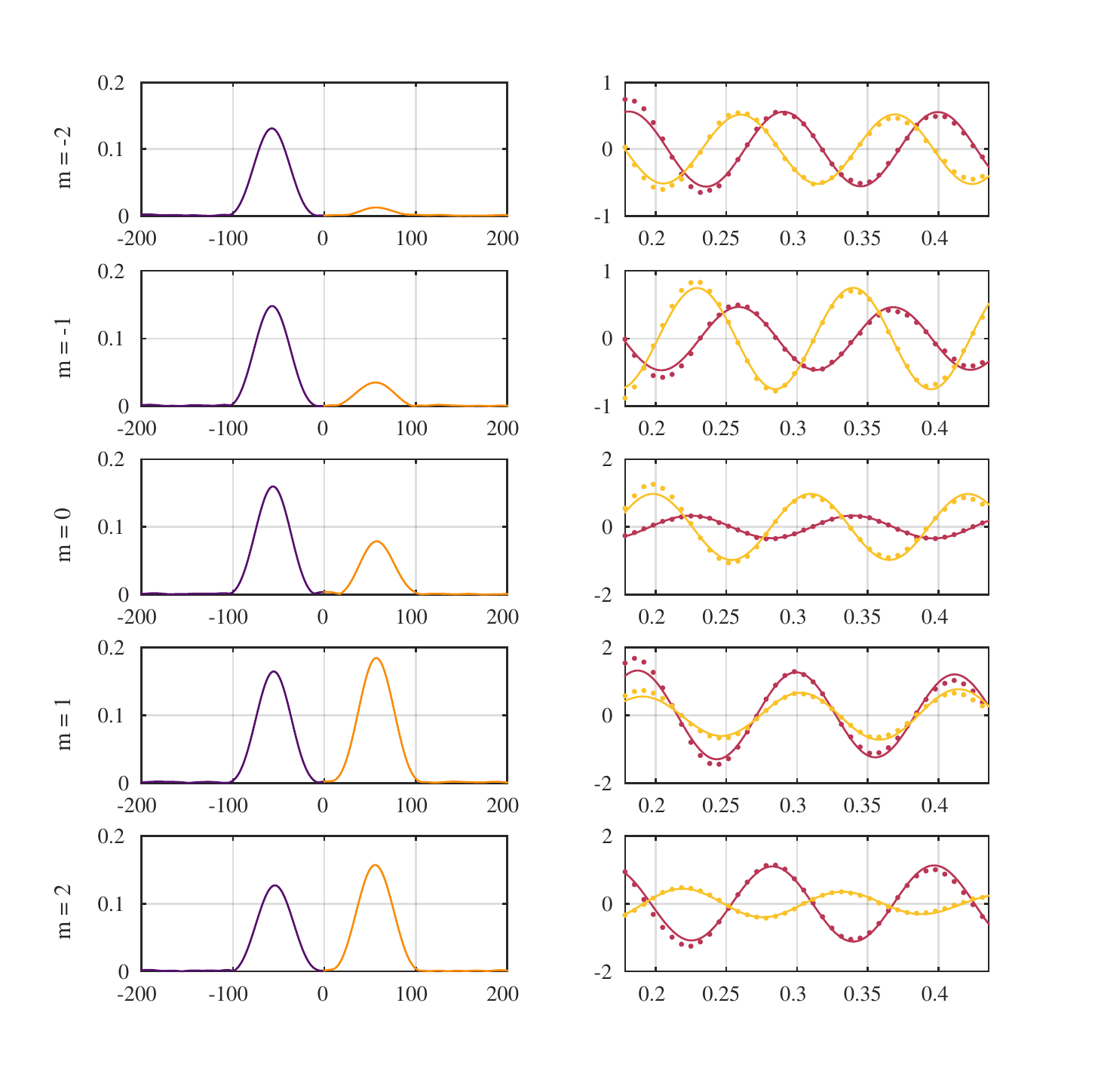}
\caption{Left side: Modulus of the Fourier profiles $|\tilde \varphi_m(k)|^2$ for various $m$. Right side: Radial profiles $\varphi_m(r)$ for various $m$ (maroon: real part, yellow: imaginary part). The vertical axis is in arbitrary units. The horizontal axes in inverse metres ($\mathrm{m^{-1}}$) on the left side, and metres ($\mathrm{m}$) on the right side. The dots are the experimental data (for clarity, only 1 out of 3 is represented), and the solid lines show the approximation of Eq.~\eqref{in_out} for the extracted values of $A_{\mathrm in}$ and $A_{\mathrm out}$.} \label{Fig:FourierAndReconstruction} 
\end{figure}

To extract the inward and outward amplitudes $A_{\mathrm in}$ and $A_{\mathrm out}$, we compute the radial Fourier transform $\tilde \varphi_m(k) = \int \varphi_m(r) e^{-i k r} dr$ over the window $[r_{\mathrm min}, r_{\mathrm max}]$. Due to the size of the window compared to the wavelength of the waves, we can only capture a few oscillations in the radial direction, typically between 1 and 3. This results in broad peaks around the values $k_{\mathrm in}$ and $k_{\mathrm out}$ of the inward and outward components. We assume that these peaks contain only one wavelength (no superposition of nearby wavelengths), which is corroborated by the fact that we have filtered in time, and the dispersion relation imposes a single wavelength at a given frequency. To reduce spectral leakage, we use a Hamming window function on $[r_{\mathrm min}, r_{\mathrm max}]$, defined as 
\be
W(n)=0.54 - 0.46\cos \left(2\pi \frac{n}{N}\right), 
\ee
where $n$ is the pixel index running from $1$ to $N$. This window is optimized to reduce the secondary lobe, and allows us to better distinguish peaks with different amplitudes~\cite{Prabhu2013}. In Fig.~\ref{Fig:FourierAndReconstruction}, 
we show the radial Fourier profiles for various $m$ for a typical experiment (left column), and the raw radial profiles and how they are approximated by a sum of oscillatory solution (right column):
\begin{equation}\label{in_out}
\phi_m(r) = A_{\mathrm in} e^{-ikr} + A_{\mathrm out} e^{ikr}.
\end{equation} 

We also extracted the signal-to-noise ratio by comparing the standard deviation of the noise to the value of our signal. It is sufficiently high to exclude the possibility that the amplification we observed is due to a noise fluctuation, and its contribution is negligible compared to other sources of error.

\section{Results}

\subsection{The analogue Aharonov-Bohm effect}

We observe that incident waves have more wave fronts on the upper half (corresponding to counter-rotating waves) of the vortex in comparison with the lower half (corresponding to co-rotating waves) - see the various wave characteristics in panels (A-F) in Fig.~\ref{wave_charac}. This is the analogue of the Aharonov-Bohm effect in water which has been previously observed experimentally~\cite{Berry,Vivanco99}.

To explain this reason for this phenomenon, we return to the eikonal equation obeyed by the local phase:
\begin{equation}
\left( \omega - \vec{v}_0 \cdot \vnab S \right)^2 - F(\vnab S) = 0.
\end{equation}
As we have seen, this equation can be solved by the method of characteristic.
This approach was followed in the previous chapter and we have computed numerically the eikonal wave. 
Here we follow a different approach in order to derive the effect analytically. 
We are going to solve the eikonal equation far from the vortex, i.e where the background velocity $\vec{v}_0 \rightarrow 0$, by means of a perturbative expansion in $\vec{v}_0$.
We therefore seek a solution for the gradient of the local phase as:
\begin{equation}
\vnab S = \vec{a} + b \vec{v}_0 + \mathcal{O}(\vec{v}_0^2).
\end{equation}
At leading order, i.e. when $\vec{v}_0 = 0$, the wave equation admits plane wave solution. 
Therefore, we simply have $\vnab S = \vec{k}$ where $F(k) = \omega^2$. 
Substituting this into the expansion of $\vnab S$ and expansion into the eikonal equation, we have that:
\begin{equation}
\omega^2 - 2\omega \vec{v}_0\cdot \vec{k} - F(k) - b \vec{v}_0 \cdot \vnab_k F + \mathcal{O}(\vec{v}_0^2) = 0.
\end{equation}
We recognise here that $\vnab_k F = 2\omega \vec{v}_g$, where $\vec{v}_g = \vnab_k \sqrt{F}$ is the group velocity.
We obtain that:
\begin{equation}
b = - \frac{k}{v_g}.
\end{equation} 
Taking the DBT model for the velocity field, we finally get the expression for the phase of the eikonal wave:
\begin{equation}
S = \vec{k}\cdot \vec{r} - \frac{kC}{v_g}\theta - \int dr  \frac{kD}{v_g r } = \vec{k} \cdot \vec{r} - \theta_{\mathrm geo}.
\end{equation}
We see that in the case of a purely rotating flow, i.e. $D = 0$, a wave going around the centre of the vortex once will pick up an extra phase of $2\pi \alpha$ where $\alpha = kC/v_g$ resulting in the presence of extra wave fronts. 
In order to have a single valued solution, the number of extra wave fronts is equal to the closest integer to $\alpha$, denoted $ \lfloor \alpha \rceil$ \cite{Berry,Vivanco99,Coste99}. 
The case of a non-draining flow in the shallow water limit has been studied in \cite{Dolan11}. 
In our experiment, the drain of the fluid is significantly smaller than the circular motion of the fluid. 
We compare in Table~\ref{AB_table} the wave front dislocation observed in our experiment with the prediction made in a purely rotating flow. 
We note two discrepancies, one at the frequency $f = 3.27$ Hz and one at the frequency $f = 3.70$ Hz. 
They can be explained by the high sensitivity of the value of the parameter $\alpha$ to the value of the circulation parameter and the fact that the number of extra wave fronts is determined by the closest integer to $\alpha$.
For example, we can see that the theoretical prediction for the frequency $f=3.27$ Hz, is close to $2.5$ which would result in a discontinuity. 
A small shift of the value of the circulation parameter make the parameter $\lfloor \alpha \rceil$ jump from 2 to 3. Such discontinuity is visible in panel (C) of Fig.~\ref{wave_charac}. 

\begin{table}
\caption{Comparison between theoretical and experimental wave front dislocation}\label{AB_table}
\centering
\makebox[\textwidth]{\begin{tabular}{||c||c|c|c|c|c|c|c||}
\hline
Frequency (Hz) & $2.87$ & $3.04$ & $3.27$ & $3.45$ & $3.70$ \\
\hline
$\alpha $ & $1.8$ & $2.1$ & $2.6$ & $3.1$ & $3.7$ \\
\hline
$\lfloor \alpha \rceil $ & $2$ & $2$ & $3$ & $3$ & $4$ \\
\hline
$N_{\mathrm wf}$ & $2$ & $2$ & $2$ & $3$ & $3$ \\
\hline
\end{tabular}}
\end{table}

\subsection{Superradiant amplification}

\begin{figure}
\includegraphics[scale=1]{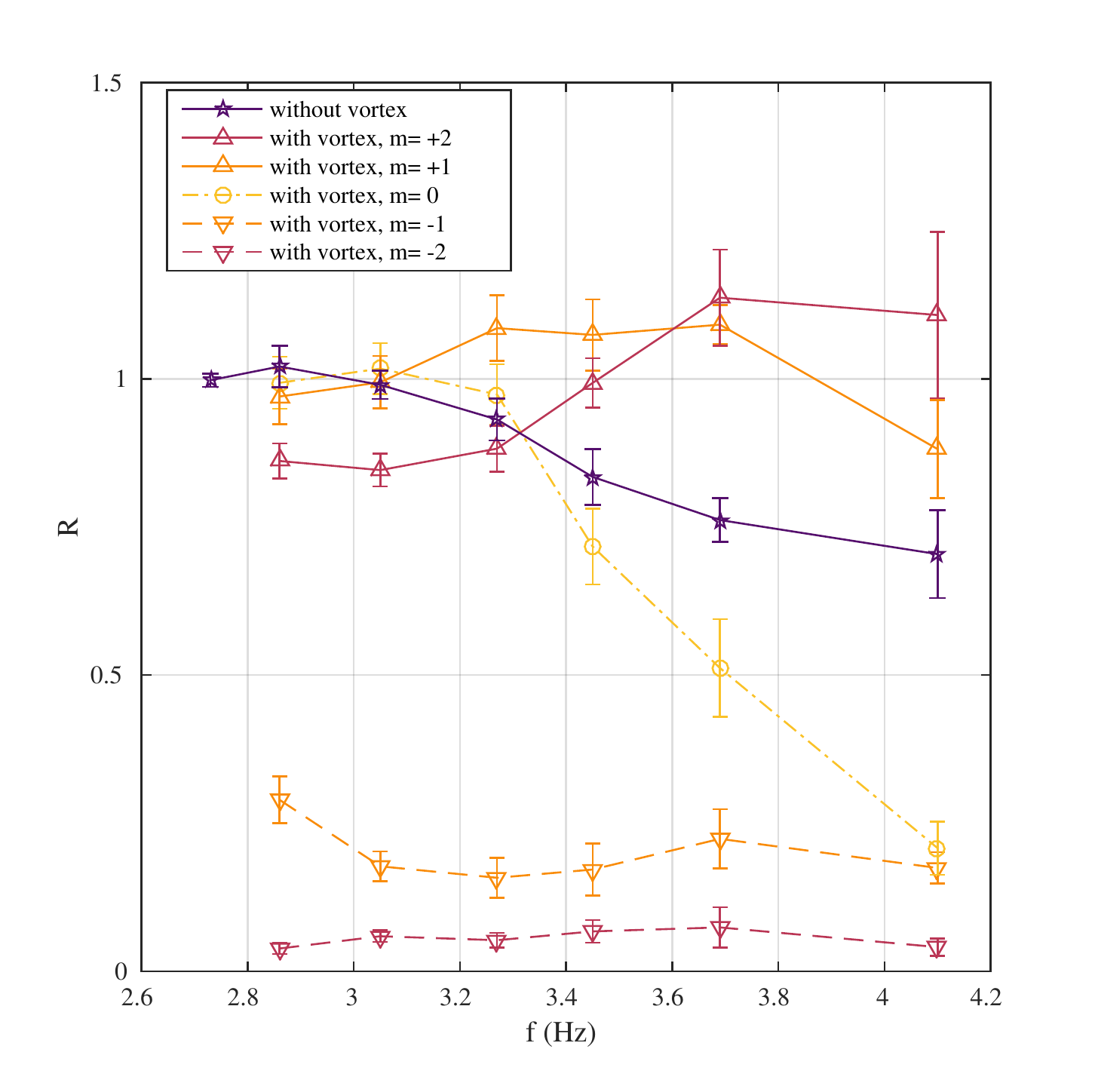}
\caption{Reflection coefficients for various frequencies and various $m$'s. For the vortex experiments the statistical average is taken over 6 repetitions, except for $f=3.70 ~\mathrm{Hz}$ where we have 15 repetitions. The purple line (star points) shows the reflection coefficients of a plane wave in standing water of the same height. We observe a significant damping for the frequencies above 3Hz. Each point is a statistical average over the $m$'s over 5 repetitions. The errors bars indicate the standard deviation over these experiments, the energy uncertainty and the standard deviation over several centre choices. The main contribution comes from the variability of the value of the reflection coefficient for different repetitions of the experiment. We have also extracted the signal-to-noise ratio for each experiment, and its contribution to the error bars is negligible.}\label{R_spectrum_plot}
\end{figure}

\begin{figure}
\includegraphics[scale=1]{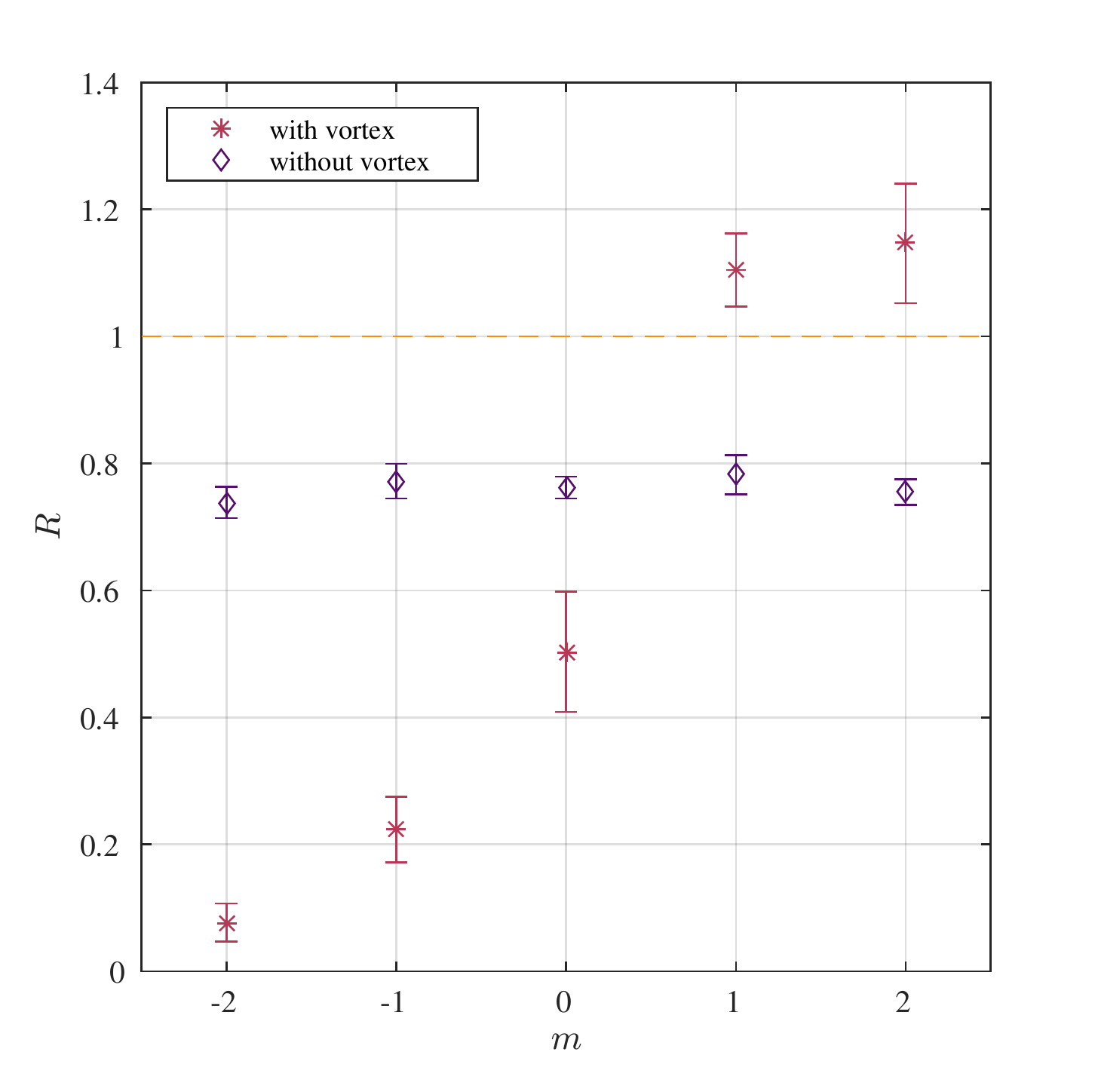}
\caption{Reflection coefficients for different $m$'s, for the frequency $f = 3.70 ~ \mathrm{Hz}$ (stars). We have also shown the reflection coefficients for plane waves without a flow, at the same frequency and water height (diamonds). We see that the plane wave reflection coefficients are identical for all $m$'s, and all below 1 (within error bars). The statistic has been realized over 15 experiments. Error bars include the same contributions as in Fig.~\ref{R_spectrum_plot}.}\label{Reflection_plot}
\end{figure}

We now turn our attention the study of the wave energy. We compare the energy current of the inward wave with respect to the outward one. 
Since energy is transported by the group velocity $v_g$, the energy current is given by 
\be
J = g \,\frac{v_g |A|^2}{f \omega_{\mathrm fluid}}
\ee 
(up to the factor $1/f$, this is the wave action, an adiabatic invariant of waves~\cite{Buhler,Richartz:2012bd,Coutant:2016vsf}). Note the factor $\omega_{\mathrm fluid}^{-1}$ which differs from the expression in the transport equation (Eq.~\eqref{transport_eq}). This is because we are dealing with height perturbation and the measured amplitude is not the amplitude of the velocity potential field. 
If the background flow velocity is zero, then the ratio $J_{\mathrm out}/J_{\mathrm in}$ is simply $|A_{\mathrm out}/A_{\mathrm in}|^2$. However, in the presence of the vortex, we observe from our radial profiles $\varphi(r)$ that the wave number of the inward and outward waves are not exactly opposite. The origin of this (small) difference is that the flow velocity is not completely negligible in the observation window. It generates a small Doppler shift that differs depending on whether the wave propagates against or with the flow. In this case, the ratio of the energy currents picks up a small correction with respect to the ratio of the amplitudes, namely, 
\be \label{Norm_eq}
 \frac{J_{\mathrm out}}{J_{\mathrm in}} = 
\left| \frac{\omega_{\mathrm fluid}^{\mathrm in} v_g^{\mathrm out}}{\omega_{\mathrm fluid}^{\mathrm out} v_g^{\mathrm in}}\right|  
\left|\frac{A_{\mathrm out}}{A_{\mathrm in}}\right|^2. 
\ee
To estimate this factor, we assume that the flow varies slowly in the observation window, such that $\omega_{\mathrm fluid}$ obeys the usual dispersion relation of water waves, $\omega_{\mathrm fluid}^2 = g k \tanh(h_0 k)$. 
This amounts to a Wentzel–Kramers–Brillouin (WKB) approximation, and capillarity is neglected. 
Under this assumption, the group velocity is the sum of the group velocity in the fluid frame, given by the dispersion relation, $v_{g,r}^{\mathrm fluid} = \p_{k_r} \sqrt{g k \tanh(h_0 k)}$, and the radial velocity of the flow $v_r$. Hence the group velocity needed for the energy ratio in Eq.~\eqref{Norm_eq} splits into two: $v_g = v_{g,r}^{\mathrm fluid} +v_r$. Since we are sufficiently far from the vortex, the first term is obtained only with the values of $k_{\mathrm in}$ and $k_{\mathrm out}$, extracted from the radial Fourier profiles. The second term requires the value of $v_r$, which we do not have to a sufficient accuracy. However, using the PIV data, we see that the contribution of this last term amounts to less than $1\%$ in all experiments (this uncertainty is added to the error bars on Figs.~\ref{R_spectrum_plot} and \ref{Reflection_plot} ). 

After extracting the various quantity of interest, we finally compute the reflection coefficient:
\be
R = \sqrt{\frac{J_{\mathrm out}}{J_{\mathrm in}}}.
\ee
The values obtained for $R$ are presented in Figs.~\ref{R_spectrum_plot} and \ref{Reflection_plot}.

On Fig.~\ref{R_spectrum_plot} we represent, for several azimuthal numbers $m$, the absolute value of the reflection coefficient $R$ as a function of the frequency $f$. 
We observe two distinct behaviours, depending on the sign of $m$. 
Negative $m$'s (waves counter-rotating with the vortex) have a low reflection coefficient, which means that they are essentially absorbed in the vortex hole. 
On the other hand, positive $m$'s have a reflection coefficient close to 1. 
In some cases, this reflection is above one, meaning that the corresponding mode has been amplified while scattering on the vortex. 
To confirm this amplification we have repeated the same experiment 15 times at the frequency $f = 3.7~\mathrm{Hz}$ and water height $h_0 = 6.25 ~\pm ~0.05 ~\mathrm{cm}$, for which the amplification was the highest. 
We present the result on Fig.~\ref{Reflection_plot}.
On this figure we clearly observe that the modes $m=1$ and $m = 2$ are amplified by factors $R_{m=1} \sim 1.09 \pm 0.03$, and $R_{m=2} \sim 1.14 \pm 0.08$ respectively. On Figs.~\ref{R_spectrum_plot} and \ref{Reflection_plot}, 
we have also shown the reflection coefficients obtained for a plane wave propagating on standing water of the same depth. Unlike what happens in the presence of a vortex, the reflection coefficients are all below 1 (within error bars). 
For low frequencies it is close to 1, meaning that the wave is propagating without losses, while for higher frequencies it decreases due to a loss of energy during the propagation,~i.e.~damping. 

The origin of this amplification can be explained by the presence of negative energy waves \cite{Stepanyants, Coutant:2012mf}. 
Negative energy waves are excitations that lower the energy of the whole system (i.e.~background flow and excitation) instead of increasing it. 
In our case, the sign of the energy of a wave is given by the angular frequency in the fluid frame $\omega_{\mathrm fluid}$. 
If the fluid rotates with an angular velocity $\Omega(r)$, in rad/s, we have $\omega_{\mathrm fluid} = 2\pi f - m\Omega(r)$. 
At fixed frequency, when the fluid rotates fast enough, the energy becomes negative. 
If part of the wave is absorbed in the hole, carrying negative energy, the reflected part must come out with a higher positive energy to ensure conservation of the total energy~\cite{Richartz:2009mi}. 
As we see on Fig.~\ref{PIV_pic}, close to the centre, the angular velocity is quite high, and the superradiant condition $2 \pi f < m \Omega$ is therefore satisfied for our frequency range.

\subsection{WKB estimate of the reflection coefficient}
\begin{figure}
\includegraphics[scale=1]{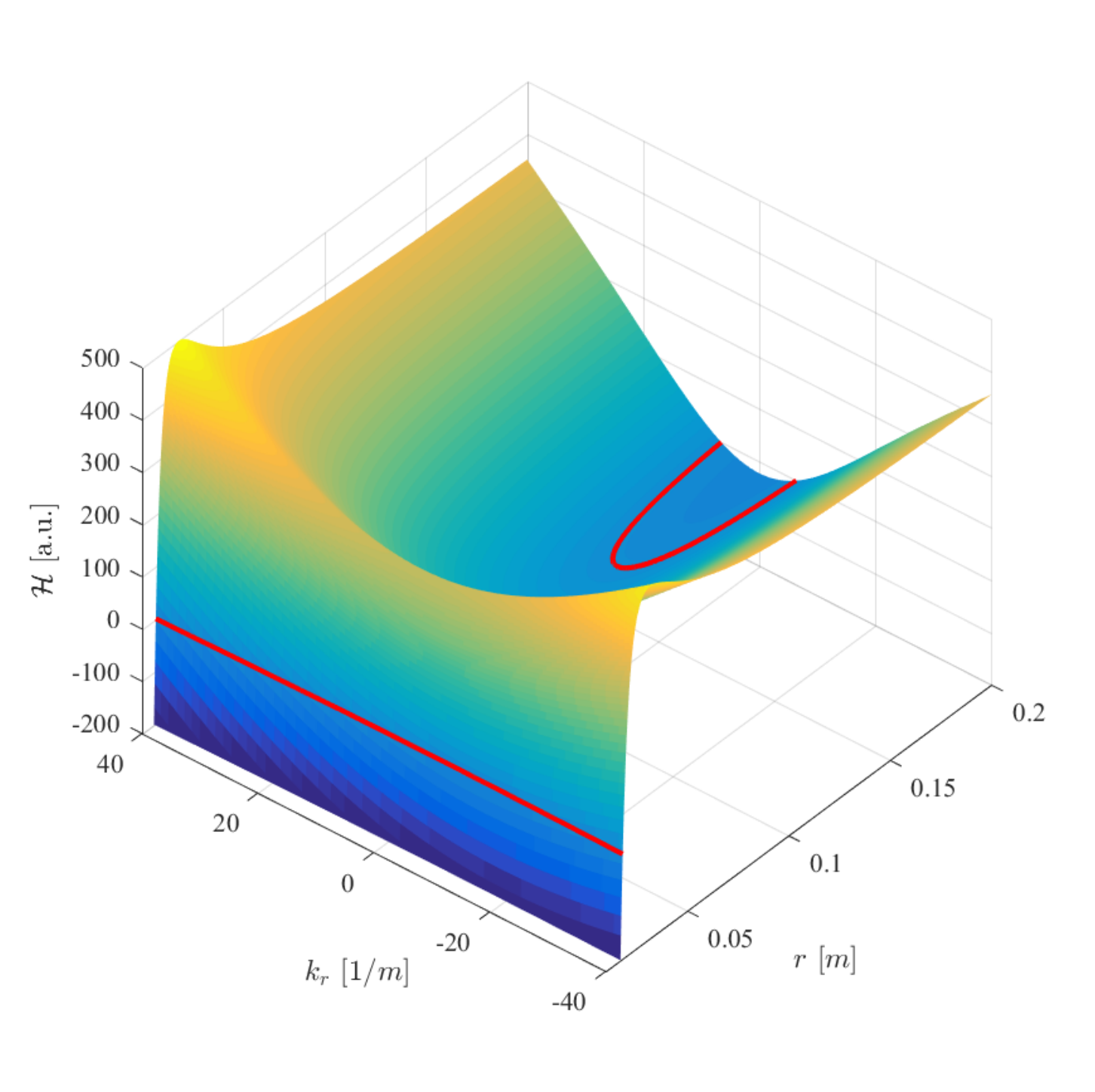}
\caption{Dispersion surface given by the Hamiltonian function $\mathcal{H}(r,k_r)$ given in arbitrary units. The flow parameters are $C = 0.01~m^2/s$, $D = 0.001~m^2/s$ and $h_0 = 6~cm$. The surface is plotted for $m=1$ and a frequency $f=1~Hz$. The red line correspond to the level line $\mathcal{H}(r,k_r) = 0$, i.e. waves satisfying the dispersion. We can see that the two lines are separated by a saddle point.}\label{dispersion_surface}
\end{figure}

We have seen that waves incident on a vortex can be understood by use of a Hamiltonian function. 
Generally, at fixed $m$ and $\omega$ the Hamiltonian function define a 2-dimensional surface parametrized by $r$ and $k_r$. 
The structure of the Hamiltonian function imposes that this surface admits a saddle point (see Fig.~\ref{dispersion_surface}). 

Pairs of $m$ and $\omega$ that are such that the Hamiltonian vanishes at the saddle point constitutes the light-ring spectrum. 
However, if one does not imposes that the Hamiltonian vanishes at the saddle point, then the condition $\mathcal{H} = 0$ will define two disconnected region in phase space (see Fig.~\ref{dispersion_surface}), located on both sides of the saddle point. This structure will allow tunnelling in between the two regions.

It is possible to relate solutions on each side of the saddle point (in the WKB limit) via a connection matrix $S$ with coefficients determined by the saddle point (see Appendix~\ref{Appendix2}).
In the case of a plane wave incident on a vortex flow, we can estimate the transmission and reflection coefficient as:

\begin{equation}
R = i\frac{\alpha^2}{\beta^*}, \quad \text{and} \quad T = i\frac{\tau\alpha^2}{\beta^*},
\end{equation}

where $\alpha$, $\beta$, and $\tau$ are constants given by the saddle point (see Eqs.~\eqref{tau_beta_def} and \eqref{alpha_def}).

These expressions can be used to understand some aspects of the experimental spectrum by noticing the following. 

First, we note that with the parameters used in the experiment (flow parameters, wave frequency, etc...) the formula for the reflection coefficient cannot be trusted when applied to the co-rotating waves. 
This is due to the fact that for these waves, the saddle point is located at a radius $r_{SP} \approx 1.6~cm$ for $m=1$ and $r_{SP} \approx 2~cm$ for $m=2$. 
At this radius, we know that the irrotational assumption breaks down. 
Therefore, in order to estimate the reflection coefficient, one needs to take into account the more complicated structure of the flow in the core, such as vorticity and a varying height. 

\begin{figure}
\centering
\includegraphics[scale=1]{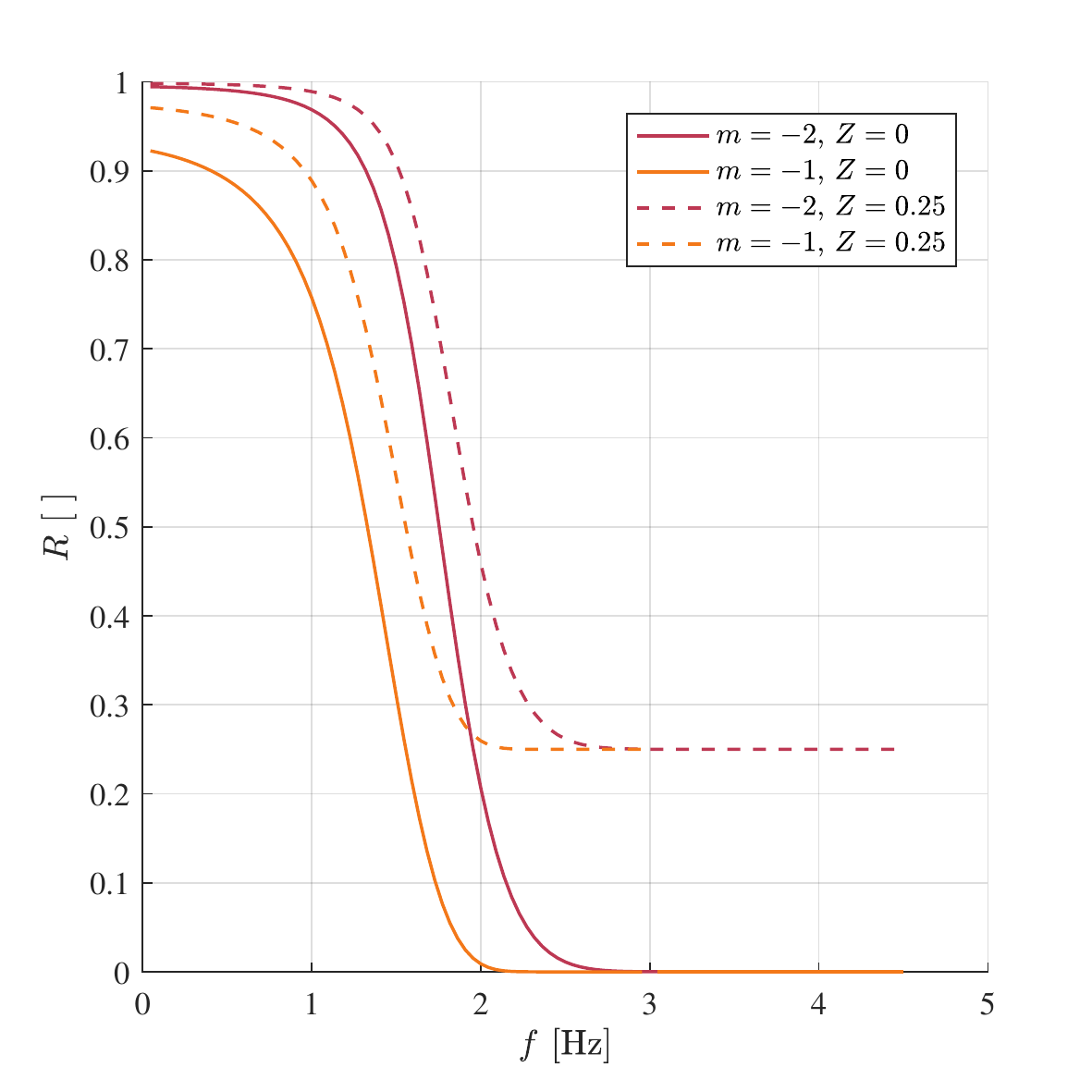}
\caption{Reflection coefficient estimated in the WKB approximation for counter-rotating waves. The blue (red) curves depicts the reflection coefficient for $m=-2$ ($m=-1$). The solid curves are the estimated reflection when the drain is assumed to be completely absorbing,~i.e. $Z=0$. 
The dashed curves show the impact of a non-purely absorbing drain on the reflection coefficient. 
In the case plotted here, we have chosen that $75\%$ of the waves was transmitted at the drain,~i.e. $Z = 0.25$. We see in particular that the reflection coefficient asymptotes the constant value of the reflection at the drain (here $25\%$).}\label{WKB_spec}
\end{figure}

On the other hand, for counter-rotating waves, the saddle point is located far away from the core. 
For the frequency range considered in the present experiment, the saddle point is located at a radius larger than the size of the experiment.
Consequently, in this approximation, counter propagating waves are not affected by the presence of the saddle point and their reflection coefficient should entirely be determined by the boundary conditions at the drain. 
If one considers that waves are purely ingoing at the drain, in analogy to what happens at the event horizon of a black hole, then counter rotating modes should be entirely absorbed. 
This is seen in the estimated spectrum plotted in Fig.~\ref{WKB_spec}.
However this is not the behaviour observed experimentally in Fig.~\ref{R_spectrum_plot}. 
Rather we see that the reflection coefficients asymptote a constant value. 
This behaviour can be explained by the fact that the drain is not a purely absorbing boundary but might reflect part the waves.
When including a non-purely absorbing drain in our WKB estimate for the reflection coefficient, we see that in the frequency range of the experiment, the reflection coefficient asymptote a constant value.
The value of this constant is directly given by the value of the reflection at the drain, that we call $Z$. 
Therefore, the experimental observation that counter-rotating waves are not completely absorbed suggests that the drain is not a purely absorbing boundary. 
In addition, since the reflection coefficient asymptotes to different values for different azimuthal numbers. 
This suggests that the absorption properties of the drain depend on the azimuthal number and not on the frequency.
In particular, our measurement suggests that the absorption at the drain for $m = -1$ is about $75\%$ and is about $90\%$ for $m=-2$.

\section{Summary and discussion}

In this section, we have presented an experiment that demonstrates that wave scattering on a vortex flow can carry away more energy than the incident wave brings in. 
This the first observation of the rotational superradiance scattering.
We have presented the experimental apparatus used to perform our observation and detailed the detection method procedure, as well as the data analysis performed to extract the reflection coefficient.
\\

While our result demonstrates the robustness of the superradiance effect, it also opens new questions.

First, the experiment was not performed in the analogue regime and we are still lacking theoretical prediction to match our observation.
We attempted to include the effect of dispersion in a WKB estimate of the reflection coefficient. 
However, we have seen that this approach is not applicable to compute the spectrum of co-rotating waves. 
This is because the connection of WKB modes must be done at a radius at which vorticity effects needs to be considered as well. 
However, we believe that this method must be valid to estimate the spectrum for the counter-rotating wave. 
If we are to believe this estimate, we must conclude that the vortex core does not behave as a purely absorbing boundary and some reflection must be included.
While the explanation behind reflection at the drain is still unclear, the presence of such a mechanism is not unreasonable. 
Indeed, as already mentioned, we know that the flow becomes more complicated than the DBT in the centre. 
For example, it has been shown that the presence of vorticity on shallow water waves introduces extra structure to the Hamiltonian (or equivalently, to the effective potential) \cite{flowmaster}. 
In particular, this introduces an extra maximum to the potential, located between the usual maximum due to the irrotational flow and the drain, on which waves can scatter. 
Combining this extra bump with a purely absorbing drain can appear at a new effective boundary which would not be perfectly absorbing. 
Moreover, since the shape of the potential depends on the azimuthal number, the value of the effective reflection could depend on the value of $m$. 
In addition to vorticity effects, we also know that the height of the water changes close to the drain. The effect of such a change is still unclear and could play a role in modifying the inner boundary.
The effect of modified boundary conditions on superradiance have been studied~\cite{Richartz:2009mi}, and it is known that the superradiance can still occur even when the system does not possess a perfect absorbing boundary.

A way to theoretically take these effects into account might be to look for new tools to describe vortex flows. 
The original analogy gave us the effective metric to better understand wave propagation on flowing fluids.  
Our experimental study shows that the superradiance effect also persists beyond the analogue regime. 
Therefore, one might hope to find new connections between curved space-time physics and vortex flows that will allow for a new theoretical understanding of wave-vortex scattering.

In the next section, we will strengthen the connection between vortex flows and BHs, outside the analogue regime. We will exhibit theoretically and experimentally another phenomena shared by both system and show that tools developed in BH physics can shine new light on the physics of vortex flows.
\chapter{Ringdown and Relaxation}\label{Ringdown_sec}
\epigraph{Whenever we proceed from the known into the unknown we may hope to understand, but we may have to learn at the same time a new meaning of `understanding'.}{Heisenberg}
\section{Introduction}

BHs are like bells; once perturbed they will relax through the emission of characteristic waves. 
According to GR, the end of the relaxation process of an astrophysical black hole is expected to depend only on its mass and angular momentum, and not on the details of its formation process~\cite{Echeverria}.
During this stage, each azimuthal component is well-approximated (in an open system) by a superposition of time-decaying modes,
called \textit{quasinormal modes} (QNMs)~\cite{Kokkotas99,Berti_review,Konoplya11}. The QNMs are a discrete set of modes with \emph{complex} frequencies, with the real part determining the oscillation frequency, and the (negative) imaginary part determining the damping rate. Formally, QNMs are defined via a pair of boundary conditions, which are purely ingoing at the horizon and outgoing far from the black hole. 

The important feature of QNM's is that the QNM spectrum depends only on the black hole parameters. This opens up the possibility of \emph{black hole spectroscopy} (BHS).
By measuring the QNM spectrum of a black hole, one should be able to extract the black hole parameters~\cite{Press72,Echeverria,Schutz09}.
With the recent detection of gravitational waves \cite{gwaves,LIGO_propBH}, we seem to be close to turning this idea into a reliable astronomy technique.

Based on the fluid-gravity analogy, it has been predicted that irrotational vortex flows, should also emit QNMs when perturbed \cite{Berti:2004ju}. 
The study of the relaxation phase of vortex flows is the subject of this chapter. 
We will start by reviewing and extending one of the methods developed in black hole physics to evaluate the QNM spectrum. 
Then, inspired by the idea of black hole spectroscopy, we will describe a scheme designed to characterise irrotational vortex flows directly from their relaxation spectrum. 
Finally, we will see an experiment in which we demonstrate the presence of analogue LR modes around vortex flows and use their spectrum to characterise the experimental fluid flow.

\section{Light-ring approximation}

\subsection{The spectrum of characteristic modes}

One of the techniques available to estimate the QNM spectrum of a black hole is based on the properties of LRs~\cite{Goebel72,Cardoso_Lyapu}. 
More precisely, the eikonal complex frequencies are given by 
\be \label{QNM_eiko}
\omega_{\mathrm QNM}(m) = \omega_\star(m) - i \Lambda(m) \left(n+\frac{1}{2} \right) , 
\ee
where $\omega_\star(m)$ is the circular orbit wave frequency of section~\ref{subsec:circorbits}, $n$ is an integer called the overtone index, and $\Lambda$ is the decay rate of these (unstable) orbits (Lyapunov exponent), to be defined below. 
The relation between QNMs and LRs modes comes from the fact that, in many (but not all~\cite{khanna}) spacetimes, QNMs can be seen as waves travelling on unstable orbits and slowly leaking out~\cite{Berti_review}.

As shown in Chapter~\ref{Rays_sec}, the draining bathtub vortex flow exhibits circular orbits analogous to the LRs of black holes.
Due to the presence of such orbits, we expect that, if one perturbs the system initially, the perturbations will spread out rapidly, except where the perturbation lingers around the circular orbit. Indeed, a wave packet passing close to such a `ring' can hover around it for a longer time before dispersing. To understand this behaviour more precisely, we consider a wave packet in the neighbourhood of a circular orbit, that is, around the family of rays $(r_\star + \delta r, k_{r \star} + \delta k, \omega_\star + \delta \omega)$ at a fixed azimuthal number $m$. To obtain the effective dynamics of these rays, we expand the Hamiltonian in the vicinity of the critical point $(r_\star, k_{r \star})$. This gives  
\be \label{LR_Ham}
\mathcal{H} = \partial_{\omega}\mathcal{H} \delta \omega + \frac{1}{2} X^{T} \cdot \left[ d^{2}\mathcal{H} \right] \cdot  X, 
\ee
where we have defined the Hessian matrix 
\be
[d^{2}\mathcal{H}] =\begin{pmatrix}
\partial_{k}^{2}\mathcal{H} & \partial_{k}\partial_r \mathcal{H} \\
\partial_{k}\partial_{r} \mathcal{H} & \partial_{r}^{2}\mathcal{H} 
\end{pmatrix} \qquad \text{and} \qquad 
X = \begin{pmatrix} \delta k \\ \delta r \end{pmatrix}.
\ee
Since $[d^{2}\mathcal{H}]$ is a symmetric and real matrix, it can be diagonalized. We call $(\mu_{1},\mu_{2})$ the eigen-values, and $(R,K)$ the components in the eigen-basis\footnote{In addition, the eigen-basis is orthogonal, which in two dimensions is enough to ensure a canonical transformation $(\delta r, \delta k) \to (R,K)$.}. In this representation, the Hamiltonian becomes
\be \label{Red_H}
\mathcal{H} = \partial_{\omega}\mathcal{H} \delta \omega + \frac{1}{2}\mu_{1}K^{2} + \frac{1}{2}\mu_{2} R^{2}.
\end{equation}
This is the Hamiltonian of a harmonic oscillator. The relative sign of the coefficients $\mu_1$ and $\mu_2$ will determine the stability of the orbits. In our case, we always have $\mu_{1}\mu_{2} < 0$ and hence all orbits are unstable. In its vicinity, wave dynamics is described by an \emph{inverted} harmonic oscillator. To obtain the Lyapunov exponent $\Lambda$, we solve Hamilton's equations with this reduced Hamiltonian. This gives  
\be \label{EoM_IHO}
\dot{R}=\mu_{1}K, \qquad \dot{K}=-\mu_{2}R, \qquad \dot{t}=-\partial_{\omega}\mathcal{H}.
\ee
The general solution is a linear superposition of decaying and growing exponentials. At late times, the growing behaviour dominates, and we have
\begin{equation}
R \sim A e^{ \frac{\sqrt{-\mu_{1}\mu_{2}}}{\partial_{\omega}\mathcal{H}}t} \qquad \text{and} \qquad K \sim B e^{ \frac{\sqrt{-\mu_{1}\mu_{2}}}{\partial_{\omega}\mathcal{H}}t},
\end{equation}
where $A$ and $B$ are constant coefficients such that Eq.~\eqref{EoM_IHO} is satisfied. The Lyapunov exponent is the rate of exponential growth away from the orbit, and is given by 
\begin{equation}\label{Lyapu_def}
\Lambda = \frac{\sqrt{-\mu_{1}\mu_{2}}}{|\partial_{\omega}\mathcal{H}|}=\frac{\sqrt{-\det[d^{2}\mathcal{H}]}}{|\partial_{\omega}\mathcal{H}|}.
\end{equation}
The Lyapunov exponent governs the decay of waves close to the circular orbit. One can see this by inspecting the amplitude governed by the transport equation, Eq.~ \eqref{transport_eq}, namely,
\be
A_0(t) \propto e^{-\Lambda t/2}. 
\ee
However this argument misses the overtone frequencies (i.e. $n \neq 0$ in Eq.~\eqref{QNM_eiko}). 

A more precise argument is to lift the reduced Hamiltonian given in Eq.~\eqref{Red_H} at the level of the wave equation. To see this, we consider a wave packet of fixed $m$, localized around the circular orbit, written as 
\be \label{LR_SVA}
\phi(t,r_\star +\delta r) \sim \psi(t,\delta r) e^{- i \omega_\star t + i k_{r \star} r}, 
\ee
where $\psi$ is a slowly varying envelope. Formally, the wave equation (Eq.~\eqref{wave_equation_dispersive}) can be written\footnote{This writing is of course only formal, since there is \emph{a priori} no unique way to promote the Hamiltonian function $\mathcal H(\omega, r, k_r)$ into an operator $\hat{\mathcal{H}}(i \partial_t; r,-i\partial_r)$. The ambiguity essentially comes from the various possible ordering between functions of $r$ and $\partial_r$. However, since we work here at the level of the eikonal approximation, different choices of ordering lead to the same result.}
\be
\hat{\mathcal H}(\omega= i \partial_t; r, k_r = -i\partial_r) \phi = 0. 
\ee
Assuming $\psi$ in Eq.~\eqref{LR_SVA} is slowly varying, one can again rescale $\partial \to \epsilon \partial$ and expand in $\epsilon$. Doing so, the effective wave equation in the vicinity of the circular orbit is given by
\be \label{Brut_Schro_eq}
- i\partial_{\omega}\mathcal{H} \partial_t \psi = -\frac{1}{2} \partial_k^2 \mathcal H \partial_r^2 \psi - \frac{i}{2} \partial_k \partial_r \mathcal H (r \partial_r + \partial_r r) \psi + \frac{1}{2} \partial_r^2 \mathcal H r^{2} \psi. 
\ee
(Note that the cross term is chosen to be symmetric.) This equation is a Schr\"odinger equation in an upside-down harmonic potential. To see this, we use the fact that there is a canonical transformation to diagonalize the Hamiltonian given in Eq.~\eqref{LR_Ham}. At the level of the Schr\"odinger equation( Eq.~\eqref{Brut_Schro_eq} ), this means that there is a unitary transformation $\psi \to \tilde \psi$ such that 
\be \label{Schro_eq}
- i\partial_{\omega}\mathcal{H} \partial_t \tilde \psi = -\frac{1}{2}\mu_{1} \partial_R^2 \tilde \psi + \frac{1}{2}\mu_{2} R^{2} \tilde \psi . 
\ee
In the case of Eq.~\eqref{Schro_eq} the time dependent response is well described with the help of QNMs, defined as the solutions of $i \partial_t \tilde \psi = \delta \omega \; \tilde \psi$ with purely out-going boundary conditions (see e.g. \cite{Kokkotas99}). Such solutions are analytically known and can be expressed using parabolic cylinder functions. This selects a discrete set of complex values for $\delta \omega$, which are the quasi-normal frequencies for $\tilde \psi$ and hence for $\psi$. Finally, using Eq.~\eqref{LR_SVA}, we see that the quasi-normal frequencies for $\phi$ are given by Eq.~\eqref{QNM_eiko}.

Alternative boundary conditions, such as those for exotic black hole mimickers, can alter the global structure of the quasinormal mode spectrum, without significantly altering the local `ringing' phenomena associated to the LRs (see e.g.~\cite{Cardoso:2017cqb, Cardoso:2017njb}).

\begin{figure}
\begin{subfigure}{.5\textwidth}
\centering
  \includegraphics[width=1\linewidth]{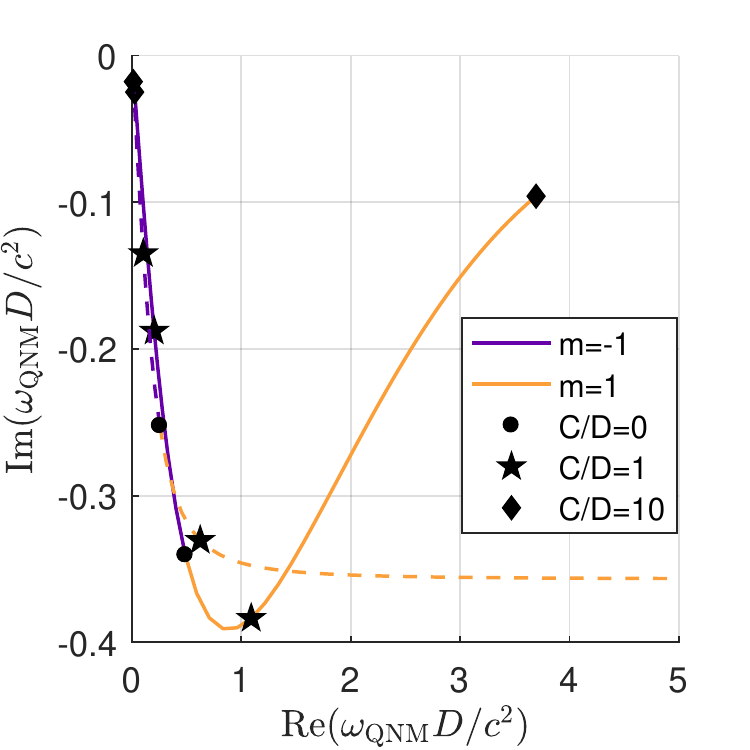}
  \caption{$m = \pm 1$}
\end{subfigure}%
\begin{subfigure}{.5\textwidth}
  \includegraphics[width=1\linewidth]{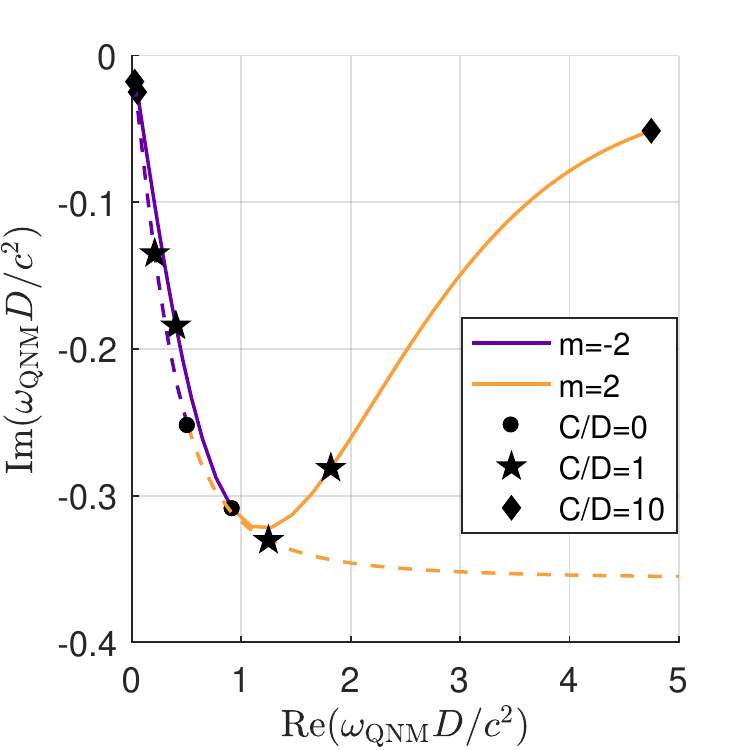}
  \caption{$m = \pm 2$}
\end{subfigure}
\caption{Fundamental quasi-normal mode frequencies $\omega_{\mathrm QNM}$ of different $m$ modes, $m = \pm 1$ (left) and $m = \pm 2$ (right), for various ratio $C/D$. The dashed line is the shallow water result, and the solid one the (general) dispersive regime. In order to make the transition more transparent, the depth is chosen so that $g h_0^{3}/D^2 = 0.04$. Each point represents a different ratio for the flow parameters. The negative $m$ modes have a similar behaviour in the two regimes, namely their lifetime increases as the ratio $C/D$ increases. On the other hand, the lifetime of the positive $m$ modes increases too with the ratio $C/D$ in the dispersive regime while it monotonically decreases in the absence of dispersion. 
}
\label{QNM_plot}
\end{figure}

We have numerically computed the fundamental ($n = 0$) eikonal quasi-normal frequencies in the full dispersive regime for a selection of azimuthal numbers and for several values of the ratio $C/D$. The result, and the comparison with the linear regime is presented in Fig.~\ref{QNM_plot}. We can see that the behaviour of the counter-rotating modes ($m<0$) is very similar in the two different regimes. As the ratio increases, the radius of the unstable orbits of these modes will increase too. For fixed $m$, the increase of the radius will increase the wavelength (decrease the frequency) and therefore will bring the mode closer to the linear regime. On the other side, the behaviour for co-rotating ($m>0$) modes is very different when dispersion is included. Indeed, in the relativistic regime, the lifetime of the co-rotating modes decreases while it appears to increase after a short drop as an effect of dispersion.

In general, it is noticeable that the co-rotating modes are damped quicker than the counter-rotating ones (see Fig.~\ref{QNM_plot}). Hence, if the initial excitation has a similar amplitude on both radii, the dominant signal at large time will be that of the counter-rotating orbit. This is the case for instance if the system is excited by sending a plane wave, which contains as much $m>0$ as $m<0$. In addition, we expect the counter-rotating mode to have a larger spatial extension than the co-rotating one since the radius of the counter-rotating orbit is larger than the co-rotating one, and hence, be more visible. This will be confirm later experimentally in the section~\ref{QNM_exp_sec}.

\subsection{Remarks on the characteristic modes}

The fact that the Hamiltonian governing the rays is re-parametrisation invariant allows us to give a new interpretation of the characteristic modes. Indeed, by noticing that:
\begin{equation}\label{LR_DR}
\frac{\p \mathcal{H}}{\p k_r} = \pm F(k) \frac{\p \omega_d}{\p k_r} \quad \text{ and } \quad \frac{\p \mathcal{H}}{\p r} = \pm F(k) \frac{\p \omega_d}{\p r},
\end{equation}
we can see that a critical point of the Hamiltonian is also a critical point of the dispersion relation. 
In particular, the LR modes will satisfy the condition $\p \omega_d/\p k_r =0$. 
This condition defines two curves, $\omega^{\pm}_{\mathrm{min}} = 2\pi f^{\pm}_{\mathrm{min}}(m,r)$ (one for each branch of the dispersion relation),
representing the minimum frequency required by a specific $m$-mode to be able to propagate at a radius $r$. They separate the $(f,r)$-plane in two regions. On the one hand, modes with a frequency $f$ above the minimum energy curve, $f > f^{+}_{\mathrm{min}}$, have a real-valued $k_r$, and are therefore able to propagate. On the other hand, modes with a frequency $f$ below the minimal energy threshold, $f < f^{+}_{\mathrm{min}}$, having imaginary $k_r$ values, correspond to evanescent modes.

The second LR condition imposes that $\p \omega_d/\p r =0$, implying that LR modes correspond to the lowest energy modes that can propagate across the entire fluid. By sitting at the top of the minimum energy curve, the LR modes possess a real valued $k_r$ everywhere, allowing them to transfer energy across the entire system.
We further note that in the shallow water regime, the minimum energy curves $\omega^{\pm}_{\mathrm{min}}$ correspond to $\omega_{\pm}$ of \cite{flowmaster} in the eikonal limit, and therefore the eikonal potential $V_{eik}$ can be reconstructed from the minimum energy curves, namely \mbox{$V_{eik} = -(\omega - \omega_{\mathrm{min}}^{+})(\omega - \omega_{\mathrm{min}}^{-})$}. 

Additionally, we note that one can also use the dispersion relation $\omega_d$ to determine the imaginary part of the LR spectrum, which is related to the Lyapunov exponent $\Lambda$.
Substituting Eq.~\eqref{LR_DR} into the formula to compute the Lyapunov exponent (Eq.~\eqref{Lyapu_def}), one finds that
\be
\Lambda = \mathrm{det}([d^2\omega_d]),
\ee
where $[d^2\omega_d]$ is the Hessian matrix of $\omega_d$ evaluated on the LR.

\section{Analogue Black Hole Spectroscopy}

The important feature that the QNM spectrum depends only on the black hole parameters opens up the possibility of BHS.
By measuring the QNM spectrum of a black hole, one should be able to recover the parameters of the black hole that produced the spectrum~\cite{Press72,Echeverria89,Schutz09}.
With the recent detection of gravitational waves \cite{gwaves,LIGO_propBH}, we seem to be close to turning this idea into a reliable astronomy technique. 
We suggest here the use of a similar approach to characterise vortex flows in fluids and superfluids. 
We will describe in this section the general procedure of what we will call \emph{Analogue Black Hole Spectroscopy} (ABHS). 
The ABHS scheme is as follow:
\begin{itemize}
\item[1)] Measure the perturbations around the vortex flow one wishes to characterise.
\item[2)] Decompose the perturbation into its azimuthal components and extract the characteristic frequency for each $m$ mode.
\item[3)] Perform a non-linear regression analysis to find the best match between the measured experimental spectrum and the theoretical predictions obtained by solving Eq.~\eqref{CO_eq}.
\end{itemize}

The key features of this process that consists in analysing the LR spectrum associated with the vortex flow are detailed below:  \\

\textit{Step 1 - Measurement:} The first step of the process is the detection of the perturbations on top of the vortex flow. 
To perform the measurement, the system needs to be in a quasi-stationary regime. 
This means that the flow should not vary over the time-scale of the measurement, but the flow should not be in its equilibrium state in order to stimulate the emission of characteristic modes. 
After this stage, the perturbations are contained in a single variable $\phi(t,\vec{x})$. 
For example, $\phi$ can represent the surface elevation in the case of surface waves in classical fluids or the atom density in a BEC.
\\

\textit{Step 2 - Characteristic spectrum:} This step consists in extracting the information contained in each azimuthal component $\phi(t,m,r)$ which is obtained from the original signal $\phi(t,\vec{x})$ through an angular Fourier transform.
Note that in order to perform this step and to be able to distinguish between positive and negative $m$'s, $\phi$ needs to be a complex field. 
The analytic representation of the real valued measurement can be constructed via the Hilbert transform. 
From $\phi(t,m,r)$ one can construct the Power Spectral Density (PSD) $PSD(\omega,r,m) \propto |\tilde{\phi}|^2$ for each azimuthal number, where $\tilde{\phi}$ is the time Fourier transform of $\phi$. 
The PSD's are then used to identify the characteristic frequencies.
As we are considering a quasi-stationary system, frequencies are conserved. 
Therefore, any signal emitted at frequency $\omega_0$ will appear as a constant line in the PSD plots at $\omega = \omega_0$ in the $(\omega, r)$ plane over a radial range corresponding to the region of propagation of the signal. 
As the characteristic modes are the lowest energy modes capable of propagating across the system, they will appear as the lowest constant frequency line stretching over the entire radial region. 
By looking for this line in every $P_m(\omega,r)$, one can define a radius independent spectrum $\omega_{\mathrm exp}(m)$.
\\

\begin{figure*}[!h]
\begin{center}
\includegraphics[scale = 0.8, trim= 1cm 0 1cm 0]{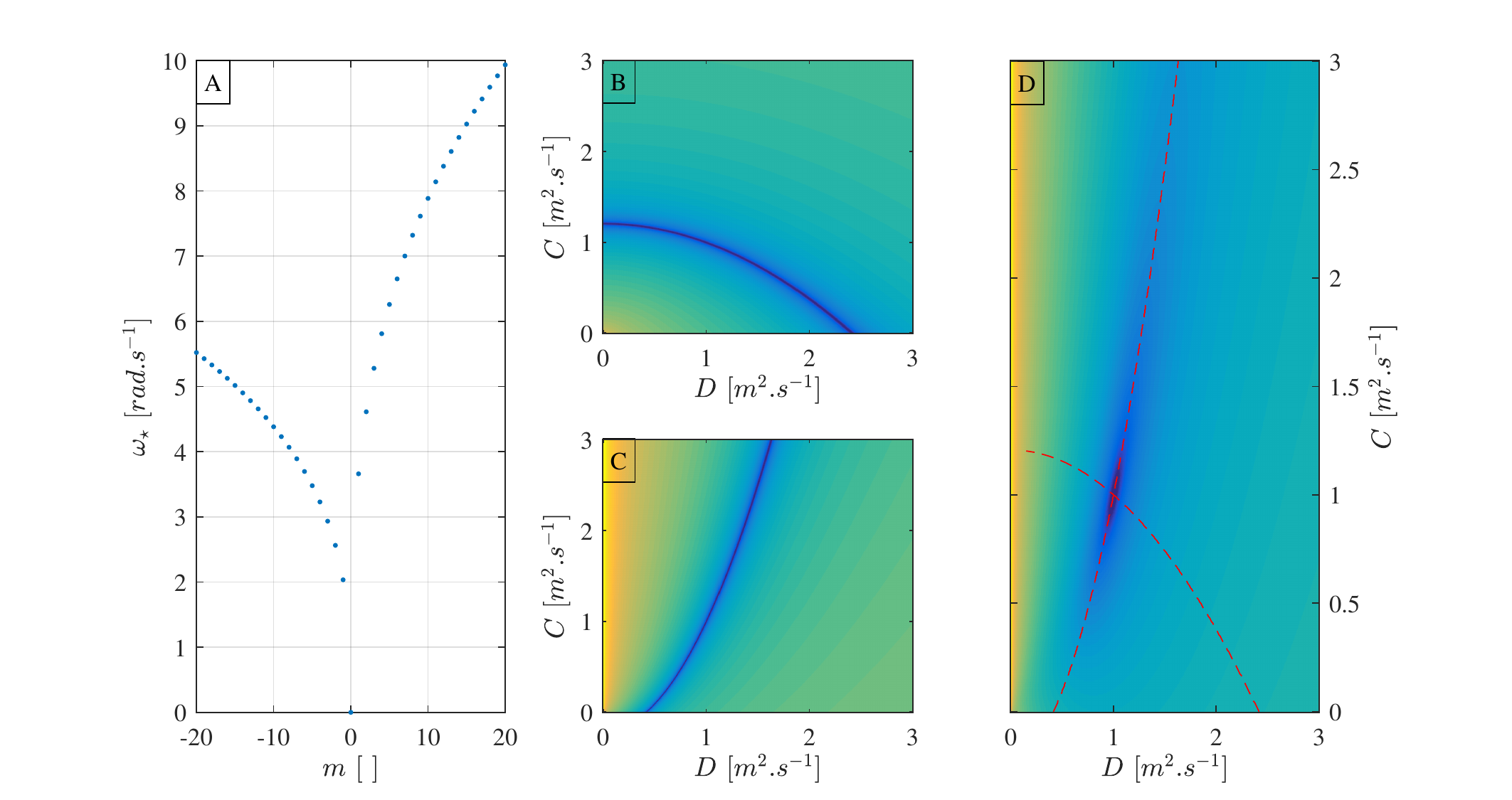}
\end{center}
\caption{\textbf{ Illustration of the Analogue Black Hole Spectroscopy in the deep water regime.}  
Panel (A) depicts the LR spectrum for surface waves in the deep water regime, i.e. $F(k) = gk$, computed with the flow parameters $C=D=1~m^2.s^{-1}$. Panel (B) (respectively (C)) represents the logarithm of the mean square error for a region of the $(C,D)$ parameter space computed using only the negative (respectively positive) $m$ part of the LR spectrum. Darker regions represent smaller values of the log mean square error. As we can see, the mean square error is not minimised for a unique pair of flow parameters but along an entire curve. This defines a family of homophonic vortices, i.e. vortex flows with the same characteristic spectrum. Panel (D) represent the log mean square error computed using the both sides of the LR spectrum. In this case, the mean square error is minimised at one point in the $(C,D)$ space which is the intersection point of the two homophonic curves obtained from panels (B) and (C). The intersection is located at $(C,D) =(1,1)$ which are the values used to produce the initial spectrum.}
\label{ABHS_deepwater}
\end{figure*}

\textit{Step 3 - Parameter extraction:} 
Once the experimental spectrum is obtained, one can estimate the flow parameters that best match the theoretical model with the experiment by minimizing the mean square error (MSE):
\begin{equation}
MSE(C,D) = \frac{1}{M}\sum_{m} \left(\omega_{\mathrm exp}(m) - \omega_\star(C,D,m) \right)^2,
\end{equation}
where $M$ is the number of azimuthal components considered.
We consider the mean square error to be a function of the irrotational flow parameters only and we assume that any other parameters entering in the characteristic spectrum (e.g.~the water depth) can be measured independently. 
We stress here that in order to single out a unique pair of flow parameters, one needs to be able to measure both the co ($m>0$) and counter ($m<0$) rotating LR modes. 
Indeed, as can be seen from panels (B) and (C) of Fig.~\ref{ABHS_deepwater}, there is an entire family of flow parameters lying on a curve that minimises the MSE if one only has access to the negative (positive) $m$ part of the spectrum. Since each point in parameter space corresponds to a different vortex flow, these curves determine a family of \emph{homophonic} vortices, i.e. vortex flows with the same characteristic frequency.
It is the intersection of these two homophonic curves that uniquely defines a pair of flow parameters, as shown in panel (D) of Fig.~\ref{ABHS_deepwater}.
In the special case of the flow being governed by a single parameter then the negative $m$ part of the spectrum is sufficient to fully characterise the flow. Indeed, in the case of a purely rotational flow (i.e. $D=0$) there are no co-rotating LR modes, while in the case of a purely draining flow (i.e. $C=0$), the co- and counter-rotating modes coincide.
\newline

We note that while the general idea of ABHS and BHS is the same, their practical application is very different. 
Indeed, the BHS procedure relies on the matching of a known waveform to the entire signal. 
Generally the waveform will be dominated by one azimuthal number (the least damped one) but the scheme doesn't make explicit use of the symmetry and the information contained in each azimuthal component. 
This is simply because in astronomy, one does not have access to the different azimuthal component experimentally. 
This would require many detectors located around the source of the signal, which is not yet realisable. 
The ABHS method makes use of the possibility to measure the perturbation at various positions. 
This allows for the decomposition into the various azimuthal components. 

Another difference is that the ABHS scheme focuses on the real part of the spectrum. 
This is because the oscillation frequency is a more robust quantity to look at than the decay rate. 
Indeed, the decay rate depends highly on the boundary conditions of the system. 
Formally, the QNM spectrum is obtained by considering perturbations which are purely outgoing at infinity and ingoing at the horizon. 
These assumptions cannot be satisfied in an experiment; the boundary conditions will modify the imaginary part of the characteristic spectrum (because of the presence of reflections or echoes). 
The real part of the spectrum will however stay unchanged, independently of the boundary conditions.
We note however that if one has access to it, the imaginary part of the characteristic spectrum can be used in conjunction with the real part to single out unique flow parameters. This is illustrated in section~\ref{Slow_drain}.
We further note that black holes are not the only objects exhibiting LRs. Because of the local nature of the LRs, the ABHS method is applicable as long as the LRs are located in the irrotational region of the flow, even when the system does not exhibit an analogue horizon.
\\

We now turn to an application of the method: shallow water waves propagating on top of an irrotational DBT flow. 
\section{Application of Analogue Black Hole Spectroscopy} \label{Application}

In this section, we apply our method to a numerically simulated perturbed vortex flow described by the velocity field given in Eq.~\eqref{DBT}. 
The wave equation to be solved is Eq.~\eqref{wave_equation} with the $\tanh$ function linearised to first order (shallow water approximation):
\begin{equation} \label{shallowWE}
\D^2 \phi - c^2 \nabla^2 \phi = 0.
\end{equation}
Exploiting the angular symmetry, this equation can be reduced to a PDE in $(r,t)$ by inserting into the wave equation the ansatz:
\begin{equation}
\phi(r,\theta,t) = \phi_m(r,t) e^{im\theta}.
\end{equation} 
The resulting equation is solved using the Method of Lines.
To implement this numerically, we define $\pi=\mathcal{D}_t\phi$ to convert the wave equation (Eq.~\eqref{shallowWE}) into a vector equation which is first order in time,
\begin{equation} \label{discreteWE}
\partial_t \begin{pmatrix}
\phi_m \\ \pi_m
\end{pmatrix} = \begin{bmatrix}
\frac{D}{r}\partial_r-\frac{imC}{r} & 1 \\
c^2(\partial_r^2+\frac{1}{r}\partial_r-\frac{m^2}{r^2}) & \frac{D}{r}\partial_r-\frac{imC}{r}
\end{bmatrix} \begin{pmatrix}
\phi_m \\ \pi_m
\end{pmatrix}
\end{equation}
We then introduce discrete radial points $r_i$ where $i$ is an integer ranging from $1$ to $N$, and approximate the derivatives using 3-point finite difference stencils,
\begin{equation} \label{stencils}
\p_rf_i = \frac{f_{i+1}-f_{i-1}}{2\Delta r}, \qquad \p_r^2f_i = \frac{f_{i+1}-2f_i+f_{i-1}}{\Delta r^2}
\end{equation}
where $f$ is a place holder for $\phi_m$ or $\pi_m$. The boundary condition at $r_N$ is a hard wall, implemented by setting $f_{N+1}=0$, with $r_N$ placed sufficiently far away that reflections do not occur. 
We place a free boundary inside the horizon which is implemented using a one-sided stencil at $r_1$ (the justification for this is that the value of the field inside the horizon cannot affect its value outside since the two regions are causally disconnected). 

We initialise the simulation with a gaussian pulse centred at $r=r_N-5\sigma$, where $\sigma$ is the spread of the gaussian pulse,
\begin{equation} \label{gaussian}
\phi_m(r,t=0) =  \frac{1}{\sqrt{2\pi\sigma^2}}\exp\left(-\frac{[r-r_0]^2}{2\sigma^2}\right),
\end{equation}
with $\pi_m$ chosen such that the perturbation propagates toward $r<0$. 

To extract the real part of the QNM frequencies, we first compute the time Fourier transform $\hat{\phi}_m(\omega)$ of  $\phi_m(t,r=5r_h)$. 
The location of the peaks in $\hat{\phi}_m$ determines the real part of the QNM spectrum: the peak at $\omega>0$ gives $\omega_\star(m)$ and the peak at $\omega<0$ gives $\omega_\star(-m)$. 
Hence, we only need to simulate for $m>0$. 
At large $m$, the co-rotating modes are much harder to excite, hence the peak at $\omega>0$ becomes difficult to resolve beneath the tail of the $\omega<0$ peak. This accounts for the poor behaviour of the blue crosses at large $m$ on the right of Fig.~\ref{shallow_spectrum}. 
Finally, we extract the imaginary part by finding the gradient of $\log|\phi_m(t,r=5r_h)|$ in the region where the signal is exponentially decaying. To do this, we must have one dominant frequency in the signal (i.e. one peak in $\hat{\phi}_m$) since the interference between two frequencies leads to a non-constant gradient.

We focus on two cases: a slowly draining flow and a flow where the circulation and drain parameters are equal. 
After the QNM spectrum is extracted from each simulation, a non-linear regression analysis is performed in order to extract the flow parameters that best match the characteristic spectrum according to the ABHS method. The obtained parameters are compared to the ones used to simulate the data. In addition, we compare our results to the standard flow measurement technique:  PIV already discussed in section~\ref{PIV_subsec}.

\subsection{Slowly rotating flow}
\begin{figure}[!h]
\begin{center}
\includegraphics[scale=1]{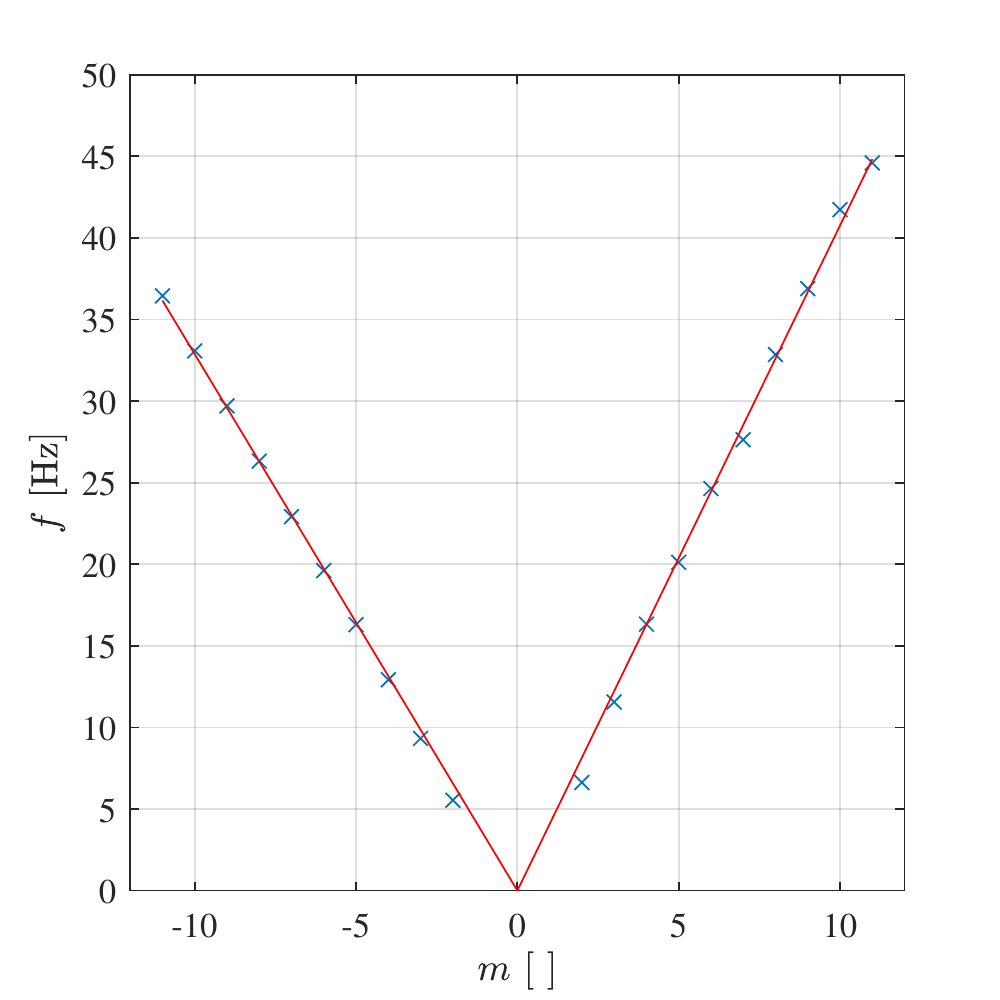}
\end{center}
\caption{ \textbf{Characteristic spectrum of DBT vortex flow in the shallow water regime.} The blue crosses represent the oscillation frequency of the characteristic modes emitted by a draining bathtub flow with \mbox{$C= 22~cm^2.s^{-1}$} and \mbox{$D=220~cm^2.s^{-}$} perturbed by a Gaussian wave packet. The red line is the spectrum obtained via the LR method with the flow parameters \mbox{$C_\star= 23~cm^2.s^{-1}$} and \mbox{$D_\star= 217~cm^2.s^{-1}$}.}\label{shallow_spectrum}

\end{figure}

Here the flow parameters are $h_0 = 10~cm$, \mbox{$C = 22~cm^2.s^{-1}$}, and 
\newline
\mbox{$D = 220~cm^2.s^{-1}$}. 
This results in a slowly rotating flow. In this regime, we can see from Eq.~\eqref{om-shallow} that the co and counter-rotating oscillatory spectrum are in the same frequency range. Therefore, if one can observe one side of the real spectrum, the other side should also be observable.  With these parameters, we are able extract the characteristic spectrum for $|m| < 11 $, see Fig.~\ref{shallow_spectrum}.

\begin{figure}[!h]
\begin{center}
\includegraphics[scale=0.8,trim= 2cm 0 2cm 0]{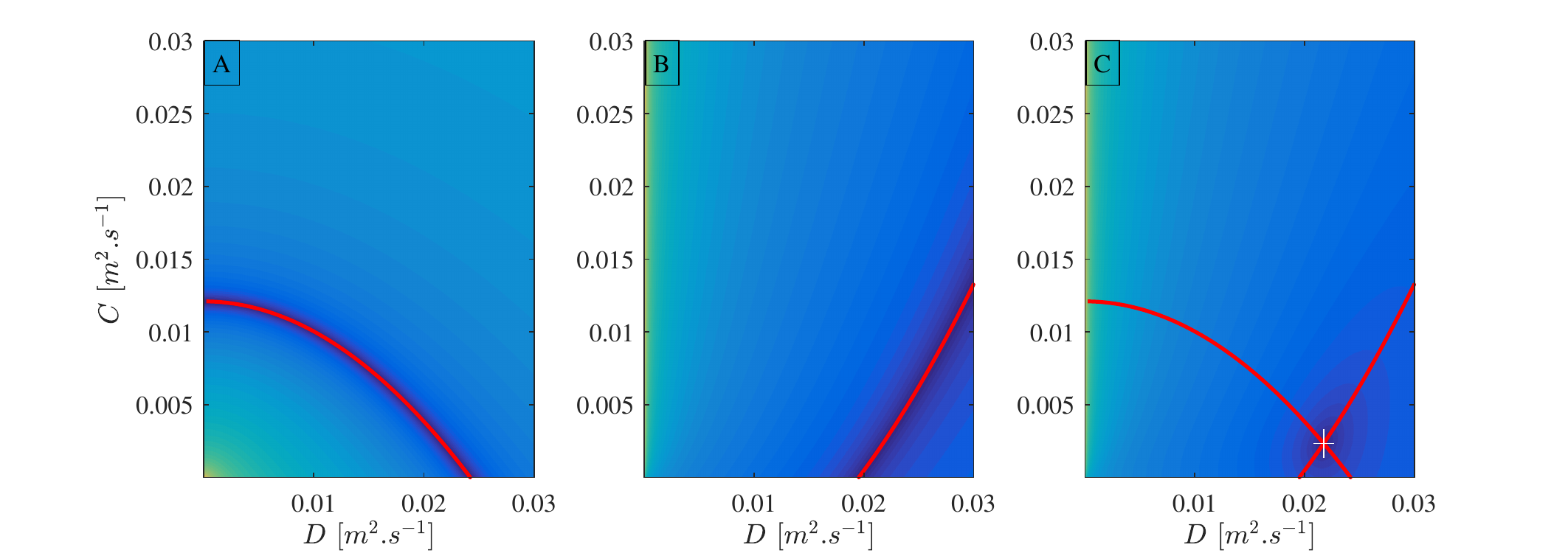}
\end{center}
\caption{ \textbf{Application of the Analogue Black Hole Spectroscopy to numerical simulation: Slowly Rotating Flow}: Panels (A), (B) and (C) represent the logarithm of the mean square error using the spectrum of negative, positive and all azimuthal numbers $m$ respectively. Darker regions represent smaller values of the log mean square error. Similarly to panels (B) and (C) of Fig.~\ref{ABHS_deepwater}, there is an entire curve in parameter space corresponding to homophonic vortices that minimises the mean square error. These two homophonic curves intersect, as seen in panel (C), to give the values of the flow parameters that best match the spectrum. The right cross on panel (C) depicts the $95\%$ confidence interval and is obtained using the covariance matrix.} \label{NLR_shallow}
\end{figure}

We then apply a non-linear regression analysis to match the numerical spectrum with the formula for the characteristic modes (Eq.~\eqref{om-shallow}), using the positive part of the spectrum, the negative part and the entire spectrum. The mean square error for each analysis is respectively plotted in panels (A), (B) and (C) of Fig.~\ref{NLR_shallow}.

Minimising the mean squared error, we find \mbox{$C_\star = 23\pm 10~cm^2.s^{-1}$} and \mbox{$D_\star = 217\pm 10~cm^2.s^{-1}$}. The error on the parameters represents the $95\%$ confidence interval and are computed using the covariance matrix. We see that the obtained values are very close to the ones used in the simulation (relative error of about $6\%$ on the estimation of the circulation parameter $C$ and about $1\%$ for the drain parameter). The discrepancy can be explained by the fact that the LR method used to estimate the characteristic frequencies is only an approximation (the higher the azimuthal number is, the better the approximation will be). 

Finally, we have extracted the flow parameters using the standard flow visualisation technique PIV. 
To create test data for the PIV method, we seed an initial image with particles of 4 pixels diameter such that the density is $0.013$ particles/pixel (these values were chosen to keep data as close as possible to that obtained in the experiments of discussed later in section~\ref{QNM_exp_sec}). 
We then evolve the position of the particles using the velocity field in Eq.~\eqref{DBT} to create a second image. 
These images are analysed using Matlab's extension PIVlab, developed in \cite{PIVlab,PIVlab2,PIVthesis}. 
We used PIVlab's window deformation tool with spline deformation to reduce the number of erroneous vector identifications. 
The analysis was performed over three iterations using an interrogation area of $256\times256$, $128\times128$ and $64\times64$ pixels respectively, each with 50\% overlapping. To obtain $C$ and $D$, we perform a $\theta$ average on the vector field found in PIVlab, then use MATLAB's curve fitting tool to fit the data with a function of the form given in Eq.~\eqref{DBT}. 
The flow parameters extracted via this method are: \mbox{$C_{\mathrm{PIV}} = 22 \pm 0.5~cm^2.s^{-1}$} and \mbox{$D_{\mathrm{PIV}} = 231\pm 1~cm^2.s^{-1}$}. Again, the errors on the parameters represent the $95\%$ confidence interval obtained via the covariance matrix. This represents a relative error on the drain parameter of about $2\%$ and on the drain parameter of about $5\%$. We note that while the confidence intervals are significantly smaller for the PIV method, a systematic error is present in the method. We further remark than the relative error on the parameters obtained using both methods are of the same order, proving the validity of ABHS as a flow measurement technique.

\subsection{Rotating and draining flow}\label{Slow_drain}

\begin{figure}[!h]
\begin{center}
\includegraphics[trim = 4cm 0 0 0]{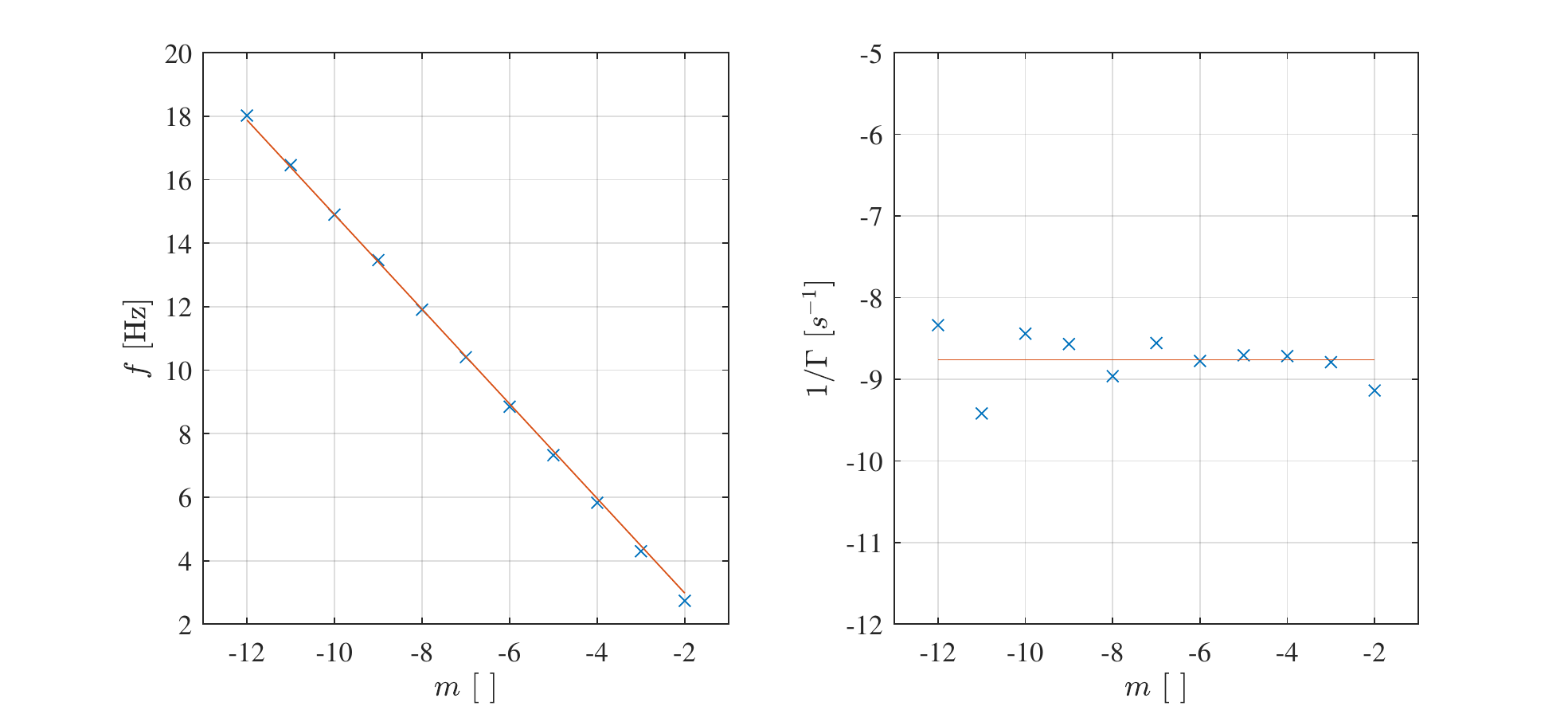}
\end{center}
\caption{ \textbf{Quasinormal mode spectrum of a draining bathtub vortex flow in the shallow water regime}. The left panel represents the oscillation frequencies of the characteristic modes of the draining vortex flow as a function of the azimuthal numbers. The right panel depicts the imaginary part of the characteristic spectrum. In both panels, the blue crosses are the frequencies obtained numerically and the red line represents the best match between the numerical spectrum and the LR prediction. The values of the flow parameters used in the simulation are \mbox{$C=D = 220~cm^2.s^{-1}$}. The values that best match the numerical spectrum are \mbox{$C_\star= 229~cm^2.s^{-1}$} and \mbox{$D_\star=201~cm^2.s^{-}$}}\label{real_and_im_spec}
\end{figure}

We now turn our attention to a flow where the drain and circulation parameters are of the same order. 
In this regime, the LR radius of the co-rotating modes will be very close the centre of the vortex. 
This will result in a very high oscillatory frequency and short decay time, making the observation of the co-rotating characteristic significantly more difficult. 
However, one can use the decay time of the counter-rotating modes in order to single out a unique pair of flow parameters. 
This is the path we adopt here in order to test our method. 
We note once more that in a real experiment, the imaginary part of the characteristic modes will be highly influenced by the boundary conditions and one should be careful when using the decay time to extract information about the set up.
The flow parameters are here set to be $C = D = 220~cm^2.s^{-1}$. With these parameters, we can extract the real and imaginary part of the QNM spectrum for $-12<m<-2$. The spectrum is shown in Fig.~\ref{real_and_im_spec}.

Using the real and imaginary part of the spectrum to perform the non-linear regression analysis we find the parameters ($C_\star$,$D_\star$) that best match the spectrum as the intersection of the two curves minimizing the mean square error of each spectra. In particular, we find that \mbox{$C_\star = 229\pm 26~cm^2.s^{-1}$} and \mbox{$D_\star = 201\pm 44~ cm^2.s^{-1}$}, which are relatively close to the values used in the simulation. The relative error on the circulation parameter is about $4\%$ and on the drain parameter is about $9\%$. We note that in this case the resolution on the parameters is worse than in the previous case. This is due to the fact that it is harder to resolve decay times than frequencies. 
 
 Similarly, we compare the ABHS approach with the standard PIV technique. Using PIV the flow parameters are found to be:
 \mbox{$C_{\mathrm{PIV}} = 229\pm 2~cm^2.s^{-1}$} and \mbox{$D_{\mathrm{PIV}} = 233\pm 1~cm^2.s^{-1}$}. This represent a relative error of about $4\%$ for the circulation parameter and about $6\%$ for the drain parameter. As in the previous case, the confidence intervals are significantly smaller when using PIV but a systematic error is present in the result. The relative errors are again comparable between the two techniques.

\section{Experimental implementation}\label{QNM_exp_sec}

We now turn our attention to the experimental implementation of the ideas discussed above by following the suggested procedure.

\subsection{Experimental setup and the Unruh vortex}

The experimental setup is essentially the same than the one used in the superradiance experiment which was described previously in section~\ref{Exp_setup_sec}. 
In particular, the draining hole is the centre is a $2~cm$ radius hole.

In order to observe the relaxation behaviour of a vortex flow, one obviously needs to perturb it.
This can be done in two ways.
The first way is to set-up and stable vortex flow, as it was successfully done in the superradiance experiment and to then perturbed by sending waves towards it. 
This is usually the path adopted in numerical simulation, in which one perturbs a stable vortex by sending a gaussian wave packet. 
The relaxation phase is then observed after the initial perturbation has passed through the region of interest.
This procedure is however not ideal to implement experimentally. 
Indeed, due to the finite size of the system any initial wave packet will be reflected and overlap with the signal emitted during the relaxation. 
Even though reflections can be attenuated by placing a beach at the end of the tank as it was done previously, the reflection is still bigger that the expected relaxation signal.

The other way to observe the characteristic relaxation modes is to directly set-up a vortex flow out of its equilibrium and let it relax to its stable configuration. This was the method we employed. We call such vortex flow an \textit{Unruh vortex}\footnote{This name comes from the German word ``Unruhe" which means restlessness and was chosen in acknowledgment of W. G. Unruh, the founder of analogue gravity.}

Water was pumped continuously from one corner at a flow rate of \mbox{$15\pm 1~\ell/\mathrm{min}$}. 
The sinkhole was covered until the water raises to a height of \mbox{$10.00 \pm 0.05~\mathrm{cm}$}. Water was then allowed to drain, leading to the formation of an Unruh vortex. 
We recorded the perturbations of the free surface (over a region of \mbox{$58~\mathrm{cm} \times58~\mathrm{cm}$} for $16$ seconds) when the flow was in a quasi-stationary state at a water depth of \mbox{$5.55 \pm 0.05~\mathrm{cm}$}. The water elevation was recorded using the Fast-Chequerboard Demodulation (FCD) method~\cite{Sanders}
and the entire procedure was repeated 25 times.
To apply the FCD, we have placed a periodic pattern at the bottom of the tank in a region of \mbox{$59~ \mathrm{cm} \times 84~ \mathrm{cm}$}. 
The pattern is composed of two orthogonal sinusoidal waves with wavelength of $6.5~\mathrm{mm}$ each.
Deformations of this pattern due to free surface fluctuations are recorded in a region of \mbox{$59~ \mathrm{cm} \times 59~ \mathrm{cm}$} over the vortex using a Phantom Camera Miro Lab 340 high speed camera at a frame rate of 40 $\mathrm{fps}$ over $16.3~ \mathrm{s}$ with an exposure time of 24000 $\mu \mathrm{s}$.

In addition to the free surface measurement, we have performed a PIV measurement. The main idea behind PIV measurement was explained in section~\ref{PIV_subsec}. Here we seeded the flow with plyolite particles that were illuminated by a laser sheet placed at a height of $5.4~cm$.
In the region of observation, i.e at a radius between $7.4~cm$ and $20~cm$, we note that the angular velocity $v_\theta$ is much larger than the radial one $v_r$. 
Namely, the maximum value of the radial velocity is about a tenth of the angular one. 
This fact is not surprising as it is expected that most of the draining in free surface vortices occurs through the bulk and through the boundary layer at the bottom of the tank, with a negligible radial flow at the surface, especially far from the vortex core~\cite{andersen1,CRISTOFANO}. This implies that we cannot determine a precise value for the drain parameter D. Nevertheless, we can determine an upper-bound for D, namely \mbox{$\mathrm{D}_{\mathrm{max}} = 39~\mathrm{cm}^2/\mathrm{s}$}.
Regarding the angular velocity, in the observation window, $v_\theta$ is found to be axisymmetric to a good approximation. This allows us to average the velocity profile over the angular range to obtain the function $v_\theta(r)$ for each experiment.

\begin{figure}[!h]
\centering
\includegraphics[scale=1]{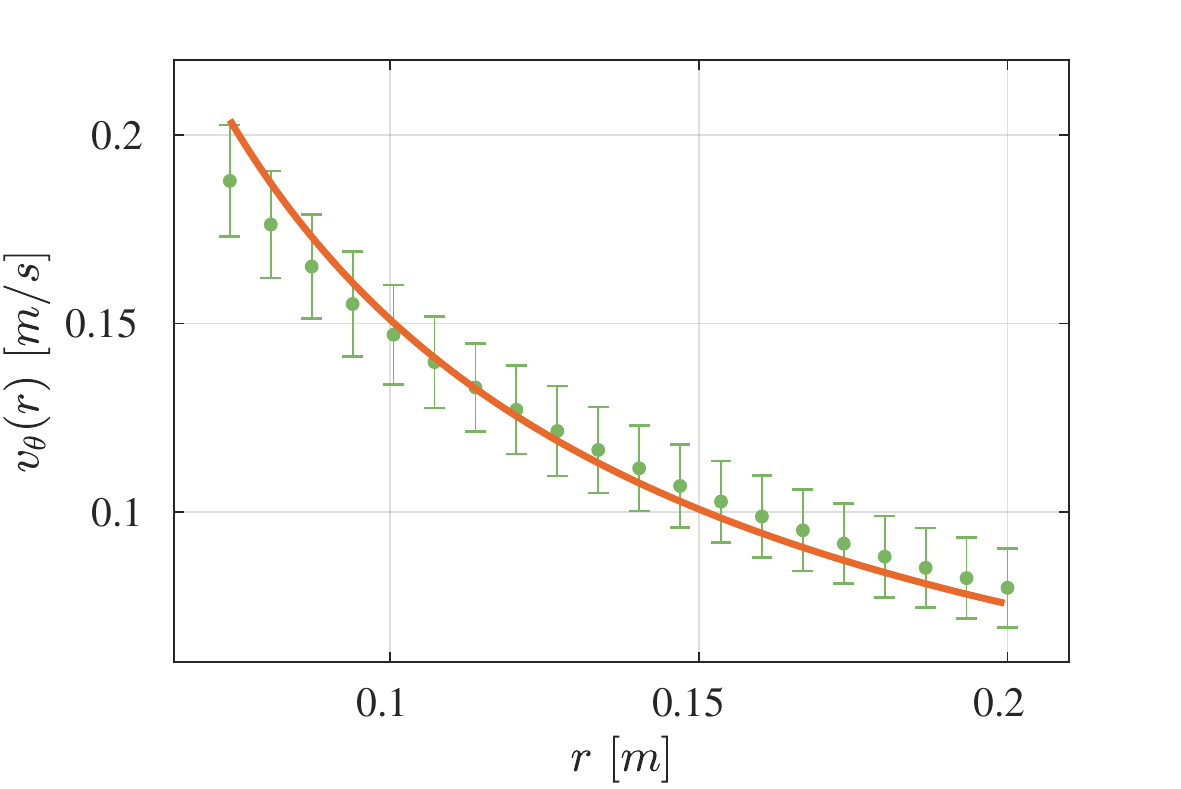}
\caption{{\textbf Angular component of the velocity profile}. Green dots correspond to the averaged (over time, angle and experiments) angular velocity. The error bars indicate the standard deviation. The orange curve is the weighted least-squares fit to the angular velocity $v_{\theta} = \mathrm{C}/r$ of the DBT model, corresponding to $\mathrm{C} = 151$ $\mathrm{cm}^2/\mathrm{s}$.}\label{PIV_QNM}
\end{figure}

The averaged (over time, angle and experiments) angular velocity profile is displayed in Fig.~\ref{PIV_QNM}. The error associated with the averaging is given by the standard deviation and indicates the spread in the data about the mean value. 

We have also checked the stationarity of the flow, by comparing the angular velocity vector field in the first and final seconds of our experimental runs, finding that the maximum difference at any point is smaller than the uncertainty due to the vector identification inherent in the PIV. Hence, we deem the velocity field to be stationary within the error of our method.
Since a draining vortex is expected to be irrotational sufficiently far from the drain hole, we fit the mean angular velocity with the function $v_\theta(r) = \mathrm{C}/r$, as shown in Fig.~\ref{PIV_QNM}. The extracted value for C is $151~\mathrm{cm}^2/\mathrm{s}$. Along with this value, we also compute the 95\% confidence interval via the likelihood function.

\subsection{The characteristic spectrum}

The aim of the data analysis is to look for a radius independent frequency spectrum $f(m)$. This is done as follows:

For each of the 651 pictures of the deformed pattern, we reconstruct the free surface in the form of an array $\xi(t_k,x_i,y_j)$ giving the height of the water at the $1600 \times 1600$ points on the free surface $(x_i, y_j)$ at every time step $t_k$\footnote{The MATLAB code used for this is available at: https://github.com/swildeman/fcd}.
The Unruh vortex created is axisymmetric to a good approximation. This allows us to perform an azimuthal decomposition.
To select specific azimuthal numbers, we choose the centre of our coordinate system to be the centre of the hole and convert the signal from Cartesian to polar coordinates. 
In addition to this change of coordinates, we discard all data points within a minimal radius $r_{\mathrm{min}}\approx 7.4~\mathrm{cm}$. 
This cropping is necessary as there is no clear pattern in the centre. 
This is either due to the hole at the bottom of the tank or due to the curvature of the vortex itself deforming the pattern too much to be detectable. 
We also discard points above a radius $r_{\mathrm{max}} \approx 25~\mathrm{cm}$, to avoid error coming from the edge of the images.
After this step, our data is in the form $\xi(t_k,r_i,\theta_j)$ with $r_{\mathrm{max}} > r_i > r_{\mathrm{min}}$.

Before selecting azimuthal components, we construct the analytic representation of the water elevation by adding to the real signal $\xi$ an imaginary part constructed of its Hilbert transform: \mbox{$\xi_\mathbb{C} = \xi + i \mathrm{Hilb}(\xi)$}. 
The Hilbert transform is computed by means of a discrete Fourier transform and by removing the redundant negative frequency components of the time-spectrum. 
We verified that the use of other methods to compute the analytic representation, e.g.~the use of wavelet transforms, gives identical results. 
This step is crucial as it will allow us to distinguish between $m > 0$ and $m < 0$, i.e.~between waves that are propagating against and with the flow. 
At this stage we are left with a complex valued array $\xi_{\mathbb{C}}(t_k,r_i,\theta_j)$, such that $\mathrm{Re}(\xi_\mathbb{C}) = \xi$. 
In the following we will discard the index $\mathbb{C}$ and keep in mind that we are now dealing with a complex array.

\begin{figure}
\centering
\includegraphics[trim = 2cm 0 0 0]{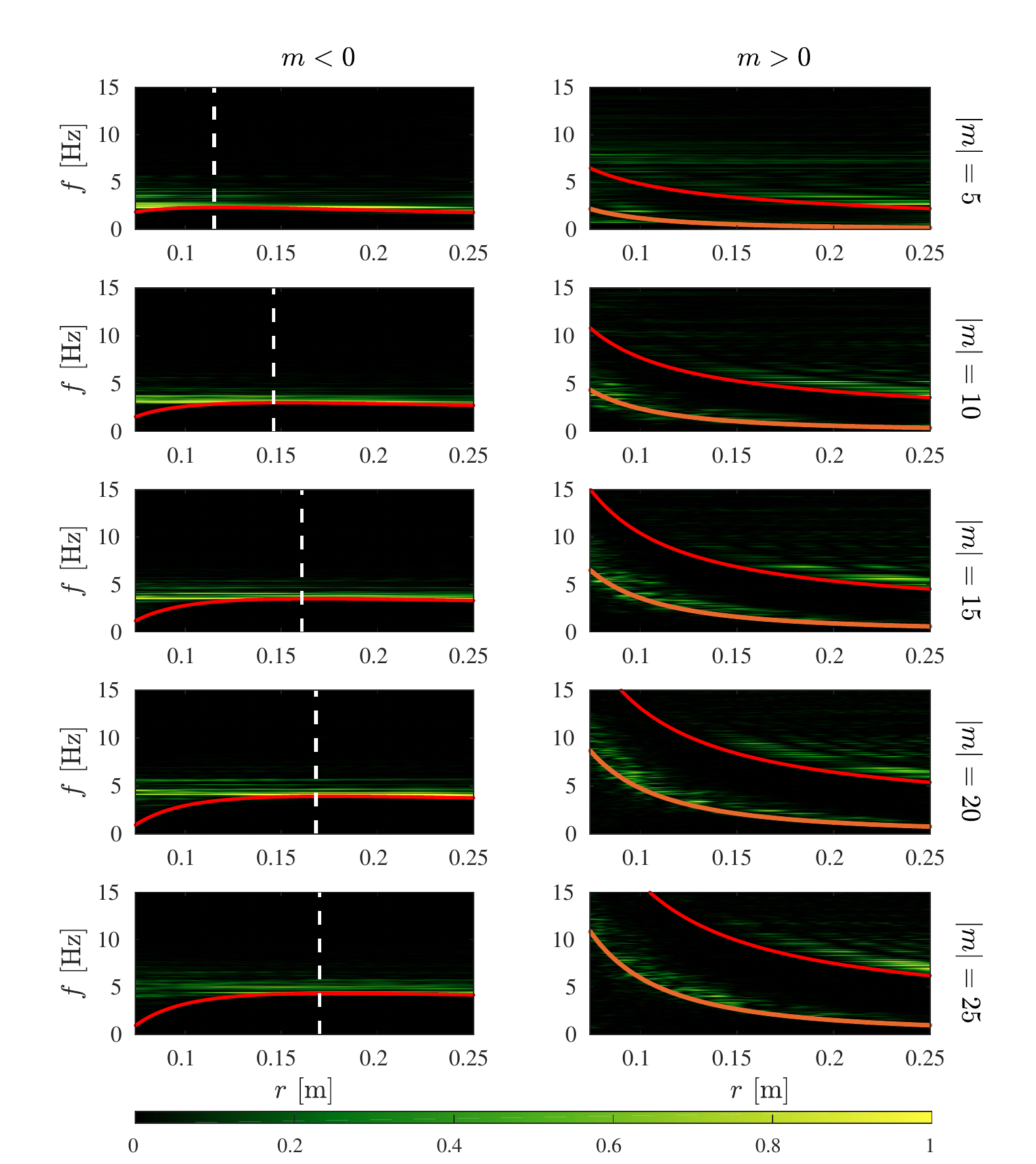}
\caption{{\textbf Normalised power spectral densities}.
The power spectral density is compared with the minimum energy curve $f_{\mathrm{min}}^{+}(m,r)$, plotted in red, for various $m$. 
The maxima of $f_{\mathrm{min}}^{+}(m,r)$ indicate the location of the LRs, $r_{LR}(m)$, which are shown in dashed white lines.
For $m<0$, the spectral density peaks and the minimum energy line are distinguishable for small radii.
For $m>0$, we observe two signals whose peaks are radius-dependent. The upper one follows the minimum energy line and corresponds, most probably, to random noise generated locally. The lower one follows the angular velocity of the fluid flow according to $f_\alpha(m,r)=mv_{\theta}(r)/(2\pi r)$ (orange curve) and is likely sourced by potential vorticity perturbations.}\label{PSD_plot}
\end{figure}

We then perform a discrete Fourier transform in the angular direction to separate the various azimuthal components:
\be
\tilde{\xi}(t_k,r_i,m) = \sum_j \xi(t_k,r_i,\theta_j) e^{-im\theta_j} \Delta \theta. 
\ee
From this, it is possible to estimate the Power Spectral Density, $PSD(f, r_i,m)$, of the waves emitted by the vortex for every azimuthal number m and every radius $r_i$:
\begin{equation}
PSD(f,r_i,m) \propto \left| \tilde{\xi}(f,r_i,m) \right|^2,
\end{equation}
where $\tilde{\xi}$ is the time Fourier transform of the height field at fixed $(m,r_i)$. 
The normalised PSDs for a range of $m$ values are presented in Fig.\ref{PSD_plot}. 
\\

We can identify two different behaviours depending on the sign of $m$.
 
\begin{figure}[!h]
\centering
\includegraphics[scale=1]{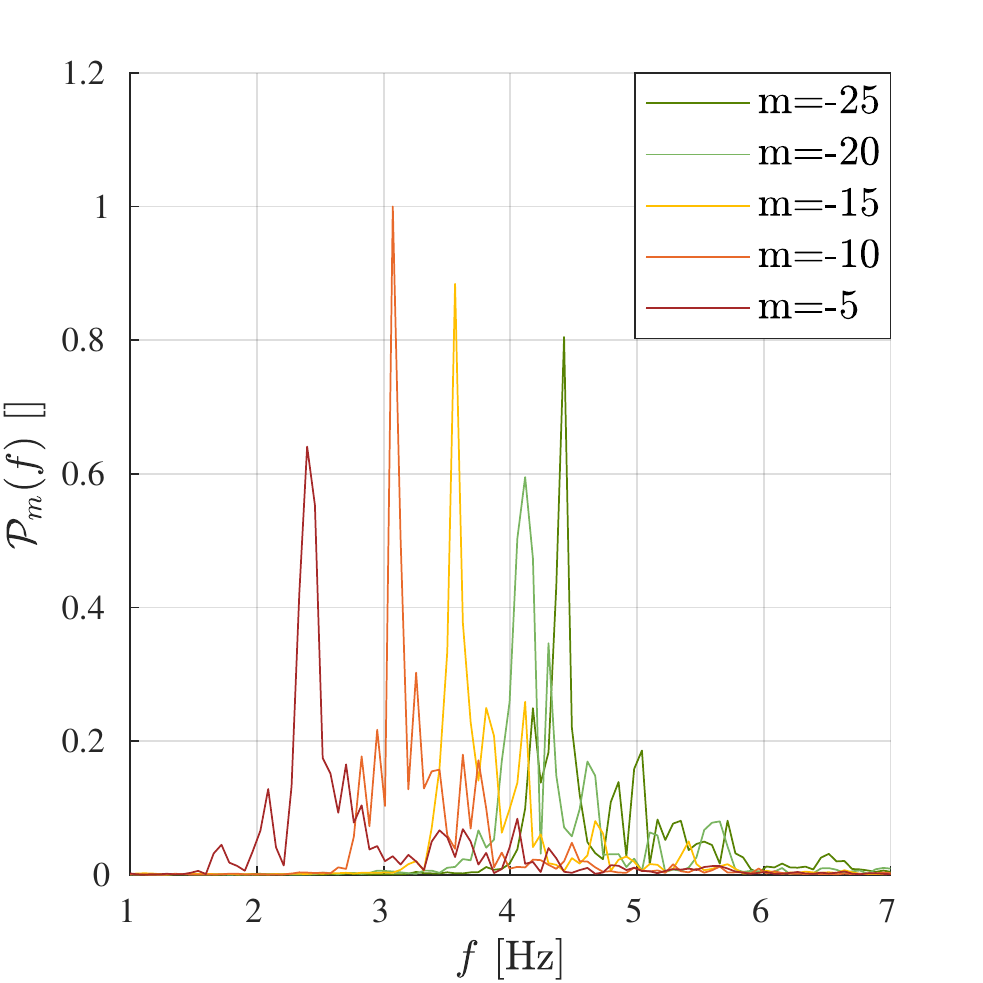}
\caption{{\textbf Typical radius-averaged power spectral densities.}}\label{peaks_figure}
\end{figure}

For negative $m$'s, the PSDs are approximately constant over the window of observation. 
The spectral density is peaked around a single frequency, which allows us to define the position-independent spectrum $f_{\mathrm{peak}}(m)$. We note that this frequency differs from the minimum energy frequency (red curves in Fig.~\ref{PSD_plot} close to the centre. This suggests that this signal is not stimulated by some noise in the system (which would lie on the minimum energy curve). 
To find these frequencies, the PSDs of counter-rotating waves are finally averaged over the radius in order to look at the r-independent frequency content, i.e.~the oscillation frequency of the LR modes.
Various averaged PSDs are presented in Fig~\ref{peaks_figure}.
For each one of them, corresponding to a different $m$, the location of the peak, $f_{\mathrm{peak}}(m)$, is obtained by a parabolic interpolation of the maximum of the power spectrum $\mathcal{P}_m$ and its nearest neighbouring points. By repeating the procedure using a different choice for the center, located 10 pixels ($\approx 4\mathrm{mm}$) away from the original, we observed that the maximum deviation in the location of a frequency peak is approximately $2\%$, attesting the robustness of the procedure against an inaccurate choice for the centre.

This approach is not possible for the positive $m$'s. 
Indeed, two radius dependent signal are present in the PSDs which prevent us to define a frequency as a function of $m$ only. 
We can understand their origin using the flow parameters previously obtained. 
By computing the minimum energy line, $f_{\mathrm{min}}^{+}(m,r)$ (red curve), we observe that one of the signals corresponds to random noise generated locally. 
The other signal is related to the angular velocity of the fluid flow and can be matched with the curve $f_\alpha(m,r)=mv_{\theta}(r)/(2\pi r)$, shown in orange. 
This peak lies below the minimum energy and, hence, corresponds to evanescent modes. 
A possible explanation for their appearence is that \textit{potential vorticity} (PV) perturbations act as a source for them \cite{Stepanyants18}.
In irrotational flows (which is the regime in which our observations are made), PV is carried by the flow as a passive tracer. 
Various $m$ components of PV will therefore source free-surface deformations which are transported at frequencies $f_\alpha(m,r)$. These observations strengthen our confidence in the flow parameters obtained.

\subsection{Results}

\begin{figure}[!h]
\centering
\includegraphics[scale=1, trim = -1cm 0 0 0]{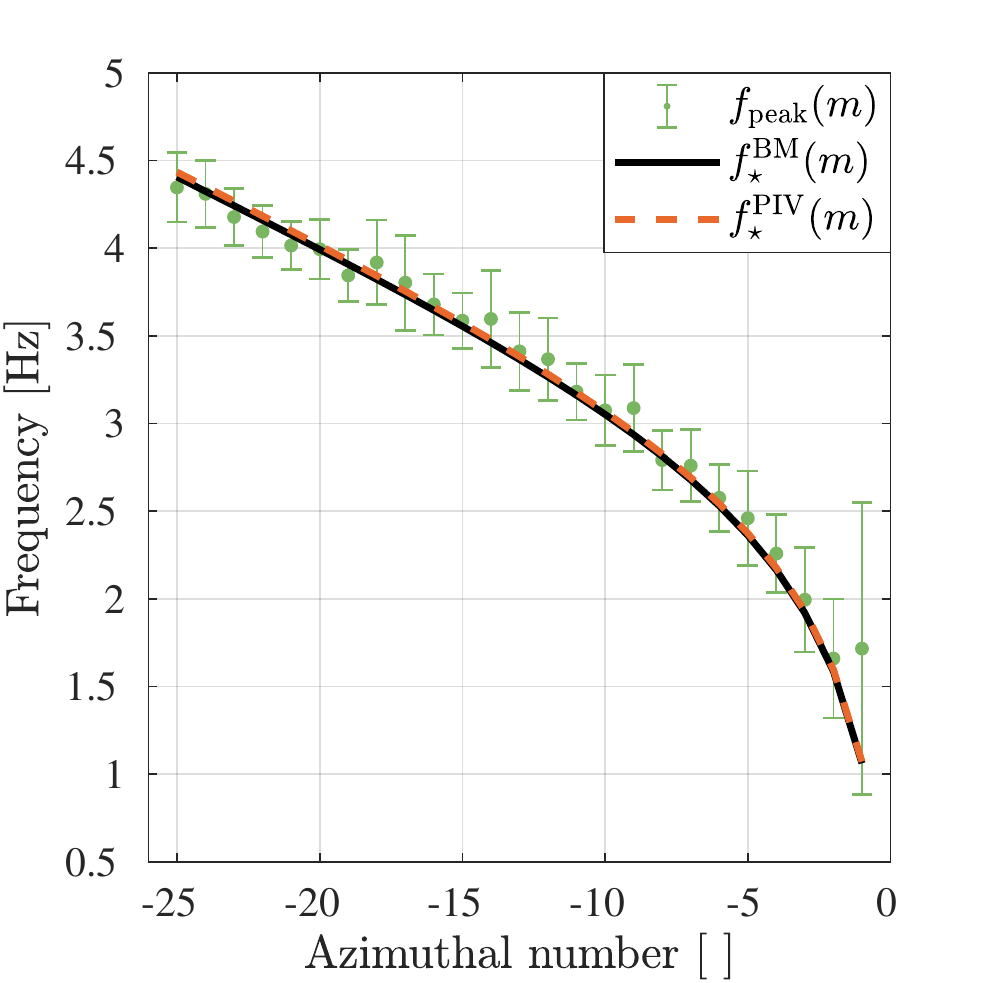}
\caption{{\textbf Characteristic spectrum of the Unruh vortex}. 
The frequency spectrum $f_{\mathrm{peak}}(m)$, extracted from the experimental data and represented by green dots, is compared with the theoretical prediction for the LR frequencies, $f_{\star}(m)$. The error bars indicate the standard deviation over 25 experiments. The dashed orange curve is the predicted spectrum, $f_{\star}^{\mathrm{PIV}}(m)$, computed for $\mathrm{C} = 151~\mathrm{cm}^2/\mathrm{s}$ and $\mathrm{D} = 0~\mathrm{cm}^2/\mathrm{s}$. These flow parameters were obtained via the independent flow measurement technique PIV. The two spectra agree, confirming the detection of LR mode oscillations. The solid black curve, $f_{\star}^{\mathrm{BM}}(m)$, is the non-linear regression of the experimental data to the draining bathtub vortex model, and provides the values for C and D presented in the red curve of Fig.~\ref{Avocado_plot}.
}\label{charac_spectrum_plot}
\end{figure}

At this stage, we have observed during the relaxation phase of a vortex flow the presence, throughout our observation window, of counter rotating modes emitted at a frequency depending only on their azimuthal number. This information is contained in the spectrum previously obtained $f_{\mathrm peak}(m)$. This spectrum is presented in Fig.~\ref{charac_spectrum_plot} and can be use twofold.

First, we can use the value of the flow parameters estimated via our PIV measurement to compute the characteristic mode spectrum $f_{\star}^{\mathrm{PIV}}(m)$ using the LRs properties, as shown by the dashed orange curve in Fig.~\ref{charac_spectrum_plot}.
We observe that the model describing the characteristic emission of an Unruh vortex as LR modes is consistent with the data.
This is the first experimental observation of the oscillatory part of the LR spectrum.

\begin{figure}[!h]
\includegraphics[scale=0.9]{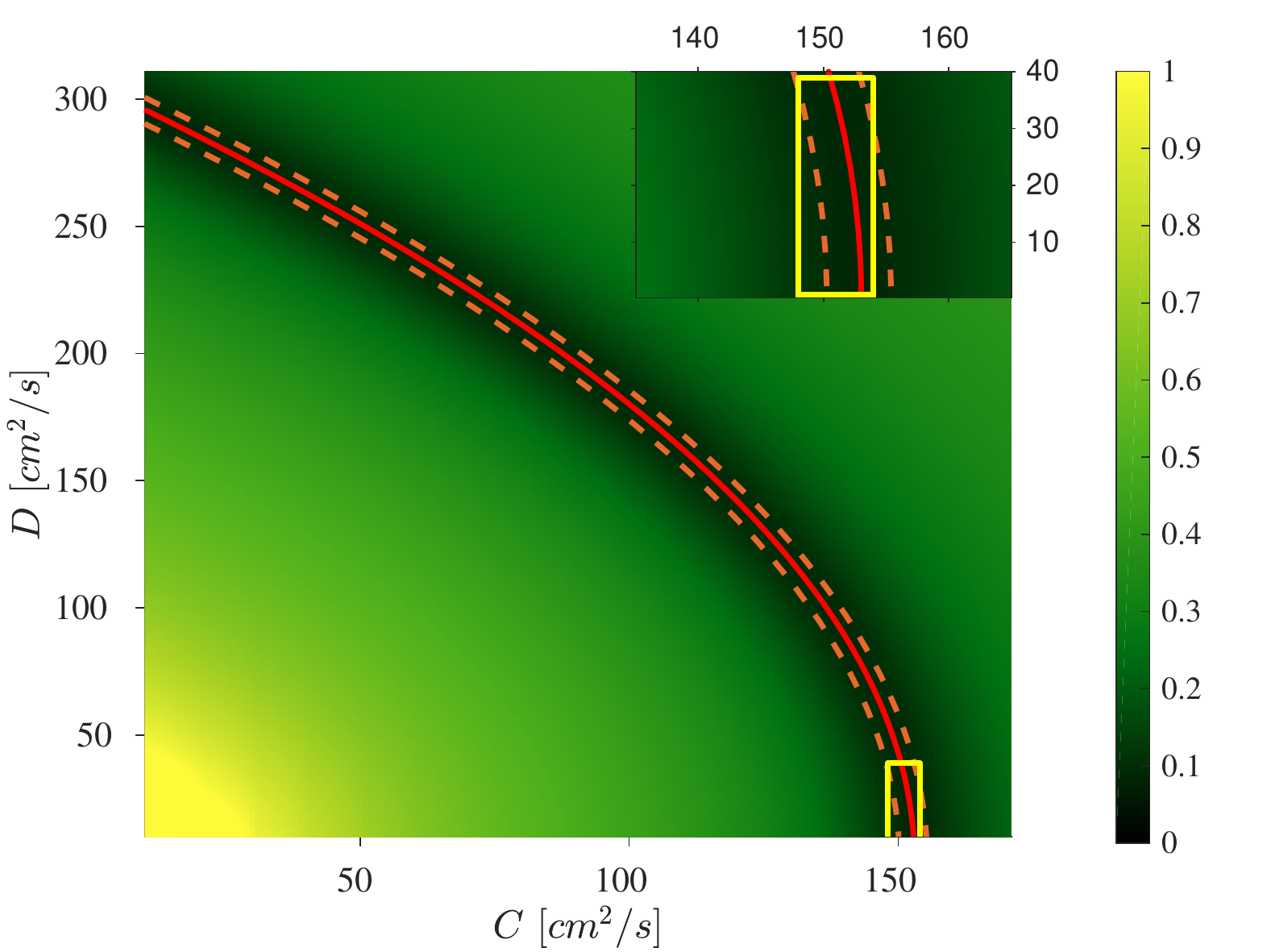} 
\caption{{\textbf Flow characterisation}.
The intensity of the background image represents the normalised weighted sum of squared residuals between the experimental spectrum, $f_{\mathrm{peak}}(m)$, and the theoretical prediction for the LR frequencies, $f_{\star}(m)$, as a function of the flow parameters. The red curve represents the family of possible values for C and D that best match the experimental data (using the method of weighted least squares). The area delimited by the dashed orange curves represents the 95\% bootstrap confidence interval. It overlaps with the yellow rectangle on the bottom right corner, which corresponds to the flow parameters obtained using PIV. The spread along the C-direction represents the 95\% confidence interval estimated via the likelihood function. The spread along the D-direction represents the extracted upper bound for D. In the top right corner, we present a detailed view of the parameter space where the two flow measurements overlap.
}\label{Avocado_plot}
\end{figure}

Second, after having validated our approach, we perform ABHS to characterise the fluid flow (as an alternative to PIV). 
By leaving the flow parameters C and D unspecified, we look for the best match (in terms of non-linear regression analysis) between the experimental spectrum $f_{\mathrm{peak}}(m)$ of counter-rotating modes and the corresponding theoretical predictions for the LR spectrum. 
This reduces the DBT parameter space from two dimensions to one, constraining the flow parameters C and D to lie on the homophonic curve shown in red in Fig.~\ref{Avocado_plot}. Any pair of points along this curve will give the same spectrum, $f_{\star}^{\mathrm{BM}}(m)$, represented by the solid black curve in Fig.~\ref{charac_spectrum_plot}. 
The region between the dashed orange curves shown in Fig.~\ref{Avocado_plot} represents the 95\% confidence intervals for the values of C and D. 
This region overlaps the yellow rectangle which represents the possible flow parameters found using PIV. Note that, in this case, the black hole spectroscopy method imposes a slightly stronger constraint on the circulation parameter than PIV.

As was previously explained, we highlight that in order to uniquely determine C and D the positive $m$ part of the LR spectrum is also needed.
However, when the flow is characterised by only one parameter (e.g.~purely rotating superfluids), the counter-rotating LR modes contain all the information about the fluid velocity. 
This is effectively the case in our experiment. 
Since $D \ll C$ in our window of observation, the vortex flow can be considered to be purely rotating and our observations are sufficient to fully characterise the flow in this region.

\section{Summary and discussion}

Based on the results obtained in Chapter~\ref{Rays_sec}, namely the presence of unstable orbits around irrotational vortex flows, we have argued that these orbits will generate a discrete set of damped resonances. 
We have characterised these modes, that we call LR modes, in the presence of dispersive effects. 
An interesting feature of dispersion is that the lifetime of the characteristic modes increases at high circulation. 
For the co-rotating mode, this behaviour significantly differs from the non-dispersive (shallow water) regime and might lead to an instability if the decay of those modes becomes smaller than their amplification via superradiance (\cite{OLI14, HOD14,BRI15}). 
\\

We have also shown that these characteristic modes depend solely on the properties of the flow, similarly to the shallow water case where the formal fluid-gravity analogy holds. 
Inspired by this analogy and the idea of BHS, we have developed a scheme, that we called ABHS, to characterise irrotational vortex flows directly from their relaxation spectrum. 
We have outlined the main steps of the method and applied it to some numerical data to show its potential in real life applications. 
We have also compared our new method with the PIV technique. 
We have seen that the two methods give similar relative errors when estimating the flow parameters. 
However, the resolution is still better when using PIV on numerical data. 
We believe however that it is possible to increase the resolution of the ABHS method by measuring the characteristic frequencies for a broader range of azimuthal modes. 
Another possibility would be to develop an alternative method applicable to dispersive systems to compute more accurately the characteristic spectrum, similar to the continuous fraction method~\cite{leaver} or via numerical simulation~\cite{pricepullin,krivan}. 
This method, applicable to fluids and superfluids alike, therefore provides a new and non-invasive technique to identify vortex flows. 

A remark should also be made about the name chosen for the method. 
While the ABHS method is a natural choice, it important to note that it is not only applicable to 'analogue black hole' flows. 
This is due to the fact that the method relies on the notion of LRs which are local objects.
While the QNMs of astrophysical and analogue black holes are strongly dependent on the boundary conditions at the horizon and at infinity, the LR modes are characterised only by the local structure around the LRs.
This is an important point for the applicability of the method. 
Indeed, the irrotational vortex flow is interesting theoretically and constitutes a good first model to describe experiments, it however remains too naive to capture real hydrodynamical vortices. 
Indeed, one expects the irrotational assumptions to break down around some radius close to the vortex core.
Below this radius, the flow becomes rotational and more parameters are needed in order to describe it.
Nonetheless, if the flow is sufficiently fast for the LRs to be in the irrotational regions, the real part of the characteristic spectrum will not be significantly affected. 
We stress here the fact that the real part will not be considerably affected. 
The imaginary part on the other hand will be modified as the inner boundary condition will be modified by the presence of extra-structures. 
A similar behaviour, manifested by the appearance of echoes, has been predicted in black hole physics when one adds structure (such as a wall) outside the event horizon~\cite{Cardoso16}. Further, if the flow presents additional structures described by a new set of parameters relevant in the effective field theory, one would expect the system to exhibit extra features. 
For example, the case of shallow water waves incident on a vortex flow with a rotational core has been studied recently~\cite{flowmaster}.
This study, using a simplified model, predicted that the characteristic response to perturbations of this more realistic vortex flow will be composed of the usual ringdown modes together with extra modes, known as quasi-bound states. 
By measuring the quasi-bound state frequencies in addition to the quasi-normal frequencies, one would have access to extra conditions in order to identify the extra parameters needed to describe the flow.

In addition, the ABHS method can be used in fluids and superfluids alike, as an alternative to the standard fluid flow visualisation techniques, such as PIV that requires tracer particles. In particular, when suitable tracer particles are hardly found or do not exist, like in superfluids~\cite{Chopra}, this is a promising non-invasive method to characterise fluid flows.
\\

We then presented an experiment in which we have put in practice the ABHS scheme. This experiment, which consisted in observation the relaxation process of a vortex flow, exhibits a new facet of the fluid-gravity analogy. 
Firstly, by providing the first observation of LR mode oscillations, we have shown the presence of a universal relaxation process between black holes and vortices. 
Secondly, we have used the idea of ABHS to infer information about the vortex flow from the measured characteristic spectrum. 
We have compared the flow parameters obtained via this scheme with the common PIV method. Our findings show that, in this case, the ABHS method gave a slightly better accuracy than the PIV method. 
This practically establishes a new non-invasive method to identify wave-current interactions and the effective field theories describing such systems.

\chapter{Conclusion}\label{Conc_sec}
\epigraph{\textit{We must, and we should, and we always do, extend as far as we can beyond what we already know, beyond those ideas that we have already obtained. Dangerous? Yes. Uncertain? Yes. But it is the only way to make progress. Althought it is uncertain,  it is necessary to make science useful.}}{Feynman}

In this thesis, we have reported experimentally and theoretically on the similarities between matter around rotating BHs and surface waves propagating around vortex flows. 
This idea, that one can mimic curved space-time physics by using a condensed matter system is the heart of the field of AG.
We started our discussion by reviewing the history of this field which started almost forty-years ago.
\\

After having re-derived the standard analogy in Sec.~\ref{surface_wave_analogy_sec}, we extended the well known concepts of geodesic of curved manifold beyond the usual analogue regime in Chapter~\ref{Rays_sec}. 
In differential geometry, geodesics are the curves of shortest length between two points on a given manifold, and they correspond to the world lines of free particles\footnote{This is true when the connection used to define "straight lines", which are the world lines of free particles, is the connection associated to the metric used to define the notion of distance and of shortest length.}.
In the analogue system, one can introduce effective massless particles (in our case called hydrons~\cite{synge1963hamiltonian}), by means of a gradient expansion method.
This procedure was carried out in the presence of dispersive effects. 
It was possible to observe and characterise the effect of a background flow on the world lines of these effective particles. In particular, we have studied the analogue of light-bending in the presence of dispersion.

Using this gradient expansion method, we have shown that irrotational vortex flows allow for the existence of closed circular orbits, analogous to the LRs present around astrophysical BHs. 
This fact was known when dispersive effects were neglected and when the analogue metric for irrotational vortex flows resembles the one of a rotating BH. 
Our findings show that the correspondence extends beyond the analogue regime, attesting of the robustness of the concept of LRs.
The main difference from the relativistic regime lies in the fact that the radius of the dispersive orbits are now frequency dependent.

By comparing our theoretical predictions with experimental data, we have shown that such orbits are to be expected in real life systems. As it is the case in BH physics, the presence of such orbits around vortices is associated with various effects. These findings were reported in~\cite{TheoLR}.
\\

In Chapter~\ref{Superradiance_sec} we presented the experimental observation of the rotational superradiance effect in a vortex flow. 
Superradiance is a radiation enhancement scattering process during which an incident waves extracts energy from the scatterer.
This process was originally described for electromagnetic radiation incident on a rotating cylinder by Zel'Dovich.
Following his seminal papers~\cite{zeldovich1,zeldovich2}, it was soon realised that the superradiance effect could occur in several systems. 
The universality of the effects lies in the very few ingredients required for the phenomena to take place.
Namely one needs a reservoir of energy, a mechanism to allow for the presence negative energies, and an absorption mechanism.
These criteria are met by a rotating BH and therefore by a draining vortex flow in the analogue regime.
Despite the large number of systems capable of exhibiting superradiance, the effect has remained unobserved for more than 40 years. 

We have reported here on the first experimental evidence for the superradiance effect. 
We have conducted a wave/vortex scattering experiment in a water tank setup. 
By comparing the energy of the incident and reflected waves, we were able to show that an amplification process was at stage. 
Parts of the initial wave, corresponding to co-rotating modes, were amplified up to $14\%$. 
On the other hand, counter rotating waves were absorbed, in agreement with the superradiance effect.

The amplification process was clearly observed, however the regime in which our experiment was performed is currently lacking a quantitative theoretical prediction. 
Indeed, the experiment was performed in a regime were the height of the water was comparable to the wavelength of the waves, implying the presence of dispersive effects. 
In addition, while the fluid flow was shown to be irrotational far away from the vortex core, a flow measurement (made using PIV) revealed that our vortex flow exhibits a rotational core. 
These additional effects brought us in a regime where it is still unclear what are the relevant equations of motion.

While the presence of these extra structures does not allow us to draw any conclusion about the superradiance effect around rotating BHs, it however accounts for the robustness of the effect. 
In addition, the persistence of the process in such an extreme regime suggests that a link between vortex flows and rotating BHs still holds outside the analogue regime. 
This link will almost certainly not be in the form of an analogue metric alone, but our experiment begs the question about the deeper nature of the relationship between vortex flows and rotating BHs.
Exploring this connection will certainly advance our understanding of vortex flows, but it might also provide us with a new understanding of BHs.
The results presented in Chapter~\ref{Superradiance_sec} were published in~\cite{Superradiance}.
\\

Motivated by the results presented in Chapters~\ref{Rays_sec} and~\ref{Superradiance_sec}, namely that BHs and vortex flows are the stage of similar processes even outside the standard analogue regime, we have studied theoretically and experimentally the relaxation process of vortices. 
This was the subject of Chapter~\ref{Ringdown_sec}.
In this chapter, we have shown that, when perturbed, a vortex will emit characteristic waves with a complex frequency. These frequencies can be estimated using the analogue LRs exhibited in chapter~\ref{Rays_sec}. 
This is analogous to the ringdown phase of BHs.
These characteristic modes are fully determined by the local properties of the flow and do not depend on the initial source of the perturbation.
Therefore, by measuring the characteristic spectrum, it is possible to infer information about the vortex flow.
This provides a non-invasive method to measure vortex flows that we called Analogue Black Hole Spectroscopy. This is in line the BHS idea which states that one can extract the BH parameters by measuring its ringdown modes.
We developed a practical scheme to apply the ABHS method that we have tested numerically.
We have shown that this method compares with a standard flow visualisation technique (PIV).

We have conducted an experimental study to confront our theoretical results. 
By setting up an \textit{Unruh vortex}, we have first established that vortices do emit characteristic modes with a frequency that corresponds to the analogue LR prediction.
This experiment shines light on the relaxation process of vortex flows as there was no theory predicting if and how these modes should be excited in such a system.
This is the first experimental confirmation of the existence of analogue LRs around vortex flows.
Secondly, we have applied the ABHS method to deduce the vortex flow parameters (circulation and drain). 
We have seen that the ABHS spectroscopy matches with PIV measurements.
These results were published in~\cite{Torres_QNM,Torres_ABHS}.
\\

While the main focus of this thesis is the study of analogue rotating BHs, we have also presented in Appendix~\ref{Appendix3} a novel system capable of mimicking field propagation during various cosmological evolutions.
The system is composed of a paramagnetic fluid immersed in a strong gradient magnetic field.
The effect of the magnetic field is to modify the gravitational force acting on the fluid which results in the possibility to modulate in time the speed of interface waves.
We have shown that this system reduces to the standard FRLW metric in the long wavelength regime and when surface tension is neglected.
The great tunability of the external magnetic field allows for the simulation of various cosmological scenario and even allow for the study of field propagation during signature change event.
Such a process can be achieved as it is possible to change the sign of the propagation speed squared by inverting the effective gravitational force, leading to magnetic levitation. 
Experimentally exploring this regime is a promising way to gain insight into this scenario where a clear theoretical understanding is still missing.

Before exploring such drastic changes in the effective space-time, we have studied the simulation of analogue inflation. 
By moving the vessel containing the fluid in a superconducting magnet, it is possible for our analogue universe to exhibit an exponential scale factor.
We have seen that our system can be used to observe mode freezing, mode amplification and the build up of correlation, typical of inflationary evolution.
Since the technology is readily available to build this system, we believe that exciting experimental results will soon be available.
This study was published in~\cite{Fifer:2018hcv}.
\\

We have seen at the beginning of our discussion, that analogue gravity experiments were, and are still, very much focused on the observation of the Hawking radiation. 
These experiments brought a new perspective on the Hawking effect and stimulated many theoretical studies that deepened our understanding of both BH radiation and condensed matter systems.
However, we have also seen, that analogue gravity is a resourceful tool that can be used to investigate many other phenomenon such as light-bending, superradiance, cosmological particle production or BH relaxation.
This well known fact has been exploited theoretically but never experimentally. 
The aim of this thesis was to start filling this gap by performing experiments around analogue rotating BHs.
By observing BH like effects in vortex flows, and using gravity inspired tools such as the LRs to describe them, we have experimentally shown the strong link that exists between vortices and BHs.
More interestingly, we have shown the persistence of this link outside of the analogue regime. 
Namely, even in rotational and dispersive fluids, vortices are capable of mimicking BH effects. 
It is important here to stress that the connection between vortices and BHs is made via the universal effects shared by both systems. 
While in the analogue regime, the existence of the analogue metric allows us to draw a clear link between vortices and BHs, such connection is not so clear for rotational and dispersive vortex flows.
However, we cannot help but think that more complicated hydrodynamical systems might correspond to more complicated gravitational scenarios. This possibilities has been explored in~\cite{Visser:2005ss,Weinfurtner:2006wt,Weinfurtner:2007br,Cropp:2015tua,Liberati:2018hfk}.

Precisely answering this question might be the next goal of analogue gravity.
This could be done by investigating unexplored regime and novel analogue effects, such as backreaction processes where the non-linear structure of the theory comes into play or including viscous effects.
With the rapid technological development, it seems that analogue experiments will be able to probe such regimes and provide physicists with new challenges.

Here lies the strength of analogue gravity, in its capacity to stimulate new ideas, and to develop an intuition and a feeling about the various ways nature expresses itself.

\appendix
\chapter{Error induced by a wrong centre}\label{Appendix1}
In this appendix, we study the influence of a wrong choice of centre for the coordinate system used to perform an angular Fourier transform (AFT).

\section{First order calculations}

We consider here a circular tank of centre C. The tank can be represented in Cartesian coordinates $(x,y)$ where the origin is chosen to be the centre of the tank. This means that the coordinates of C are $(x=0,y=0)$. After having chosen the origin for coordinates, one can go from Cartesian to polar coordinates $(r,\theta)$ as follow :
\be
(r,\theta) = (\sqrt{x^2 + y^2}, \atan(y/x)).
\ee

Let's now choose a different origin for our coordinates C' where C and C' are separated by a small distance $\epsilon$. This new centre correspond to different Cartesian coordinates $(x',y')$ and different polar coordinates $(r',\theta')$. 
We further denote by $\alpha$ the angle between the horizontal axis and $CC'$.
\be
x = x' + \epsilon \cos(\alpha) \quad \mathrm{ and } \quad y = y' + \epsilon \sin(\alpha)
\ee
We can relate the two polar coordinates at first order in $\epsilon$:
\begin{eqnarray}
 r &=& \sqrt{x^2 + y^2} \\
   &=& \sqrt{\left( x' + \epsilon \cos(\alpha)\right)^2 + \left(y' + \epsilon \sin(\alpha) \right)^2} \\
   &\approx & r' + \epsilon \cos(\theta' - \alpha), 
\end{eqnarray}
and
\begin{eqnarray}
\theta &=& \atan\left(\frac{y}{x}\right) \\
       &=& \atan \left(\frac{y' + \epsilon \sin(\alpha)}{x' + \epsilon \cos(\alpha)}\right) \\
       &\approx & \theta' - \frac{\epsilon}{r'}\sin(\theta' - \alpha)
\end{eqnarray}
We note here that for these expansions to be valid we need to have $\epsilon \ll x'$. This will not be satisfied near the centre.

We now considered a single spherical harmonic in two dimensions, described by a single azimuthal number $M$, spiralling around the real centre C of the system. This wave is given by:
\be
\phi(r,\theta) = f(r)e^{iM\theta},
\ee
where $f(r)$ is an arbitrary radial function.

Performing an AFT in the centred polar coordinates gives naturally:
\be
\tilde{\phi}(r,m) = \int_0^{2\pi} \phi(r,\theta) e^{-im\theta} d\theta = f(r)\delta (M-m).
\ee

Suppose now that one chooses a centre of coordinates that is not the true centre of the system. 
Therefore, the wave $\phi$ in this coordinate system is:
\be \label{mode_nc}
\phi(r',\theta') = f(r)e^{iM\theta} = f\left(r' + \epsilon \cos(\theta' - \alpha) \right) e^{iM\theta'} \exp\left\{ -iM\epsilon \frac{\sin(\theta' - \alpha)}{r'}\right\}
\ee
Expanding again the second exponential to first order in $\epsilon$ as well as the radial function $f$ we have:
\begin{eqnarray}
\phi(r',\theta') &\approx& \left[ f(r') + \epsilon \cos(\theta' - \alpha) f'(r') \right]
e^{iM\theta'} \nonumber\\
& & \times
\left[ 1 - iM\frac{\epsilon}{r'}\sin(\theta' - \alpha) \right] \\
&=& f(r')e^{iM\theta'} \\
& & + \frac{\epsilon}{2}e^{-i\alpha}\left[ f'(r') - \frac{f(r')}{r'}M \right] e^{i(M+1)\theta'} \nonumber \\
& & + \frac{\epsilon}{2}e^{i\alpha}\left[ f'(r') + \frac{f(r')}{r'}M \right] e^{i(M-1)\theta'}. \nonumber
\end{eqnarray}

Performing the AFT in this uncentred coordinates system therefore gives:
\begin{eqnarray}
\tilde{\phi}(r',m') &=& \int_0^{2\pi} \phi(r',\theta') e^{-im'\theta'} d\theta' \\
                    &=& f(r')\delta(M-m') \\
& & + \frac{\epsilon}{2}e^{-i\alpha}\left[ f'(r') - \frac{f(r')}{r'}M \right] \delta(M+1 - m') \nonumber \\
& & + \frac{\epsilon}{2}e^{i\alpha}\left[ f'(r') + \frac{f(r')}{r'}M \right] \delta(M-1-m'). \nonumber
\end{eqnarray}

As we can see, the spectrum is not composed by only one azimuthal component but by three. 
The dominant one has still the original shape, namely $f$, but this signal has leaked into the neighbouring azimuthal modes. 
This is expected as picking a wrong centre will break the angular symmetry and we therefore expect the different azimuthal modes to mix. 
We find the amplitude for the three modes, respectively $m = M-1$, $M$, and $M+1$ to be:
\begin{eqnarray}
\phi_1(r',M-1) &=& \frac{\epsilon}{2}e^{i\alpha}\left[ f'(r') + \frac{f(r')}{r'}M \right],\\
\phi_1(r',M) &=& f(r'),\\
\phi_1(r',M+1) &=& \frac{\epsilon}{2}e^{-i\alpha}\left[ f'(r') - \frac{f(r')}{r'}M \right].
\end{eqnarray}
where the subscript 1 stands for the fact that this is an expression to first order in $\epsilon$.

These expressions are very useful as one can test its choice of centre. By placing a unique and known azimuthal mode in a system and performing an AFT, one can check if its choice of the centre is correct by observing a possible leakage. Furthermore, by matching the leaked signal at $M\pm1$ using the signal for the original azimuthal $M$ mode, one can deduce the parameters $\epsilon$ and $\alpha$ in order to find the true centre.
Let $r'_0$ denotes a radius such that $\phi_1(r'_0,M) = f(r'_0) = 0$. Then the parameter $\alpha$ can be extracted by evaluating the following combination:
\be
\alpha = \frac{1}{2i}\ln\left(\frac{\phi_1(r'_0,M-1)}{\phi_1(r'_0,M+1)}\right).
\ee
Once the angle $\alpha$ has been determined, one can obtained the distance to the true centre $\epsilon$ by evaluating:
\be
\epsilon = \frac{2r'_1}{M}\frac{\phi_1(r'_1,M-1)}{\phi_1(r'_1,M)}e^{-i\alpha},
\ee
where $r'_1$ satisfies $f'(r'_1) = 0$.

\section{Numerical verification}

\begin{figure}
\begin{center}
\includegraphics[trim=3cm 0cm 3cm 0cm]{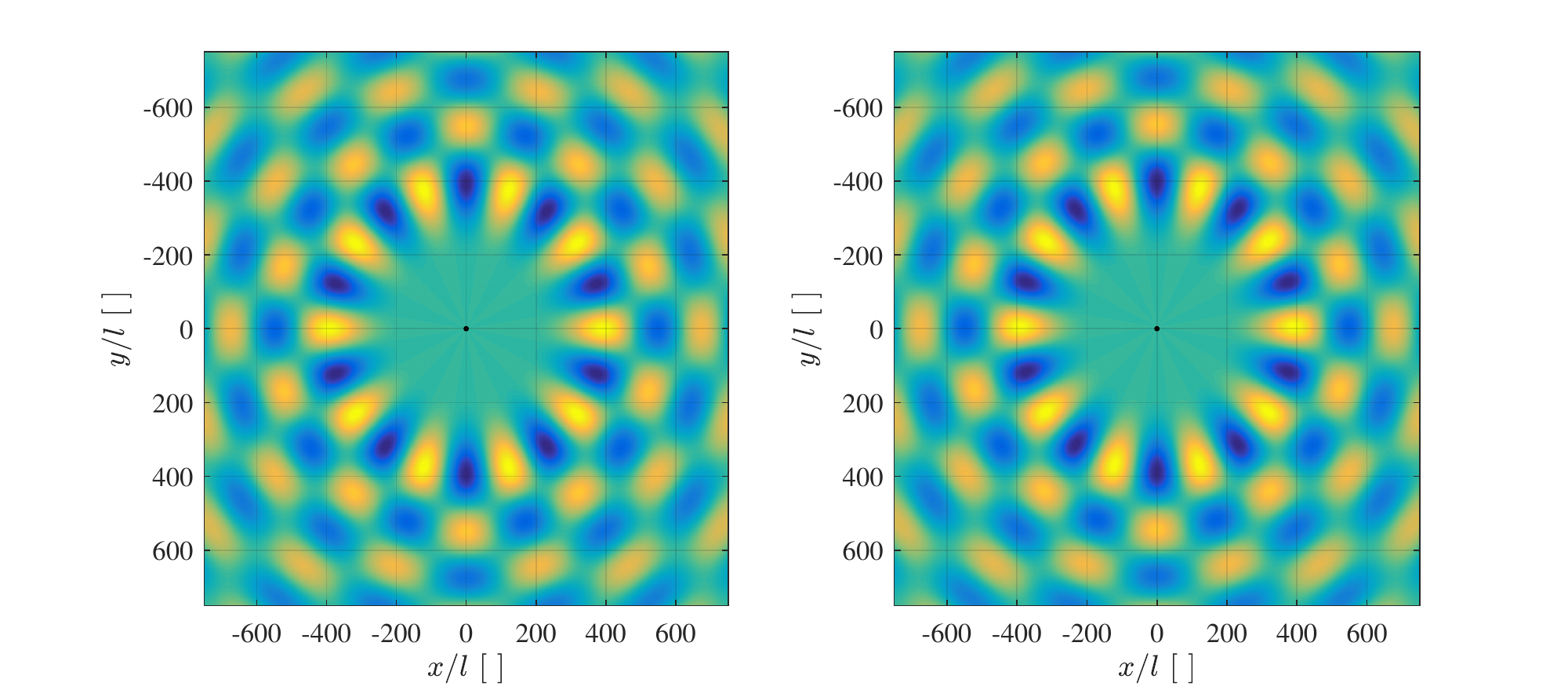}
\end{center}
\caption{Centered wave $\phi$ for $M=10$ and $k=0.03~\mathrm{l}^{-1}$ on the left and uncentered on the right. $l$ is the pixel size and is used to adimensionalise the axis. Here $\epsilon = 5l$.}\label{waves}
\end{figure}

We present here numerical simulation to test the validity of the expression derived above.
We consider an azimuthal wave of the form:
\be
\phi(r,\theta) = J_M(kr)e^{iM\theta}.
\ee
centred at ($0$,$0$) (see Fig.~\ref{waves}) and where $J_M$ is a Bessel function. 

\begin{figure}[!h]
\begin{center}
\includegraphics[trim=3cm 0 3cm 0]{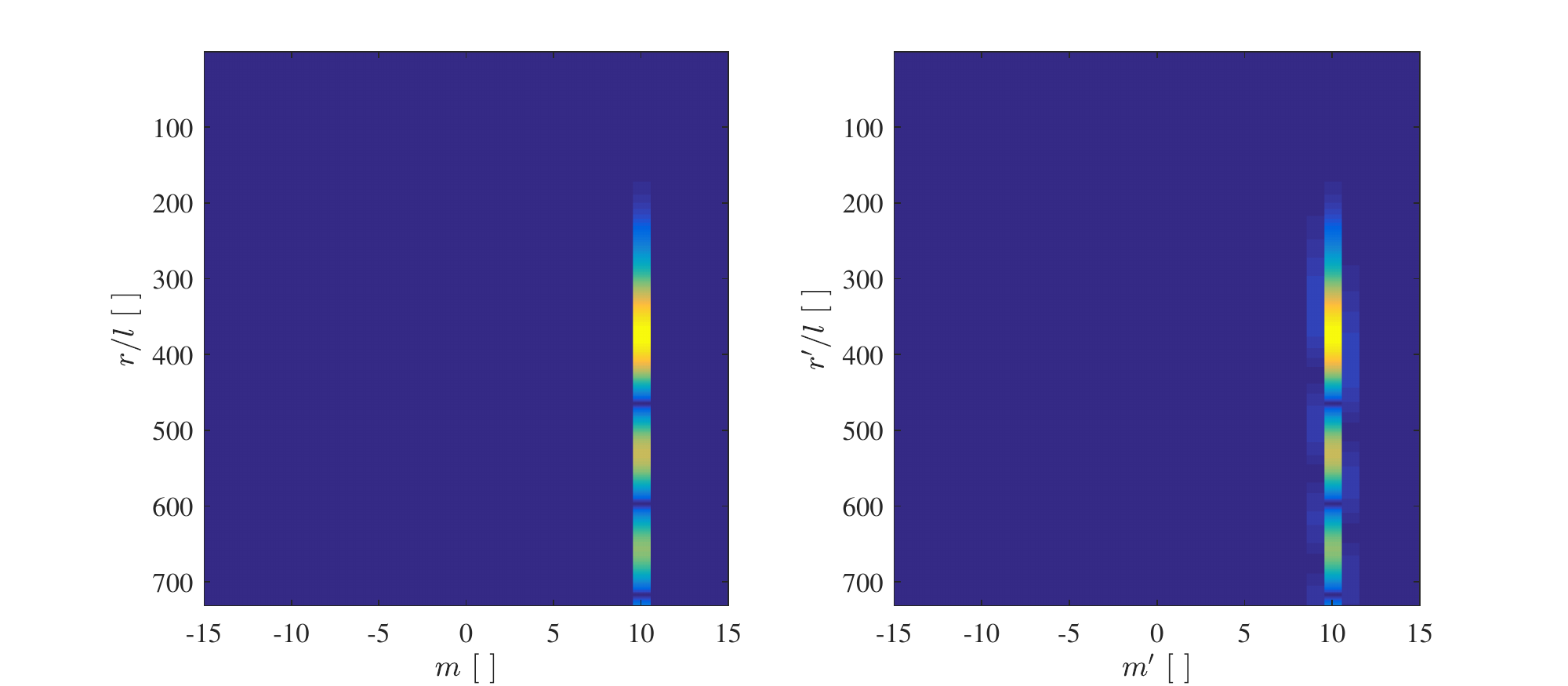}
\end{center}
\caption{Angular Fourier transform in the centred (uncentred) on the left (right) polar coordinate.}\label{spectrum}
\end{figure}

Fig.~\ref{spectrum} present the angular Fourier transform for the two waves, i.e. $\tilde{\phi}(r,m)$ and $\tilde{\phi}(r',m')$.

\begin{figure}[!h]
\begin{center}
\includegraphics[trim=0 0 0 0]{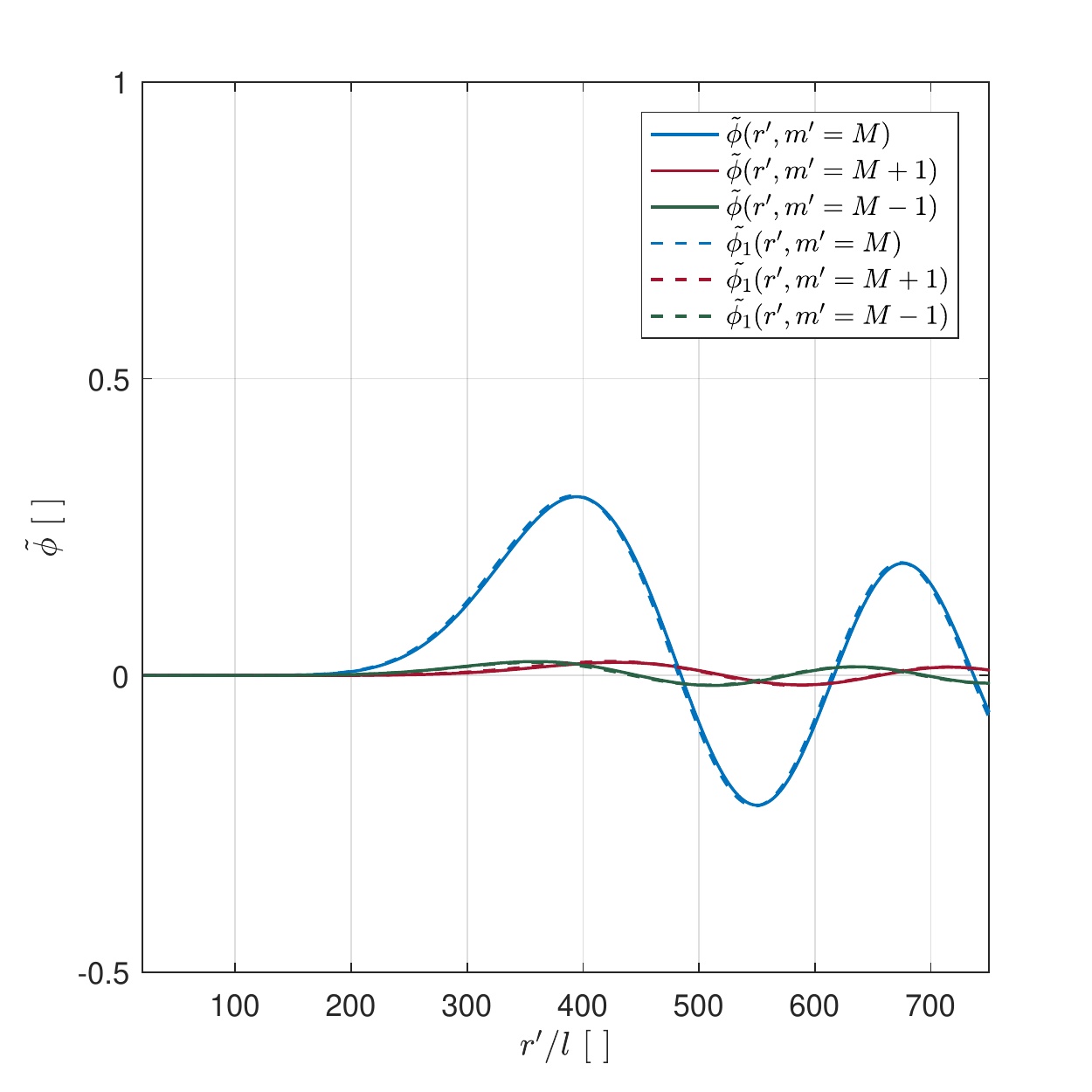}
\end{center}
\caption{Different azimuthal bands and theoretical prediction to first order in $\epsilon$}\label{match_order1}
\end{figure}

Below we plot the specific profile and we compare them with the first order estimate derived earlier. In Fig.~\ref{match_order1} we can see that the predictions reproduce the behaviour observed in the numerical simulation but don't match perfectly. We also note that to first order in $\epsilon$ the mode $m=M$ in unaffected. This can be corrected by going to next in $\epsilon$. It can be shown that the signal detected in the $M\pm n$ band is proportional to $\epsilon^n$. This means that the amplitude of the leakage decreases as $m$ gets away from the main band at $m=M$.

\chapter{Tunneling of WKB modes through a saddle point}\label{Appendix2}
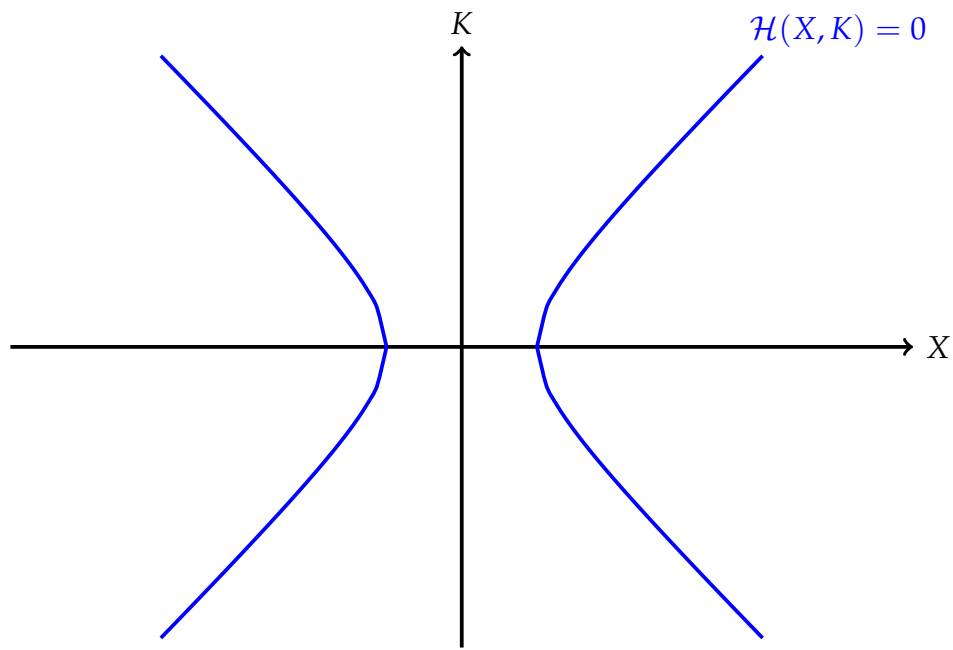
\begin{figure}[!h]
\centering
\begin{tikzpicture}
      \draw[line width=0.5mm,->] (-6,0) -- (6,0) node[right] {$X$};
      \draw[line width=0.5mm,->] (0,-4) -- (0,4) node[above] {$K$};
      \draw[line width=0.5mm,scale=1,domain=-4:-1,smooth,variable=\x,blue] plot ({\x},{sqrt{abs{(\x*\x- 1}}});
      \draw[line width=0.5mm,scale=1,domain=-4:-1,smooth,variable=\x,blue] plot ({\x},{-sqrt{abs{(\x*\x- 1}}});
      \draw[line width=0.5mm,scale=1,domain=1:4,smooth,variable=\x,blue] plot ({\x},{sqrt{abs{(\x*\x- 1}}});
      \draw[line width=0.5mm,scale=1,domain=1:4,smooth,variable=\x,blue] plot ({\x},{-sqrt{abs{(\x*\x- 1}}});
      
      \draw[blue] (5,4.2) node {$\mathcal{H}(X,K) = 0$};

\end{tikzpicture}
\caption{Sketch of the phase-space diagram. The blue curves represent the values of $X$ and $K$ satisfying the condition $\mathcal{H}(X,K)=0$ and correspond to the rays defined by the Hamiltonian $\mathcal{H}$. The saddle point is located at $(X,K)=(0,0)$}\label{phase_space_diag1}
    \end{figure}

Here we assume that the Hamiltonian of the system $\mathcal{H}(X,K)$ has a saddle point. We further assume that the saddle point is located at $(X,K) = (0,0)$. In phase-space this will appear as shown in Fig. \ref{phase_space_diag1}.
 
This means that 
\begin{eqnarray}
\p_X \mathcal{H} (0,0) &=& 0, \\
\p_K \mathcal{H} (0,0) &=& 0. 
\end{eqnarray}
In the vicinity of the saddle point, the Hamiltonian can be expanded as:
\begin{equation}\label{Ham}
 H(X,K) = \eta^2  - \frac{1}{2}(X^2 - K^2).
\end{equation}
The relative $-$ sign comes from the saddle structure, i.e. the Hessian matrix has a negative determinant.
This Hamiltonian can be `lifted' at the level of a local wave equation:
\begin{equation}\label{local_we}
\left[ \eta^2  - \frac{1}{2}\left(X^2 + \frac{d^2}{dx^2}\right) \right] \phi = 0.
\end{equation}
Note here that one needs to be careful when lifting a Hamiltonian to the level of a wave equation, as $x$ and $\p_x$ do not commute. This is rigorously done by means of the Weyl symbols~\cite{tracy2014ray}. However, the ordering is of no importance when working at the eikonal level.

In order to match WKB modes on both side of the saddle point, we need to find an exact solution of Eq.~\eqref{local_we} valid globally. We then need to match the asymptotic expansions of this global solution to approximate WKB solutions sufficiently far away from the saddle point.

\section{Exact solutions}

We present here the global solution of Eq.~\eqref{local_we} and the properties needed in order to perform the matching.

Eq.~\eqref{local_we} can be solved exactly by means of parabolic cylinder function, $U$,~\cite{Handbook}. Indeed, it admits as a general solution:
\begin{equation}\label{global_sol}
\phi(x) = A U(i \eta^2, \sqrt{2} X e^{-i\pi/4})  + B U( -i \eta^2, \sqrt{2} X e^{i\pi/4}).
\end{equation}
The function $U$ has the following asymptotic behaviour for $z\rightarrow \infty$:
\begin{equation}\label{asym1}
U(a,z) \approx e^{-z^2/4} z^{-a-1/2}   \quad \text{ for } \quad|arg(z)|<3\pi/4
\end{equation} 
\begin{eqnarray}\label{asym2}
U(a,z) \approx & & e^{-z^2/4} z^{-a-1/2} \pm i\frac{\sqrt{2\pi}}{\Gamma(1/2 + a)} e^{\mp i \pi a} e^{z^2/4} z^{a-1/2}  \\ 
&\text{ for }& \quad \pi/4 < |arg(z)| <5\pi/4 \nonumber
\end{eqnarray}

We define the following functions to simplify the notation:
\begin{eqnarray}
U_1 &=& U\left(-i\eta^2, \sqrt{2}|X|e^{i\pi/4} \right), \\
U_2 &=& U\left(i\eta^2, \sqrt{2}|X|e^{-i\pi/4} \right), \\
U_3 &=& U\left(i\eta^2, -\sqrt{2}|X|e^{-i\pi/4} \right), \\
U_4 &=& U\left(-i\eta^2, -\sqrt{2}|X|e^{i\pi/4} \right),
\end{eqnarray}

\begin{figure}
\centering
\begin{tikzpicture}
      \draw[line width=0.5mm,->] (-6,0) -- (6,0) node[right] {$X$};
      \draw[line width=0.5mm,->] (0,-4) -- (0,4) node[above] {$K$};
      \draw[line width=0.5mm,scale=1,domain=-4:-1,smooth,variable=\x,blue] plot ({\x},{sqrt{abs{(\x*\x- 1}}});
      \draw[line width=0.5mm,scale=1,domain=-4:-1,smooth,variable=\x,blue] plot ({\x},{-sqrt{abs{(\x*\x- 1}}});
      \draw[line width=0.5mm,scale=1,domain=1:4,smooth,variable=\x,blue] plot ({\x},{sqrt{abs{(\x*\x- 1}}});
      \draw[line width=0.5mm,scale=1,domain=1:4,smooth,variable=\x,blue] plot ({\x},{-sqrt{abs{(\x*\x- 1}}});
      
      \draw[blue] (5,4.2) node {$\mathcal{H}(X,K) = 0$};
      \draw[black] (0,-4.5) node {\Large $ AU_3 + BU_4 \leftarrow \phi \rightarrow AU_2 + BU_1$};
\end{tikzpicture}
\caption{Phase-space diagram and asymptotic expansions of the solution to the local wave equation (Eq.\eqref{local_we}).}\label{phase_space_diag2}

    \end{figure}
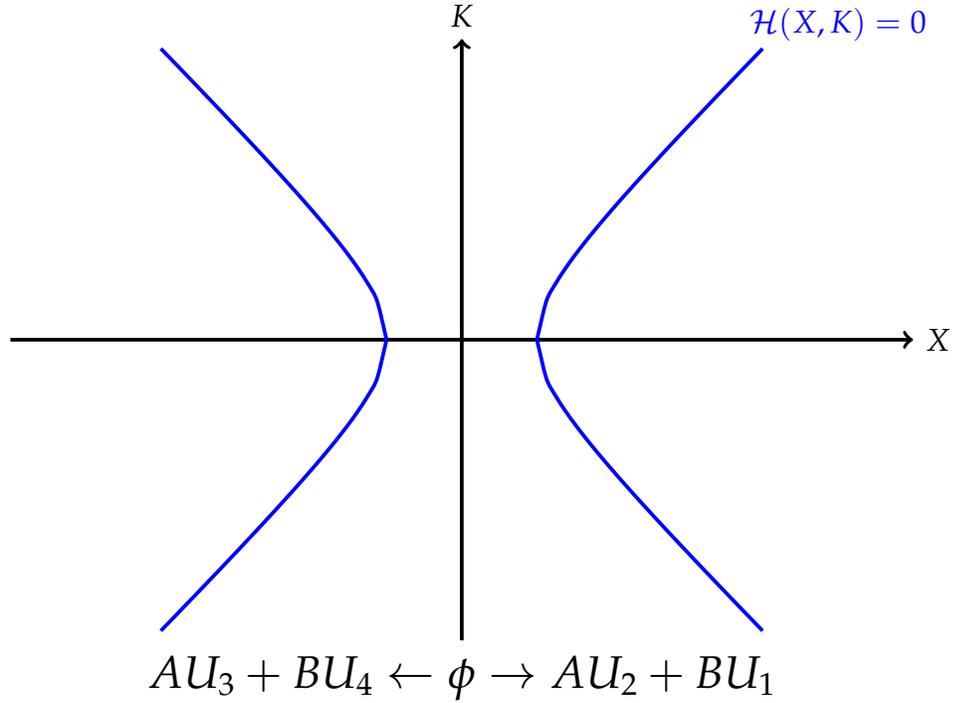

We can rewrite the solution given in Eq.~\eqref{global_sol} as:
\begin{equation} 
\phi = 
     \begin{cases}
       AU_2 + BU_1 &\quad\text{if } X>0\\
       AU_3 + BU_4 &\quad\text{if } X<0 \\
     \end{cases}
\end{equation}
The asymptotic expansion are added to the phase-space diagram in Fig. \ref{phase_space_diag2}.
We now explicitly write down the asymptotic expansion of the functions $U_1$, $U_2$, $U_3$, and $U_4$.
The asymptotic expansion of $U_1$ and $U_2$ are obtained by setting $a = \pm i\eta^2$ and $z = \sqrt{2} |X| e^{\mp i \pi/4}$, and using the expansion given in Eq.~\eqref{asym1} as $|arg(z)| = \pi/4$. We obtain:
\begin{equation}\label{global_exp1}
U_1 \approx \left( \sqrt{2} |X| \right)^{i\eta^2 - 1/2}\exp{\left[-iX^2/2\right]} \exp{\left[-\frac{\pi}{4}(\eta^2 + i/2)\right]},
\end{equation}
and
\begin{equation}\label{global_exp2}
U_2 \approx \left( \sqrt{2} |X| \right)^{-i\eta^2 - 1/2}\exp{\left[iX^2/2\right]} \exp{\left[-\frac{\pi}{4}(\eta^2 - i/2)\right]}.
\end{equation}

The expansion for $U_3$ and $U_4$ is obtained in the same manner with $a = \pm i\eta^2$ and $z = -\sqrt{2} |X| e^{\mp i\pi /4}$. Since $|arg(z)| = 3\pi / 4$, the asymptotic expansion given in Eq.~\eqref{asym2} should be used. After some algebra, we obtain:
\begin{equation}\label{global_exp3}
U_3 \approx -i \tau^{-1} U_2 + \beta \tau^{-1} U_1,
\end{equation}
and
\begin{equation}\label{global_exp4}
U_4 \approx i \tau^{-1} U_1 + \beta^* \tau^{-1} U_2,
\end{equation}
where we have defined the following parameters:
\begin{equation}\label{tau_beta_def}
\tau = e^{-\pi \eta^2} \quad \text{and} \quad 
\beta = \frac{\sqrt{2\pi i \tau}}{\Gamma\left(i \eta^2 + 1/2 \right)}.
\end{equation}
Note that we have the following relation: $|\beta|^2 - \tau^2 =1$.
    
\section{WKB modes around the saddle point}
Here, we construct the WKB modes in a region where the quadratic approximation for the Hamiltonian is valid.
Since the WKB approximation breaks down at the saddle point, this construction must be done on both sides of the saddle point separately. 

We first start by computing the eikonal phase using the local Hamiltonian given in Eq.~\eqref{Ham}. We have:
\begin{equation}
K = \p_X S = \pm \sqrt{X^2 - 2\eta^2}.
\end{equation}
After integration, we obtain the phase $S$:
\begin{equation}
S(X) = S_0 \pm \frac{1}{2} \left( X\sqrt{X^2 - 2\eta^2 } - 2\eta^2 \ln\left(\sqrt{X^2 - 2\eta^2} + X\right) \right),
\end{equation}
where $S_0$ is a constant of integration.
For $|X| \gg \eta$, this approximates to:
\begin{equation}\label{phase_WKB}
S(X) \approx \pm \left( \frac{X^2}{2} - \eta^2 \ln X \right) + S_0.
\end{equation}

We, then, compute the amplitude of the WKB mode. To do so, we first express the group velocity as:
\begin{equation}
v_g = \frac{dX}{d\sigma} = -\p_K H = K = \pm \sqrt{X^2 - 2\eta^2}.
\end{equation}
Inserting this result into the transport equation (Eq.~\eqref{transport_eq}), we get that:
\begin{equation}
A_{\mathrm{WKB}} \propto \frac{1}{\left( X^2 - 2\eta^2\right)^{1/4}}.
\end{equation}
For $|X| \gg \eta$, this reduces to:
\begin{equation}\label{amp_WKB}
A_{\mathrm{WKB}} \approx |X|^{-1/2}.
\end{equation}
Combining Eqs.~\eqref{amp_WKB} and~\eqref{phase_WKB}, we obtain the WKB modes:
\begin{equation} \label{WKB_exp}
\phi_{\mathrm R,L}^{\rightleftharpoons} \propto  \big(|X| \big)^{\mp i \eta^2 - 1/2} \exp{\left[\pm i X^2/2 \right]}.
\end{equation}
The subscripts R and L denote that the solution is valid either to the right or to the left of the saddle point but not across. The superscript $\rightleftharpoons$ denotes the direction of propagation of the wave. To determine which direction corresponds to which sign, we look at the group velocity. 
If $v_g > 0$ the wave propagates towards the right, while if $v_g < 0 $ the wave propagates towards the left.
Therefore, we have explicitly the four WKB solutions. To the right of the saddle point, they are given by:
\begin{eqnarray}
\phi_{\mathrm R}^{\leftarrow} &\propto & \big( X \big)^{-i\eta^2 - 1/2} \exp{\left[ \frac{i X^2}{2} \right]} \label{WKB_R1}\\
\text{ and } \quad
\phi_{\mathrm R}^{\rightarrow} &\propto & \big( X \big)^{i\eta^2 - 1/2} \exp{\left[ \frac{-i X^2}{2} \right]}, \label{WKB_R2}
\end{eqnarray}
while to the left of the saddle point, they are given by:
\begin{eqnarray}
\phi_{\mathrm L}^{\rightarrow} &\propto & \big( |X| \big)^{-i\eta^2 - 1/2} \exp{\left[ \frac{i X^2}{2} \right]}
\quad \label{WKB_L1}\\
\text{ and } \quad
\phi_{\mathrm L}^{\leftarrow} &\propto & \big( |X| \big)^{i\eta^2 - 1/2} \exp{\left[ \frac{-i X^2}{2} \right]}.\label{WKB_L2}
\end{eqnarray}

\begin{figure}
\begin{tikzpicture}
      \draw[line width=0.5mm,->] (-6,0) -- (6,0) node[right] {$X$};
      \draw[line width=0.5mm,->] (0,-4) -- (0,4) node[above] {$K$};
      \draw[line width=0.5mm,scale=1,domain=-4:-1,smooth,variable=\x,blue] plot ({\x},{sqrt{abs{(\x*\x- 1}}});
      \draw[line width=0.5mm,scale=1,domain=-4:-1,smooth,variable=\x,blue] plot ({\x},{-sqrt{abs{(\x*\x- 1}}});
      \draw[line width=0.5mm,scale=1,domain=1:4,smooth,variable=\x,blue] plot ({\x},{sqrt{abs{(\x*\x- 1}}});
      \draw[line width=0.5mm,scale=1,domain=1:4,smooth,variable=\x,blue] plot ({\x},{-sqrt{abs{(\x*\x- 1}}});
      \draw[blue] (5,4.2) node {$\mathcal{H}(X,K) = 0$};
      \draw[black] (0,-4.5) node {\Large $ AU_3 + BU_4 \leftarrow \phi \rightarrow AU_2 + BU_1$};
      \draw[line width=0.5mm] (-2,1.73-0.2) -- (-2,1.73) -- (-2+0.2, 1.73);
      \draw[black] (-3, 1.73) node {\Large $\phi_{\mathrm L}^{\leftarrow}$};
      
      \draw[line width=0.5mm] (-2-0.2,-1.73) -- (-2,-1.73) -- (-2, -1.73 -0.2);
      \draw[black] (-3, -1.73) node {\Large $\phi_{\mathrm L}^{\rightarrow}$};
      
      \draw[line width=0.5mm] (2 -0.2,-1.73) -- (2,-1.73) -- (2, -1.73 + 0.2);
      \draw[black] (3, -1.73) node {\Large $\phi_{\mathrm R}^{\rightarrow}$};
      
      \draw[line width=0.5mm] (2,1.73+0.2) -- (2,1.73) -- (2+0.2, 1.73);
      \draw[black] (3, 1.73) node {\Large $\phi_{\mathrm R}^{\leftarrow}$};
      \draw[black] (2.7,-5.5) node {\Large $\rightarrow a_{\mathrm R}^{\leftarrow}\phi_{\mathrm R}^{\leftarrow} +
a_{\mathrm R}^{\rightarrow}\phi_{\mathrm R}^{\rightarrow} $};
      \draw[black] (-2.7,-5.5) node {\Large $ a_{\mathrm L}^{\leftarrow}\phi_{\mathrm L}^{\leftarrow} +
a_{\mathrm L}^{\rightarrow}\phi_{\mathrm L}^{\rightarrow} \leftarrow $};

\end{tikzpicture}
\caption{Phase-space diagram and WKB modes on both sides of the saddle point. The arrow on the blue curves indicate the direction of propagation of the WKB modes.}\label{phase_space_diag3}
    \end{figure}
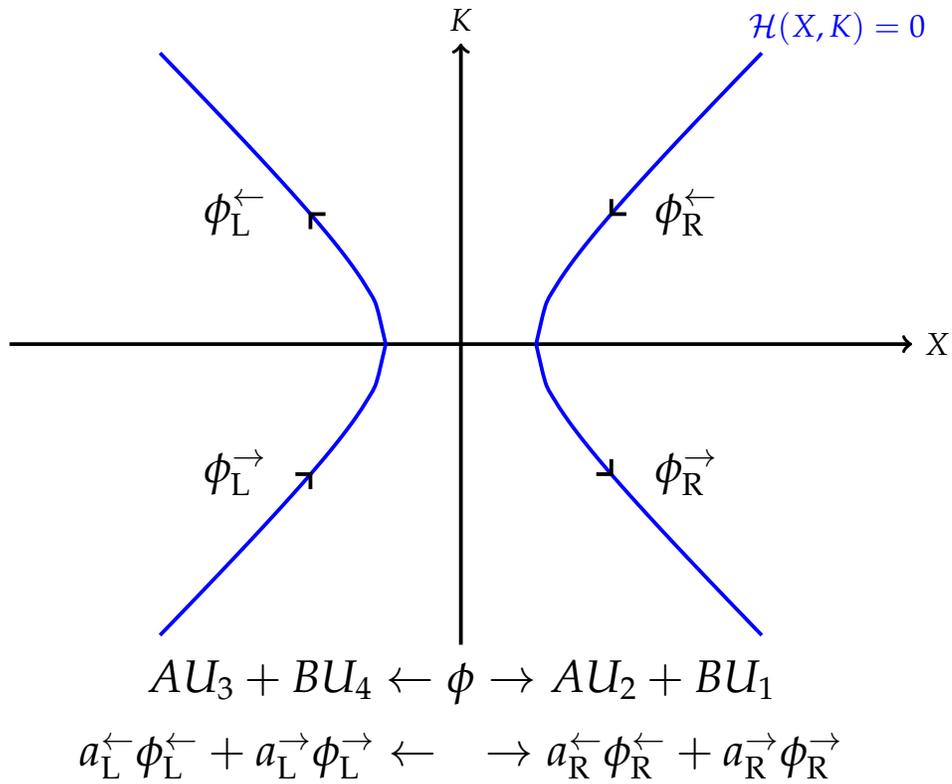

The WKB modes are added to the phase-space diagram in Fig. \ref{phase_space_diag3}.

    \section{Matching conditions}
    
    We have solved, both exactly and by means of a WKB expansion, the local wave equation (Eq.~\eqref{local_we}). The two solutions should match sufficiently far away from the saddle point as can be seen from the asymptotic expansions of both solutions. We now work out the matching conditions and relate the WKB modes on both sides of the saddle point.
    
    First we need to identify the correspondence between the asymptotic expansion of the exact solutions $U_j$ and the WKB modes $\phi_{\mathrm R,L}^{\rightleftharpoons}$.
    
    For $X\gg \eta$, using the asymptotic expressions given in Eqs.~\eqref{global_exp1}, \eqref{global_exp2}, as well as Eqs.~\eqref{WKB_R1}, and \eqref{WKB_R2} ,we have that:
\begin{eqnarray}
U_1 &=& \alpha \gamma \phi_{\mathrm R}^{\rightarrow} 
\quad \label{match_R1}\\
 \text{and} \quad 
U_2 &=& \alpha^{-1} \gamma \phi_{\mathrm R}^{\leftarrow}.\label{match_R2}
\end{eqnarray}
where the parameters $\alpha$ and $\gamma$ are defined as:
\begin{equation} \label{alpha_def}
\alpha = \sqrt{2}^{i\eta^2} e^{-i\pi /8} 
\quad \text{and} \quad
\gamma = 2^{-1/4} e^{-\pi \eta^2 /4}.
\end{equation}

Now using Eqs.~\eqref{global_exp3} and~\eqref{global_exp4}, in addition to Eqs.~\eqref{WKB_L1} and \eqref{WKB_L2}, we obtain:
\begin{eqnarray}
U_3 &=& -i \tau^{-1} \alpha^{-1} \gamma \phi_{\mathrm L}^{\rightarrow} + \beta\tau^{-1} \alpha \gamma \phi_{\mathrm L}^{\leftarrow}
\quad \label{match_L1}\\
\text{and} \quad
U_4 &=& i\tau^{-1}\alpha\gamma\phi_{\mathrm L}^{\leftarrow} + \beta^*\tau^{-1} \alpha^{-1} \gamma \phi_{\mathrm L}^{\rightarrow}. \label{match_L2}
\end{eqnarray}

Eqs.~\eqref{match_R1} and \eqref{match_R2} give the matching conditions to the right of the saddle point. It allows us to relate the amplitude of the right WKB modes with the amplitude of the exact solution. This leads to:
\begin{equation}
A = \alpha \gamma^{-1} a_{\mathrm R}^{\leftarrow} 
\quad \text{and} \quad
B = \alpha^{-1} \gamma^{-1} a_{\mathrm R}^{\rightarrow}.
\end{equation}

The same procedure using Eqs.~\eqref{match_L1} and~\eqref{match_L2} allows us to connect $a_{\mathrm L}^{\rightleftharpoons}$ with $A$ and $B$:
\begin{eqnarray}
a_{\mathrm L}^{\rightarrow} &=& A \beta \tau^{-1} \alpha \gamma + B i \tau^{-1} \alpha \gamma
\\
a_{\mathrm L}^{\leftarrow} &=& -A i \tau^{-1} \alpha^{-1} \gamma + B \beta^* \tau^{-1} \alpha^{-1} \gamma.
\end{eqnarray}

Combining the previous relations, we finally obtain the connection formula between the right and left WKB mode amplitudes:
\begin{equation}\label{match_SP}
\binom{a_{\mathrm L}^{\leftarrow}}{a_{\mathrm L}^{\rightarrow}} = 
S  
\binom{a_{\mathrm R}^{\leftarrow}}{a_{\mathrm R}^{\rightarrow}}
\end{equation}
where S is the matrix given by:
\begin{equation}
S=
  \begin{bmatrix}
      \beta \tau^{-1} \alpha^{2} & i\tau^{-1} \\
    -i \tau^{-1} & \beta^*\tau^{-1}\alpha^{-2} \\
  \end{bmatrix}.
\end{equation}

\chapter{Hydrodynamic simulations of cosmological scenarios}\label{Appendix3}

Throughout the main chapter of this thesis, we have focused our attention on hydrodynamic simulations of rotating BHs. 
In this appendix, we will discuss a hydrodynamic system capable of mimicking cosmological scenarios. 

Our current understanding of the Universe on very large scales is based on the $\Lambda$-CDM (lambda - Cold Dark Matter) model~\cite{dodelson2003modern}.
This model is based on the cosmological principles which assumes that the Universe is isotropic and homogeneous.
The most general metric satisfying these conditions is known as the Friedmann-Lema\^{i}tre-Robertson-Walker (FLRW) metric:
\be
ds^2 = -dt^2 + a(t)^2 d\Sigma^2,
\ee
where $d\Sigma$ is the metric on the 3-dimensional space with uniform curvature.
The function $a(t)$ is called the scale factor. It describes the expansion or contraction of the Universe. 
Its form is given by the Einstein equations.
To complete the model, an inflationary epoch is added to these assumptions. 
During this brief period, which lasted from about $10^{-36}$ seconds to around $10^{-33}$ seconds after the Big Bang singularity, the Universe is thought to have expanded in a exponential fashion~\cite{Guth}.
This violent expansion was initially designed to solve important issues coming from cosmological observations such as the horizon and the flatness problem. 
In addition of bringing solutions to these problems, inflation provided a mechanism to explain the large-scale structures in the Universe, which would be the result of the gravitational collapse of quantum mechanical fluctuations. 
Inflation predicted that these fluctuations have a nearly scale invariant power spectrum, a feature that was observed through the cosmic microwave background.
Despite its success, inflation is not fully accepted and suffers from various criticisms. The most important is a fine-tuning problem for inflation to occur. 
This led to the development of alternative models such as cyclic or ekpyrotic scenarios~\cite{CyclicSc,Cyclic,Khoury,Lehners,Gasperini,Veneziano}, or even more drastic changes in the space-time geometry, involving transitions for a Lorentzian to an Euclidean signature of the space-time metric~\cite{HawkMoss,Vilenkin1,Vilenkin2,LindeQC,HartHawk,HartHawkHert}. 
Since these scenarios could have only been realised at the very earliest periods in the evolution of our Universe -- corresponding to energy scales unattainable in a laboratory experiment -- we do not have direct experimental access to their fascinating physics.
However, we have seen that analogue systems may offer a way to experimentally investigate these elusive processes that might have occurred during the early Universe. We have already pointed out that superfluids have been suggested as a promising analogue cosmology simulator~\cite{Fedichev03,Barcelo:2003et,Weinfurtner:2004mu,PhysRevA.70.063615,PhysRevA.76.033616,cha2017probing}. Here, we propose to add to this list an analogue simulator based on surface waves in a strong gradient magnetic field.

\section{Governing equations}
\begin{figure}[!h] 
\centering
  \includegraphics[scale=1]{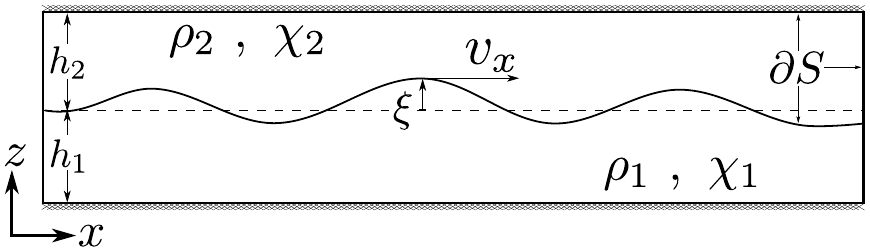}  
  \caption{Schematics of the two-liquid system. Two immiscible liquids, with densities $\rho_{1,2}$, magnetic susceptibility $\chi_{1,2}$, and height $h_{1,2}$, separated by the equilibrium interface at $z_0$ (dotted line). Gravity interface waves distort the interface layer (solid line) where each point is characterized by its amplitude $\xi$ and velocity $v_x$ ($v_y$). $\partial S$ are the boundaries (here depicted for fluid 2) given by a hard wall (upper), an interface boundary (right), and a moving boundary at the interface of the two liquids (lower).     } \label{system_cosmo}
\end{figure}

The system we will consider here is composed two immiscible fluids with densities $\rho_1 \!>\! \rho_2$, heights $h_{1,2}$, small magnetic susceptibilities $|\chi_{1,2}| \ll 1$, and flow velocities $v_{1,2}$ (see Fig.~\ref{system_cosmo}).
The system is subjected to a magnetic field $\vec{B}(\vec{x},t)$ with a large vertical gradient. The flow of an inviscid and incompressible fluid is described by the continuity equation and Euler's momentum equation, presented in Chapter~\ref{Intro_sec}, with the inclusion of the magnetic potential energy \cite{berry_frog, geim}:
\begin{align} 
\vec{\nabla} \cdot \vec{v}_i &= 0  \label{continuity} \\
\rho_i \left( \partial_{t} + \vec{v}_i \cdot\vec{\nabla} \right) \vec{v}_i &= \vec{\nabla} \left( -p_i + \frac{\chi_i}{2\mu_0}B^2 \right) + \rho_i \vec{g} \label{euler} ~.
\end{align}
 The index $i=1$ ($i=2$) labels the lower (upper) fluid, $p_i$ is the fluid pressure, $\mu_0$ the vacuum  permeability, and $\vec{g} = (0,0,-g)$ is the acceleration due to gravity. The kinematic and, in case of a free surface, dynamic boundary conditions are 
\begin{align}
\vec{v}_i \cdot \vec{n} &= \vec{V} \cdot \vec{n} \qquad \text{on} \qquad \partial S \label{BC_kinematic} \\
 [p] &= \sigma (R_1^{-1} + R_2^{-1}) \label{BC_dynamic} ~,
\end{align}
on a boundary $\partial S$ with velocity $\vec{V}$. The angled bracket $\left[ * \right]$ denotes the jump in value across the interface, here the jump in pressure $p$ according to the Young-Laplace law \cite{LandauV6} with surface tension $\sigma$, and principal radii of curvature $R_{1,2}$. 

We assume an irrotational velocity field $\vec{v}_i = \vec{\nabla} \phi_i$, and a liquid-liquid interface $\xi(x,y,t)$. 
We choose the $z$-axis such that the unperturbed interface is located at $z=0$.
 We then linearise Eq.~\eqref{euler} around a steady background flow \mbox{ $\vec{v}_0 = \vec{\nabla} \phi_0$} with \mbox{$\phi_i = \phi_0 + \varphi_i$}.
 We note that the flows can be different in the two liquids, however they should be equal at the interface due to the boundary condition (Eq.~\eqref{BC_kinematic}). 
 In particular, at the interface $\varphi_1=\varphi_2=\varphi$ 
We further take $\partial_z B \gg \partial_x B$. Using the fact the curvature for small deformation $\xi$ is given by $(R_1^{-1} + R_2^{-1}) \simeq - \nabla^2 \xi$~\cite{LandauV6}, the linearised equation of motions together with the boundary conditions leads, at the interface, to: 
\begin{align}
 \rho_1 \mathcal{D}_{t} \varphi_1 - \rho_2 \mathcal{D}_{t} \varphi_2 &= \left( \sigma \nabla^2 - [\rho] g_0 + \frac{[\chi]}{\mu_0} B \partial_z B \right) \xi \label{EoM1}\\
 \mathcal{D}_t \xi &= \frac{1}{2} \partial_z \left( \varphi_1 + \varphi_2 \right) \label{EoM2} ~,
\end{align}
where $\mathcal{D}_t = \partial_t + \vec{v}_0 \! \cdot \! \vnab$ is the material derivative at the interface.

We now restrict ourselves to the case of liquids with equal depths $h_1=h_2=h_0$, and vanishing background flow, i.e. $\mathcal{D}_t \rightarrow \partial_t$ (time derivative will be denoted by a dot).
By solving Eq.~\eqref{continuity} together with the boundary conditions at the solid wall and the interface, we combine Eqs.~\eqref{EoM1} and \eqref{EoM2} to get:
\begin{equation} \label{wave_equation}
 \ddot{\varphi}_k + \omega_k^2 \varphi_k = \frac{\dot{G}_k}{G_k} \dot{\varphi}_k ~,
\end{equation}
where
\begin{equation} \label{Gkdef}
G_k = \left( [\rho] g_\mathrm{eff} + \sigma k^2 \right) / \tilde{\rho} \,, 
\end{equation}
with $\tilde{\rho} = \rho_1 + \rho_2$, 
\begin{equation} \label{full_dispersion}
 \omega_k^2 = G_k k \tanh(k h_0)\, ,
\end{equation}
and the effective gravity $g_\mathrm{eff}$~\cite{poodt2006using},
\begin{equation} \label{g_eff}
  g_\mathrm{eff} =  g - \frac{[\chi]}{[\rho] \mu_0} B \partial_z B .
\end{equation}
We can see that a time-dependent external magnetic field $\vec{B}(t)$,  
will introduce an explicit time dependence of the frequency $\omega_k = \omega_k(t)$, as well as an additional friction term (right hand side of Eq.~\eqref{wave_equation}).

\section{Shallow water limit}

\subsection{Analogue cosmology}
We know that the analogue regime appears in the long wavelength limit, $h_0k\ll 1$, and when surface tension effects are negligible compare to the effect of gravity, $[\rho]g_\mathrm{eff}\gg \sigma k^2$. In this limit, all waves travel at the same speed and the dispersion relation, given in Eq.~\eqref{full_dispersion}, reduces to:
\be
\omega_k = \frac{[\rho]}{\tilde{\rho}}g_\mathrm{eff} h_0 k^2.
\ee
The wave speed is therefore:
\be
c(t) = \sqrt{\frac{[\rho]}{\tilde{\rho}} g_\mathrm{eff}h_0}.
\ee
Introducing an analogue scale factor via $a(t) = c(t)^{-1}$, the wave equation becomes:
\be
\ddot{\varphi_k} + 2\frac{\dot{a}}{a} \dot{\varphi_k} + \frac{k^2}{a^2}\varphi_k = 0.
\ee
This wave equation is equivalent to the wave equation governing the propagation of a massless scalar field on a FLRW-metric, with flat spatial section. 

The evolution of our analogue universe is governed by the scale factor which can be tuned externally by changing the applied magnetic field. 
This fact, which is in contrast with GR where the scale factor is determined by the dynamical equations, gives the system a lot of tunability.
It is possible to use this system to mimic various cosmological evolution, such as inflation, cyclic evolutions or even signature change scenarios.
Indeed, by changing the sign of $g_\mathrm{eff}$, it is possible to change the sign of $c^2$ and to go from an hyperbolic to an elliptic wave equation.

\subsection{Analogue rainbow cosmology}

When modifying the effective gravity, one should be careful about the validity of the assumptions used to derive the dynamical equations. 
In particular when considering cases where $g_\mathrm{eff} \rightarrow 0$, surface tension effects will not be negligible anymore. 
In this case, dispersion will be present in the system, resulting in a $k$-dependent scale factor:
\be
a_k^{-2} = c_k^2 = G_k h_0.
\ee
The wave equation will in addition correspond to a massless scalar field on a rainbow FLRW Universe~\cite{weinfurtner2009cosmological}:
\begin{equation} \label{EoM_Rainbow}
 \ddot{\varphi}_k + 2 \frac{\dot{a}_k}{a_k} \dot{\varphi}_k + \frac{k^2}{a_k^2} \varphi_k = 0 ~.
\end{equation}

\section{Experimental proposal}

One strength of the proposed system is that it can be implemented with the current technology. 
Namely, there exists superconducting solenoid magnets capable of creating sufficiently strong magnetic fields in order to mimic cosmological evolution. 
These type of magnet have been used to study magnetic levitation~\cite{BAL15, liao2017shapes}.

\begin{figure}	
\centering
  \includegraphics[scale=1.2]{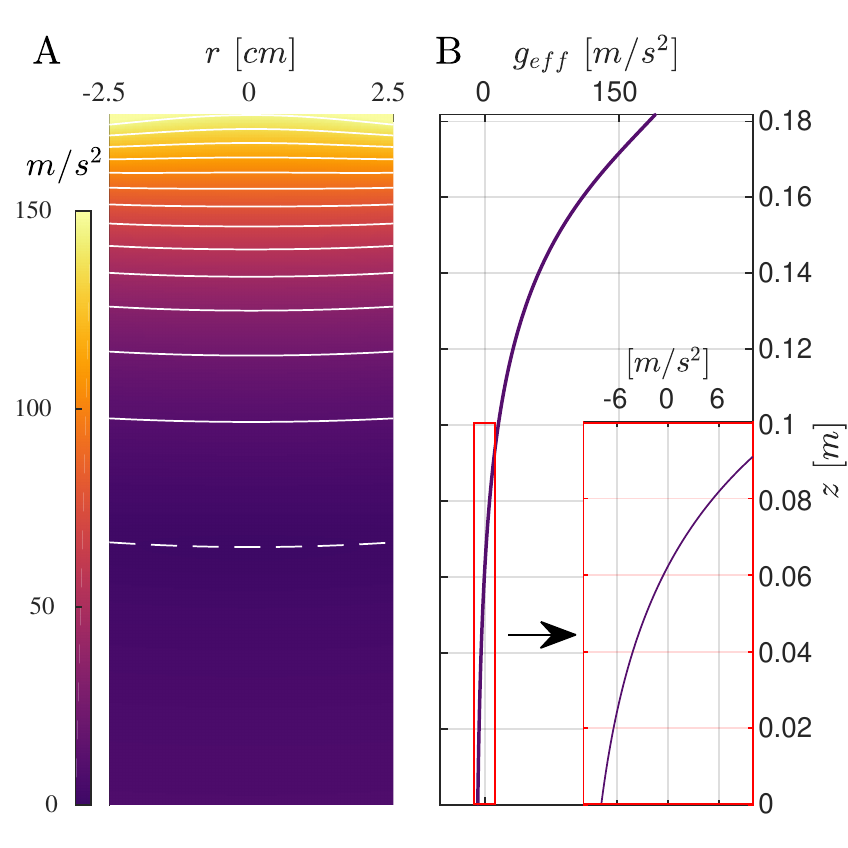}
  \caption{
Panel A depicts the effective gravity $g_{\mathrm{eff}}$ in the bore of the magnet for the butanol-aqueous solution.
The vertical axis gives the vertical position in the magnet $z$, and the horizontal axis the radial position $r$.
The magnitude is given by the colorbar. 
The solid white lines are contours of the effective gravity, with the dashed line depicting the region over which $g_\mathrm{eff}$ changes sign.
Panel B is the effective gravity along the axis of the solenoid ($r=0$). 
The inset rescales the horizontal axis to better demonstrate the sign change of $g_\mathrm{eff}$.
}
\label{g_eff_pic}
\end{figure}

We will now focus on the particular magnet used in the experiment of magnetic levitation mentioned above (see Fig.~\ref{g_eff_pic}) and consider a realistic analogue cosmology set-up.

The system consists of a small tank moving vertically in the bore of the superconducting magnet. 
In order to reduce the effect of surface tension in comparison to gravity, we choose a two-fluids system composed of a layer of butanol lying atop a denser layer of a weak aqueous paramagnetic solution. 
This choice allow for a relatively small interfacial surface tension, $\sigma=1.8 \mathrm{mN/m}$ \cite{doi:10.1021/ed060pA322.2}. Note that we cannot completely remove the effect of surface tension as the condition that the two fluids be immiscibility requires $\sigma \neq 0$.
The liquids fill completely a transparent cylindrical vessel with diameter $d=4 \mathrm{cm}$, designed to fit within the bore of the $18 \mathrm{T}$ superconducting solenoid magnet used in~\cite{BAL15, liao2017shapes}.
We focus on azimuthal waves in the vessel and in particular on the mode with longest wavelength in order to reduce dispersive effects.
We can see in panel (A) of Fig.~\ref{g_eff_pic} that one can effectively increase the effect of gravity by a factor fifteen and even change its sign (at the dashed white line). 
In particular, we can mimic various cosmological scenarios by tuning the movement of the vessel inside of the bore. 
For example, we can design the vessel trajectory such that the scale factor will be the one of a Universe which content is dominated either by radiation with $a(t) \propto t^{1/2}$, or by matter with $a(t) \propto t^{2/3}$.
While the aim of analogue simulators is to experimentally explore regimes where a clear theoretical understanding is limited, it is interesting and necessary to first explore well known regimes. This step is needed in order to establish the robustness of the analogue system.
For that reason, we will now focus on cosmological inflation. 

\subsection{Analogue inflation}

As already mentioned, the inflationary era is a time of exponential expansion of the Universe. The time of that expansion can be expressed as the number of times, $N$, the Universe increases by a factor of $e$. $N$ is called the number of e-folds and can be defined for each mode in our system as:
\be
 N = \ln\left(\frac{a_k(t_\mathrm{f})}{a_k(t_\mathrm{i})}\right) = \frac{1}{2} \ln\left(  \frac{ \sigma k^{2} +[\rho]g_\mathrm{eff}(t_\mathrm{i}) }{\sigma k^{2} +[\rho] g_\mathrm{eff}(t_\mathrm{f})}  \right),
\ee
where $t_i$ and $t_f$ are the initial and final time of the expansion.
Since it is possible to invert the sign of the effective gravity, we can see that it is, in principle, possible to mimic an inflationary period with an arbitrary large number of e-folds. 
During this period, space expands so quickly that fluctuations are stretched beyond the characteristic scale of the expansion (known as the Hubble horizon), at which point they stop propagating in time (the modes are said to be frozen).
By this process small initial perturbations get amplified and converted to density fluctuations, eventually leading to the observed large-scale structure of our Universe.
We demonstrate below that our system is suitable to observe mode freezing and the amplification process, characteristic of an inflationary epoch.

\begin{figure}[!t]	
\centering
  \includegraphics[scale=1.3]{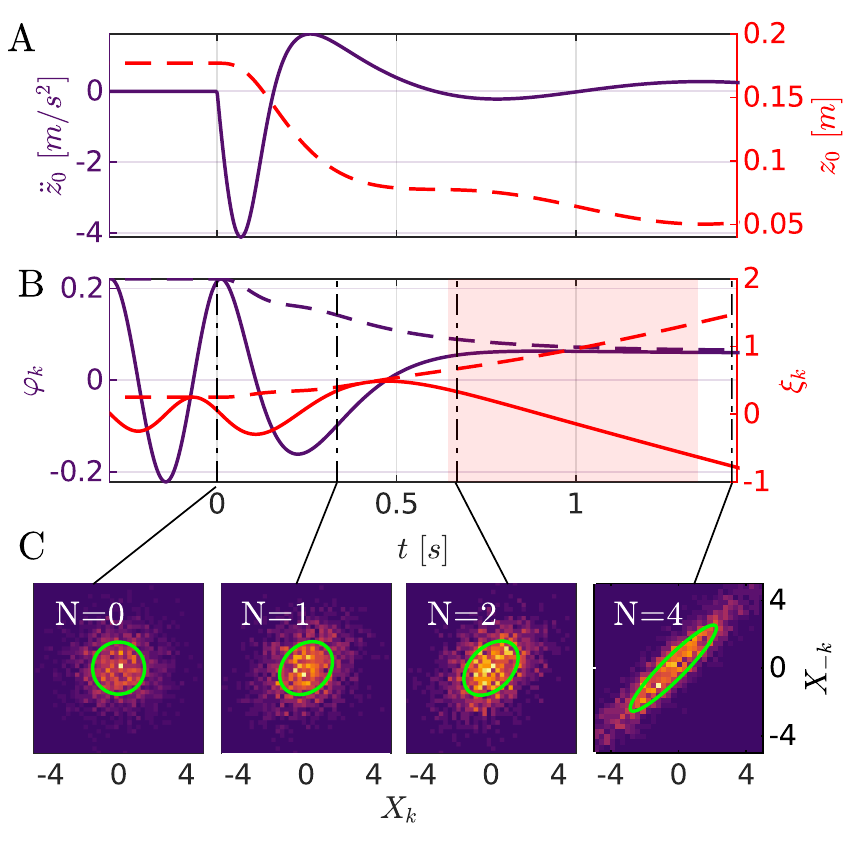}
  \caption{
Panel (A) shows the evolution of the vertical position $z_0(t)$ (dashed red) and acceleration $\ddot{z}_0(t)$ (solid purple) of the vessel in the magnet corresponding to an exponentially inflating analogue universe with Hubble parameter $H=\dot{a}/a = 3 \, s^{-1}$. 
Panel (B) depicts the solution of Eq.~\eqref{EoM_Rainbow}. 
The solid (dashed) line is the real part (absolute value) of the velocity potential $\phi_k$ (purple) and of the height field $\xi_k$ (red). 
The black dash-dotted lines represent different number of e-folds $N=0,1,2$ and $4$ for the mode. 
The shaded region indicates where the mode is outside the Hubble horizon. 
Panel C depicts the maximal two-mode squeezing of the system projected onto the instantaneous eigenbasis at the times (equivalently number of e-folds) indicated.
The solid line indicates the theoretical full width at half maximum for each distribution. 
The intensity plots represent the probability of a given measurement of $X_k$ and $X_{-k}$ after 2000 simulated experimental runs.
  } \label{inflation_pic}
\end{figure}

We tune the scale factor $a_k(t)$ to be exponential in the linear dispersion limit.
The path through the magnet $z_0(t)$, as well as the acceleration to produce this expansion are shown in Fig.~\ref{inflation_pic} (note the magnitude of $\ddot{z}_0$, as compared to $g_\mathrm{eff}$ in Fig.~\ref{g_eff_pic}). 

To understand the behaviour of the field, it is common to introduce the auxiliary field $\mathcal{X}_k = a_k \, \varphi_k$ for which the wave equation (Eq.~\eqref{EoM_Rainbow}) takes the form of a time-dependent harmonic oscillator:
\be
\ddot{\mathcal{X}_k} + \Omega_k^2(t) \mathcal{X}_k = 0,
\ee \label{EoMX}
with frequency
\begin{equation} \label{Omegak}
 \Omega_k^2(t) = \frac{k^2}{a_k^2} - \frac{\ddot{a}_k}{a_k} ~.
\end{equation}

Horizon crossing occurs at $\Omega_k^2 = 0$, separating the oscillating solution dominated by the first term on the right hand side of Eq.~\eqref{Omegak} from the exponentially growing\,/\,decaying solutions at late times, dominated by the time-independent second term. 
The essence of inflationary dynamics is fully captured for late times after a mode has crossed the horizon, since the dynamics of the physical field $\varphi_k$ freezes and becomes trivial, obeying a constant solution in time. 

Panel (B) of Fig.~\ref{inflation_pic} depicts the numerical solution of the field equation (Eq.~\eqref{wave_equation}), including the full non-linear dispersion relation given in Eq.~\eqref{full_dispersion}, for the longest wavelength of our system.
For $t<0$ the system is evolved in flat space, reducing Eq.~\eqref{wave_equation} to a simple harmonic oscillator. 
At $t=0$ the system begins to expand, leading to an oscillatory, damped time evolution of the field.
Upon crossing the effective horizon, the field rapidly approaches a nearly constant solution. 

The full model (including dispersive effects) exhibits minor differences compared to a completely frozen field solution. These are caused by the surface tension. 
The surface tension adds a small time-dependence to $\ddot{a}_k/a_k$ and in turn a further slow evolution of the field outside the horizon. 
Apart from a different effective expansion experienced by high momentum modes, dispersive effects lead to the mode re-entering the Hubble horizon, exhibiting oscillatory behaviour at later times. 
Nevertheless, our system is able to simulate mode freezing.

We further present the evolution of the surface height $\xi_k = a_k^2 \dot{\varphi}_k$. 
The surface height exhibits a growing, non-oscillatory solution after horizon crossing.

As already mentioned, an inflationary evolution leads to wave amplification, which is the analogue of cosmological particle production. This amplification comes in the form of a creation of perfectly correlated pairs of waves with opposite momenta. This is in analogy with the squeezed-state formulation of inflation \cite{Grishchuk:1990cm,Albrecht:1992kf}.
Mode amplification and correlations are the result of the rapid effective expansion of our analogue universe connecting two flat regions of spacetime $a(t_\mathrm{i}) \to a(t_\mathrm{f})$~\cite{BIR82}. 
To each flat regions of space-time we can associate a well-defined set of normalised functions $(f_i,f_i^*)$ and $(f_f,f_f^*)$ which form a basis for the solution space of Eq.~\eqref{EoMX} (Note that the functions $f_i$ and $f_f$ depend on $k$ but we omitted this dependence to simplify notation). The functions are normalised using the Wronskian:
\be
\langle f \, ; \, g \rangle \equiv \mathrm{i} \left( f^{*} \partial_t g - (\partial_t f^{*}) g \right).
\ee
We can then expand the field as:
\be
\mathcal{X}_k = b_k f_i(t) + b_{-k} f_i^{*}(t) = d_k f_f(t) + d_{-k} f_f^{*}(t).
\ee
Since $(f_i,f_i^*)$ and $(f_f,f_f^*)$ are two bases of the same space, we can relate them via a Bogoliubov transformation:
\be 
f_i = \alpha_k f_f + \beta_k f_f^*. 
\ee
and the conservation of the Wronskian implies that:
\be 
|\alpha_k|^2 - |\beta_k|^2 = 1.
\ee 
We can therefore relate the amplitudes of the final modes $(d_k,d_{-k})$ to the one of the initial modes $(b_k,b_{-k})$:
\be
d_k = \alpha_k b_k + \beta_k^* b_{-k}
\ee
Therefore, the mode intensity after the expansion can be expressed in terms of the initial mode intensity:
\be \label{amplification}
 \langle {d_k}^* d_k \rangle = \left( 2 |\beta_k|^2 + 1 \right) \langle {b_k}^* b_k \rangle ~,
\ee
where $\langle ... \rangle$ denotes the arithmetic mean taken over sufficiently many measurements, allowing us to assume $\langle {b_k}^* b_{k} \rangle = \langle {b_{-k}}^* b_{-k} \rangle$.
We also note that the evolution causes modes of opposite momenta to mix:
\be \label{correlator}
\langle {d_k}^* d_{-k} \rangle = 2\alpha_k \beta_k^* \langle {b_k}^* b_k \rangle ~,
\ee
To illustrate this point, we define:
\be 
X_{k} = \left( f_f(t)d_{k} + f_f(t)^*d_{k}^* \right)/|f_f|,
\ee
and the conjugate quadratures $\left( X_k \pm X_{-k} \right)$. 
We find from Eq.~\eqref{correlator} that the variance of a quadrature is lowered below its initial value while the variance of its conjugate is raised so that their product remains constant.
Note that while the freezing of the modes is sensitive to $\ddot{a}_k / a_k$ being nearly constant, creation of counter-propagating pairs is generic (see e.g.\ \cite{time_mirror}).

\begin{figure}[!t]
\centering	
  \includegraphics{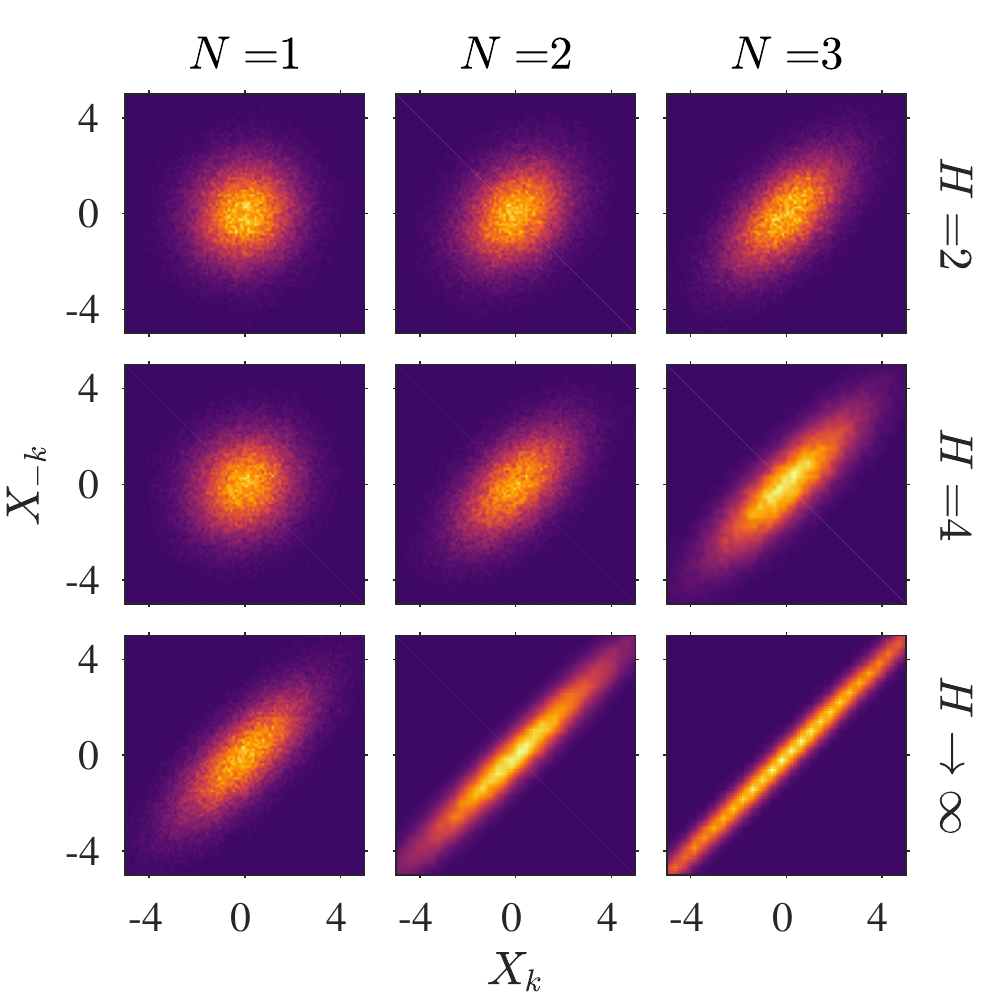}
  \caption{Two-mode squeezing of the system projected onto the instantaneous eigenbasis at the times (equivalently number of e-folds) indicated for various values of the Hubble parameter $H$.} \label{squeeze_pic}
\end{figure}

Panel (C) of Fig.~\ref{inflation_pic} shows the build-up of correlations for an initial uncorrelated Gaussian state for varying numbers of e-folds $N$ during the expansion.
The maximum squeezing is seen to increase with $N$. A similar behaviour is found by increasing the effective Hubble parameter $H$, as depicted in Fig.~\ref{squeeze_pic}. This means that the faster and the longer the expansion is, the stronger the correlations will be.

\section{Summary}

In this appendix we have presented a novel system capable of mimicking field propagation during a cosmological expansion. 
The system consists in a paramagnetic fluid immersed in a strong gradient magnetic field. 
The effect of the magnetic field is to change the gravitational force felt by the fluid. This results in a change of the propagation speed of the waves at the interface of the fluids. 
In the linear regime, surface waves can be described as a massless scalar field in a spatially-flat Universe. 
The analogue universe is characterized by a scale factor $a$, such that $a^2$ is inversely proportional to the effective gravitational constant. 
By tuning the external magnetic field, we can mimic various scale factors describing different cosmological scenarios.
Our system also allows for a sign change of the effective gravity. 
In the cosmological picture, this corresponds to a signature change of the space-time metric.
Therefore, our system offers a promising way to study experimentally this regime in which we still lack a clear theoretical understanding.

Before exploring this extreme regime, we have studied and presented here a more conventional cosmological evolution, namely inflation. 
We have shown that our system is capable to mimic field evolution during an inflationary epoch.
In particular, it is possible to observe the mode freezing as well as the mode amplification and the building up of correlations due to the cosmological evolution. 
Performing this experiment will allow us to test the effective field theory needed to describe this system and to establish the robustness of our analogue simulator for future investigation of cosmological scenarios.

The main challenge to overcome to perform such experiment is to have an accurate enough detection method to observe the perturbation while staying in the linear regime.

\chapter{List of acronyms}
\begin{tabular}{cp{0.6\textwidth}}
  ABHS & Analogue Black hole Spectroscopy \\
  AFT & Angular Fourier Transform \\
  AG & Analogue Gravity \\
  BEC & Bose-Einstein Condensate \\
  BH & Black Hole \\
  BHS & Black Hole Spectroscopy \\
  DBT & Draining Bathtub \\
  FCD & Fast-Chequerboard Demodulation \\
  FLRW & Friedmann-Lema\^{i}tre-Robertson-Walker \\
  GR & General Relativity \\
  HR & Hawking Radiation \\
  LR & Light-Ring \\
  MSE & Mean Square Error \\
  ODE & Ordinary Differential Equation \\
  PDE & Partial Differential Equation \\
  PIV & Particle Imaging Velocimetry \\
  PSD & Power Spectral Density \\ 
  PV & Potential Vorticity \\
  QG & Quantum Gravity \\
  QNM & Quasi-Normal Mode \\
  QT & Quantum Theory \\
  WKB & Wentzel–Kramers–Brillouin
\end{tabular}


\bibliographystyle{utphys}
\bibliography{bibliography_PhD}

\end{document}